\documentclass[%
superscriptaddress,
 amsmath,amssymb,
 aps, physrev,
prb,
longbibliography
]{revtex4-2}

\usepackage[letterpaper, margin=1in]{geometry}

\usepackage{graphicx,bm}
\usepackage{subcaption}
\usepackage{color}
\usepackage{enumitem}
\usepackage{amsmath}
\usepackage{hyperref}
\usepackage{float}
\hypersetup{
    colorlinks=true,
    linkcolor=blue,
    filecolor=green,
    urlcolor=cyan,
    citecolor=magenta,
}
\usepackage{comment}
\usepackage{listings}
\usepackage{xspace}

\usepackage{float}
\makeatletter
\let\newfloat\newfloat@ltx
\makeatother
\usepackage{algorithm}
\usepackage{algpseudocode}

\newcommand{\sumdd}[3]{\ensuremath{\sum_{\substack{#1 \\ #2}}^{#3}}}

\newcommand{\eqn}[1]{Eq.~(\ref{#1})}
\newcommand{\eqns}[1]{Eqns.~(\ref{#1})}
\newcommand{\eqnx}[1]{(\ref{#1})}

\newcommand{\sect}[1]{Section~\ref{#1}}

\newcommand{\app}[1]{Appendix~\ref{#1}}
\newcommand{\fig}[1]{Fig.~\ref{#1}}
\newcommand{\figs}[1]{Figs.~\ref{#1}}
\newcommand{\figx}[1]{\ref{#1}}
\newcommand{\alg}[1]{Algorithm~\ref{#1}}

\newcommand{\be} { \ensuremath{\boldsymbol {e}}}
\newcommand{\bc} { \ensuremath{\boldsymbol {c}}}

\newcommand{\cB} { \ensuremath{\mathcal {B}}}
\newcommand{\cC} { \ensuremath{\mathcal {C}}}

\newcommand{\cS} { \ensuremath{\mathcal {S}}}

\newcommand{\dirac} { \ensuremath{\delta_{\rm D}} }

\newcommand{\comp}{Fe$_{(1-x)/2}$Ni$_{(1-x)/2}$Cr$_x$\xspace}

\newcommand{\nixal}{Ni$_{x}$Al$_{1-x}$\xspace}

\begin{document}

\preprint{APS/123-QED}

\title{\textbf{Local Order Average-Atom Interatomic Potentials}
}%

\author{Chloe A. Zeller}
\affiliation{
Department of Aerospace Engineering and Mechanics,
University of Minnesota, Minneapolis, MN 55455, USA}
\author{Ronald E. Miller}
\affiliation{
Department of Mechanical and Aerospace Engineering,
Carleton University, Ottawa, Canada}
\author{Ellad B. Tadmor}
\email{Contact author: tadmor@umn.edu}
\affiliation{
Department of Aerospace Engineering and Mechanics,
University of Minnesota, Minneapolis, MN 55455, USA}

% figures path
\graphicspath{{figures/}{./}}

\date{\today}% It is always \today, today,

\begin{abstract}
This article describes an extension to the effective Average-Atom (AA) method for random alloys to account for local ordering (short-range order) effects by utilizing information from partial radial distribution functions. The new Local Order Average-Atom (LOAA) method is rigorously derived based on statistical mechanics arguments and validated for non-stoichiometric binary 2D hexagonal crystals and 3D FeNiCr and NiAl alloys whose ground state is obtained through Monte Carlo sampling. Material properties for these alloys computed from atomistic simulations using standard interatomic potentials (IPs) exhibit a strong dependence on local ordering that is captured by simulations with effective LOAA IPs, but not the original AA method. The advantage of LOAA is that it requires smaller system sizes to achieve statistically converged results and therefore enables the simulation of complex materials, such as high-entropy alloys, at a fraction of the computational cost of standard IPs.
\end{abstract}

\maketitle

\section{Introduction}
\label{sec:intro}
The average-atom (AA) approach to modeling random alloys is based on an effective medium approximation in which the species of atoms are taken to be independent and consistent with their concentrations in a crystalline alloy. In this method, each atom is identical and typeless, with an effective atomic interatomic potential (IP) in which interactions between atoms are computed as an average over interactions that account for species identities weighted by their probability of occurrence assuming an ideal random alloy. The AA approach was originally introduced by Smith and Was \cite{smith:was:1989}, Ackland and Vitek \cite{ackland:vitek:2009}, and Najafabadi et al.\ \cite{najafabadi:wang:1991}, and revived by
Curtin and coworkers \cite{varvenne:luque:2016, nohring:curtin:2016}. The method has primarily been applied to alloy systems to study the effect of species concentrations on material properties \cite{jian:xie:2020, zhao:xiong:2021, cheng:yuan:2021, seoane:farkas:2022, liang:goodelman:2023, fang:li:2024}. A recent application area for AA is the class of high-entropy alloys (HEAs) \cite{George2019-pl} or multi-principal element alloys \cite{Yeh2004-dy, Shittu2024-wb}. These are alloys that typically contain five or more species of approximately equal concentrations and which are of interest due to their high strength and high fracture toughness relative to traditional alloys \cite{yeh:chen:2004, gludovatz:hohenwarter:2014, zhang:liu:2024}. Such systems require large simulation cells in order to correctly capture species fractional occupancy and hence benefit from AA, which can provide converged results with much smaller system sizes.

While the AA approach provides an excellent model for the ideal case of random alloys, in practice all materials exhibit some level of ``local ordering'' (LO) (also called ``short-range order''), i.e.\ a preferential arrangement making certain species combinations at certain distances more or less favorable than the ideal random case \cite{Sheriff2024-jk, Islam2025}. Bragg and Williams \cite{bragg:williams:1934} first noticed the effects of temperature on atomic arrangements in the 1930s. Since then, it has become clear that LO exists in alloys, and that it can have important effects on material properties \cite{mcmanus:1965}. For example, in the brittle-to-ductile  transition, specific LO arrangements of atoms can directly result in increased stiffness while maintaining ductility \cite{Wang2023-qy}. Liquid alloys are another example where LO becomes important. During the cooling process of an alloy, chemical short-range order plays a critical role in both the nucleation mechanism and the specific phases and microstructures formed \cite{Ruppersberg1975, Liang2020-ql, Greer2016, Simonet1998-rm}. More generally, ordering in liquid alloys affects the electrical, thermodynamic, and structural properties \cite{Saboungi1990}. LO is also very important for HEAs, as it can dramatically change a material's strength, hardness, ductility, and defect evolution \cite{zhang:zhao:2020, chen:aitken:2021, zhao:2021}. Understanding LO in HEAs and other alloys is therefore important for predicting their response to changes in stress and temperature for real-world applications.

Here we extend the AA formalism to explicitly account for LO effects by including information from partial radial distribution functions (RDFs) in the definition of the effective IP. The new \emph{local order average-atom (LOAA)} method is as efficient as AA, but correctly captures the effect of LO on alloy properties. In LOAA, the effective interaction between atoms is weighted based on the probabilities of observing pairs of species at a given distance as described by the partial RDFs, as opposed to ideal random alloy probabilities. We validate the approach by studying the elastic properties of a two-dimensional (2D) binary hexagonal crystal with Lennard-Jones (LJ) interactions, a three-dimensional (3D) \comp alloy with embedded-atom method (EAM) interactions, and surface energies of \nixal with EAM interactions. We compare the accuracy of LOAA with AA, and with ``true species'' (TS) computations, calculations using standard IPs with the actual atomic species identities, averaged over many realizations of the system. We find that LOAA provides a more accurate estimate for the TS case than AA with no increase in computational cost except for the initial one-time generation of the partial RDFs. We observe that while the AA method can only accurately predict properties such as lattice constants, energies, and elastic constants for an ideal random alloy, the LOAA formulation correctly predicts these properties with increasing LO.

The remaining sections of this paper are organized as follows. \sect{sec:loaa} introduces the basic definitions and derivation of the LOAA method. \sect{sec:results} includes results for a 2D binary system where the effect of LO is explored by tuning the LJ parameters, and 3D alloy systems (FeNiCr and NiAl) where the effect of LO on mechanical properties and surface energies is studied. \sect{sec:conclude} concludes with a summary and a discussion of possible extensions to the LOAA method. The appendices contain additional derivations and results referred to in the main sections of the paper.

\section{Local Order Average-Atom Potentials}
\label{sec:loaa}

\subsection{Preliminaries}
\label{sec:random}
Consider an infinite multilattice crystal (i.e.\ a lattice with a basis) with species $A$, $B$, \dots
appearing in concentrations $c_A$, $c_B$, \dots, such that
\begin{equation}
\lim_{N\to\infty} \frac{1}{N} \sum_{i=1}^N s_i^X = c_X, \quad
x=A,B,\dots,
\label{eq:cxconstraint}
\end{equation}
where $N$ is the number of atoms in the crystal, and
\begin{equation}
s_i^X =
\begin{cases}
1 & \text{if atom $i$ is of species $X$}, \\
0 & \text{otherwise}.
\end{cases}
\label{eq:siX}
\end{equation}

We define the ensemble average over all possible species arrangements
$\cS$ satisfying the constraints in \eqn{eq:cxconstraint}, for a given
\emph{geometric configuration} $\cC$ specifying the positions of the atoms, as
\begin{equation}
\langle f(i,j,\dots); \cC \rangle_{\cS}
= \lim_{M\to\infty} \frac{1}{M} \sum_{m=1}^M
f(i,j,\dots; \cS_m),
\label{eq:specave}
\end{equation}
where $f(i,j,\dots; \cS_m)$ is some function related to atoms $i,j,\dots$
that may depend on their species, $\cS_m$ is a particular species
arrangement (realization) on configuration $\cC$, and $M$ is the size of the ensemble
taken to be infinite since the system size is infinite.

\eqn{eq:cxconstraint} is the concentration constraint for a single
species arrangement for the specified configuration $\cC$.
This constraint implies that the probability
$p(s_i^X=1 \,|\, \cC)$ that atom $i$ is of type $X$ given geometric
configuration $\cC$ is $c_X$. This is evident if one considers
drawing atom $i$ out of the pool of all possible realizations where the
fraction of atoms of species $X$ is $c_X$.
The expectation (average value) of $s_i^X$ follows by applying
\eqn{eq:specave}:
\begin{equation}
\langle s_i^X; \cC \rangle_{\cS} = 1 \times c_X + 0 \times (1-c_X) = c_X.
\label{eq:aves}
\end{equation}

\subsection{Pairwise Interactions}
The energy of a particular species arrangement on the crystal assuming pairwise
interactions is
\begin{equation}
E = \frac{1}{2} \sum_i \sum_{j\ne i} \sum_X \sum_Y V_{ij}^{XY} s_i^X s_j^Y,
\end{equation}
where $i$ and $j$ run over all atoms, and $X$ and $Y$ over all species,
and $V_{ij}^{XY}$ is the energy in a bond connecting pair $ij$
of species $XY$,
\begin{equation}
V_{ij}^{XY} = V_{XY}(r_{ij}),
\end{equation}
in which $V_{XY}(r)$ is the pair potential function for
species $X$ and $Y$, and $r_{ij}$ is the distance between atoms
$i$ and $j$.

First, we consider the \emph{random alloy case} where atomic species are randomly assigned to atoms consistent with a specific set of concentrations.  The expectation of the energy follows by applying the ensemble average in \eqn{eq:specave}:
\begin{align}
\langle E; \cC \rangle_{\cS}
&= \left\langle
\frac{1}{2} \sum_i \sum_{j\ne i} \sum_X \sum_Y V_{ij}^{XY} s_i^X s_j^Y; \cC
\right\rangle_{\cS} \nonumber \\
&= \frac{1}{2} \sum_i \sum_{j\ne i} \sum_X \sum_Y V_{ij}^{XY}
\left\langle s_i^X s_j^Y; \cC \right\rangle_{\cS} \nonumber \\
&= \frac{1}{2} \sum_i \sum_{j\ne i} \sum_X \sum_Y V_{ij}^{XY}
\left\langle s_i^X; \cC \right\rangle_{\cS}
\left\langle s_j^Y; \cC \right\rangle_{\cS} \nonumber \\
&= \frac{1}{2} \sum_i \sum_{j\ne i} \sum_X \sum_Y V_{ij}^{XY} c_X c_Y,
\label{eq:aveE}
\end{align}
where in passing from the second to the third line we assume that
the variables $s_i^X$ are independent random variables, i.e.\
that the species of different atoms are independent \footnote{A
necessary (but not sufficient) condition in assuming the variables $s_i^X$ are independent random variables is that the system
is infinite. As a demonstration of this requirement,
consider a crystal with just two atoms ($N=2$)
with species $A$ and $B$, so that the concentrations are $c_A=c_B=0.5$.
According to the reasoning in \eqn{eq:aveE}, the probability that both
atoms are of species $A$ is $c_Xc_Y=0.25$, but this is impossible
since there is only one atom of species $A$.
The discrepancy is because in a finite crystal, occupation of sites
is coupled. In this case, if one atom is of type $A$, the other
can only be of type $B$, so the probability of it being of type $A$
is zero, not $c_A$. Even in an infinite system, it is possible that there
is a correlation between atoms due to LO of species.},
and in passing from the third to fourth line,
\eqn{eq:aves} is used.

In \eqn{eq:aveE}, we recover the AA formalism in which the
expectation of the energy of the system is expressed in terms of an effective pair potential between typeless atoms (i.e.\ with no specified species):
\begin{equation}
\label{eq:E_AA}
\langle E \rangle_{\cS}
= \frac{1}{2} \sum_i \sum_{j\ne i} V_{\rm eff}(r_{ij}),
\end{equation}
where
\begin{equation}
V_{\rm eff}(r)
= \sum_X \sum_Y V_{XY}(r) c_X c_Y.
\label{eq:Veff}
\end{equation}
 In general,
\eqn{eq:Veff} has to be evaluated explicitly, making this approach
somewhat more expensive computationally compared to the underlying pair potentials \footnote{For a system containing $n$ species,
there are $n(n+1)/2$ function evaluations (accounting for symmetry).  Pair potentials with a finite cutoff scale linearly with the number of atoms, $N$.  Thus, the AA formulation scales as $O(n^2N)$.}.  However, as it is more general (and more efficient) to store pair potentials on a discrete grid that is interpolated on-the-fly at runtime, it is usually possible to pre-compute $V_{\rm eff}$ and therefore recover the same speed and scaling of the original pair potential model.  The key benefit of the AA approach is that better statistics are
obtained than in a calculation of a single realization of a finite-sized random alloy.  One is effectively calculating the average behavior over all such possible realizations.

In the \emph{local order average-atom (LOAA) method} introduced here, we relax the assumption in \eqn{eq:aveE}
that the variables $s_i^X$ are independent.  In general,
\begin{align}
\left\langle s_i^X s_j^Y; \cC \right\rangle_{\cS}
&= (1 \times 1) \, p(s_i^X=1,s_j^Y=1 \,|\, \cC) \nonumber \\
&+ (1 \times 0) \, p(s_i^X=1,s_j^Y=0 \,|\, \cC) \nonumber \\
&+ (0 \times 1) \, p(s_i^X=0,s_j^Y=1 \,|\, \cC) \nonumber \\
&+ (0 \times 0) \, p(s_i^X=0,s_j^Y=0 \,|\, \cC) \nonumber \\
&= p(s_i^X=1,s_j^Y=1 \,|\, \cC),
\label{eq:sxsy}
\end{align}
where $p(s_i^X=1,s_j^Y=1 \,|\, \cC)$ is the probability that atom $i$ is of
species $X$ and atom $j$ is of species $Y$ given the geometric configuration $\cC$.
In the special case that the species assignments are independent, then
\begin{equation}
p(s_i^X=1,s_j^Y=1 \,|\, \cC)=c_X c_Y,
\label{eq:pind}
\end{equation}
so that from \eqn{eq:sxsy},
\begin{equation}
\left\langle s_i^X s_j^Y; \cC \right\rangle_{\cS} = c_X c_Y,
\end{equation}
and \eqn{eq:aveE} is recovered. Otherwise the probability in \eqn{eq:sxsy} must
be known. To proceed, based on objectivity requirements, we postulate that the
probability does not depend on the identities of the atoms, but only on their
species and on the distance between them \footnote{In general, the probability in \eqn{eq:sxsy} can depend not just on the two atoms forming the bond,
but on distances to and between surrounding atoms and their species. However,
given the pair potential interactions considered in this section, we postulate
a simple pairwise form.}. We assume the following functional form:
\begin{equation}
p(s_i^X=1,s_j^Y=1 \,|\, \cC)
= c_X c_Y \Phi_{XY}(r_{ij}; \cC).
\label{eq:sxsypostulate}
\end{equation}
The function $\Phi_{XY}(r; \cC)$ characterizes the probability of finding
a bond of length $r$ connecting species $X$ and $Y$ in geometric configuration
$\cC$ relative to a random alloy where the species are independent.
$\Phi_{XY}(r; \cC)$ has the following properties:
\begin{enumerate}
\item The dependence of $\Phi_{XY}$ on $\cC$ affects the functional form.
For example if $\cC$ is a perfectly crystalline configuration, then atoms
are only found at an infinite set of discrete distances (the neighbor
shell radii). Therefore $\Phi_{XY}(r; \cC)=0$ for all distances not in $\cC$.
\item $\Phi_{XY}(0; \cC) = 0$. This indicates that two atoms cannot occupy the same position.
\item For a system with LO, i.e.\ short-range order,
$\lim_{r\to\infty}\Phi_{XY}(r; \cC) = \delta(r; \cC)$,
where $\delta(r; \cC)=1$ for all distances present in $\cC$ and
zero otherwise.  This indicates that
for atoms in $\cC$ that are far apart, the probability that they are
of species $X$ and $Y$ reverts to the independent case \footnote{We expect
$\lim_{r\to\infty}\Phi_{XY}(r; \cC)$ to decay quickly with a dependence on the cutoff radius
$r_{\rm cut}^{XY}$ of the pair potential $V_{XY}(r)$.}.
\item $\Phi_{XY}(r; \cC)=\Phi_{YX}(r; \cC)$. Symmetry with respect to atomic species is a consequence of objectivity.
\end{enumerate}

Under conditions of thermodynamic equilibrium, the functional form of
$\Phi_{XY}(r; \cC)$ reflects the nature of bonding in the material and
the thermodynamic constraints imposed by a statistical
mechanics ensemble. We relate the probability of finding atoms of a given species at a given separation to the partial RDFs of the system
(see \app{app:pcf} for the derivation). The result is
\begin{equation}
p(s_i^X=1, s_j^Y=1 \,|\, \cC)
= \frac{c_X c_Y g_{XY}(r_{ij})}{\sum_A \sum_B c_A c_B g_{AB}(r_{ij})},
\label{eq:sxsyg}
\end{equation}
where $g_{XY}(r_{ij})$ is the partial RDF between species
$X$ and $Y$.  In general, the RDF is defined in terms
of a phase average evaluated over the relevant ensemble,
such as the canonical ensemble \cite{tuckerman:2010}.
This phase average is evaluated for \emph{one} specific species arrangement
satisfying the concentration constraint in \eqn{eq:cxconstraint}.
It can be obtained from a molecular simulation or experiment for a given
system at a given temperature $T$. Thus
the left-hand side of \eqn{eq:sxsyg} is the probability over all
species arrangements for a \emph{single} geometric configuration $\cC$,
whereas the right-hand side involves all geometric
configurations with a \emph{single} species arrangement $\cS$ (i.e.\ each atom is assigned a species such that \eqn{eq:cxconstraint}
is satisfied, and the phase average is over all possible arrangements of these atoms).
The requirement that this equality holds is in the nature of an
ergodic hypothesis. Since $\cC$ is a system of infinite
extent (see \sect{sec:random}), it is assumed that the probability
over all species arrangements for a single geometric configuration snapshot
at finite temperature (left-hand side of \eqn{eq:sxsyg}) is equal to
the probability over all geometric configurations for a specified set of
species (right-hand side of \eqn{eq:sxsyg}).

Comparing \eqn{eq:sxsyg} with \eqn{eq:sxsypostulate}, we have
\begin{equation}
\Phi_{XY}(r) =
\frac{g_{XY}(r)}{\sum_A \sum_B c_A c_B g_{AB}(r)},
\label{eq:phi}
\end{equation}
where we have dropped the explicit dependence of $\Phi_{XY}(r)$
on the geometric configuration $\cC$ since it is now assumed to apply to all
configurations (positions and species) consistent with the canonical ensemble
that defines the RDF.
For the case where the species are independent, all partial RDFs are equal (see \app{app:pcf}) and \eqn{eq:phi} gives
$\Phi_{XY}(r)=1$ as expected (see \eqn{eq:probxyT:indep}).

The derivation leading to \eqn{eq:sxsyg} applies to the configuration
at which the RDF is evaluated, which is typically
a reference equilibrium state. To define an effective LOAA potential,
we therefore apply \eqn{eq:sxsyg} in the reference configuration.
Thus referring to \eqns{eq:aveE} and \eqnx{eq:sxsy}, we have
\begin{equation}
V_{\rm eff}(r,R)
= \sum_X \sum_Y V_{XY}(r) G_{XY}(R),
\label{eq:Veffg}
\end{equation}
where $r$ and $R$ are the distance between a pair of atoms in the deformed
and reference configurations, respectively, and
\begin{equation}
G_{XY}(R)
= \frac{c_X c_Y g_{XY}(R)}{\sum_A \sum_B c_A c_B g_{AB}(R)}.
\label{eq:G}
\end{equation}
Thus if the partial RDFs for an alloy are known (either from experiment or from computation as shown in \sect{sec:results}),
it is possible to construct an effective potential that accounts for species ordering.
We make the following observations regarding $V_{\rm eff}(r,R)$:
\begin{enumerate}
\item The effective potential remains pairwise. It is a combination of the species-dependent
pair potentials weighted by the species probabilities determined in the reference state.
\item The effective potential can be used in arbitrary deformed states corresponding to the same reference configuration \footnote{Consider a reference configuration $\cC$ that is mapped by some deformation to a new deformed configuration $\cC'$. Given two atoms $i$ and $j$, the probability that $i$ is of species $X$ and $j$ is of species $Y$ is the same in $\cC$ and $\cC'$, i.e.\ $\left\langle s_i^X s_j^Y; \cC \right\rangle_{\cS} = \left\langle s_i^X s_j^Y; \cC' \right\rangle_{\cS}$. This is because for any arrangement of species sampled on $\cC$, atoms $i$ and $j$ have the same species in $\cC'$ since the applied deformation does not change species identities.}.
This statement applies at sufficiently low temperatures where species arrangements do not re-equilibrate through diffusion over the time scale being simulated. In other words, the species probabilities between pairs of atoms prior to deformation are ``frozen in''.
\item The effective potential is temperature-dependent since the RDFs are obtained within the canonical ensemble at
a temperature $T$, so that $G_{XY}=G_{XY}(R; T)$.
However, the temperature dependence of the radial distribution is
rather weak (see for example the numerical results in \cite{tuckerman:2010}),
so $V_{\rm eff}$ can be used for a range of ``close'' temperatures.
\item Unlike a standard IP, $V_{\rm eff}(r,R)$
requires knowledge of both deformed and reference coordinates:
\begin{equation}
\langle E \rangle_{\cS}
= \frac{1}{2} \sum_i \sum_{j\ne i} V_{\rm eff}(r_{ij},R_{ij}).
\end{equation}
In a simulation, this requires extra storage for the atom reference coordinates (or displacement
vectors) and modifications to molecular dynamics codes.
\item Computationally,
 LOAA is not significantly slower than the AA approach in \eqn{eq:Veff} since $G_{XY}(R)$
can be precomputed \footnote{In cases where the reference configuration is a perfect crystal, LOAA can be made virtually as efficient as the underlying pair potential by precomputing a discrete set of $V_{\rm eff}$ functions, one for each of the discrete pair distances in the lattice up to a conservative neighbor distance allowing for atomic motion.  Then during a run, the computation only requires a pair to be mapped to its reference configuration, the value of $R$ computed, and the appropriate $V_{\rm eff}$ function accessed.}. We note that if the only objective is to determine ground state and the partial RDFs are not known from experiments, then LOAA does not provide a computational benefit over brute force TS simulations since the RDF needs to be obtained. However once obtained, the effective LOAA potential can be used to perform a variety of computations on smaller systems efficiently.
\end{enumerate}

\subsection{EAM-style Embedding Function}
Next consider an energy with an EAM-style nonlinear embedding term,
\begin{equation}
E = \sum_i \sum_X U_X(\rho_i) s_i^X,
\end{equation}
where $U_X$ is the embedding function for an atom of type $X$,
and $\rho_i$ is the electron charge density at the position of atom $i$ given by
\begin{equation}
\rho_i = \sum_{j\ne i} \sum_Y \rho_{ij}^Y s_j^Y,
\label{eq:rhoi}
\end{equation}
where $\rho_{ij}^Y$ is the contribution from atom $j$ of species $Y$ to the charge density at atom $i$,
\begin{equation}
\rho_{ij}^Y=\rho_Y(r_{ij}),
\label{eq:rhoY}
\end{equation}
in which $\rho_Y(r)$ is the charge density function. The average embedding energy is
\begin{equation}
\langle E \rangle_{\cS}
= \left\langle \sum_i \sum_X U_X(\rho_i) s_i^X \right\rangle_{\cS}
= \sum_i \sum_X \left\langle U_X(\rho_i) s_i^X \right\rangle_{\cS}.
\label{eq:EAMexpectedE}
\end{equation}
An approximate expression for $\langle E \rangle_{\cS}$ can be obtained by expanding the embedding energy function $U_X$ about the average charge density $\bar\rho_i$:
\begin{equation}
U_X(\rho_i) = U_X(\bar\rho_i)
+ U'_X(\bar\rho_i) (\rho_i - \bar\rho_i)
+ \frac{1}{2} U''_X(\bar\rho_i)
(\rho_i - \bar\rho_i)^2 + \dots,
\label{eq:Uexp}
\end{equation}
where
\begin{equation}
\bar\rho_i
= \langle \rho_i \rangle_{\cS}
= \left\langle \sum_{j\ne i} \sum_Y \rho_{ij}^Y s_j^Y \right\rangle_{\cS}
= \sum_{j\ne i} \sum_Y \rho_{ij}^Y \langle s_j^Y \rangle_{\cS}
= \sum_{j\ne i} \sum_Y \rho_{ij}^Y c_Y.
\label{eq:barrho}
\end{equation}
We will also need 
\begin{equation}
\langle \rho_i s_i^X \rangle_{\cS} 
= \left\langle \sum_{j\ne i} \sum_Y \rho_{ij}^Y s_j^Y s_i^X \right\rangle_{\cS}
= \sum_{j\ne i} \sum_Y \rho_{ij}^Y \langle s_i^X s_j^Y \rangle_{\cS}
= \sum_{j\ne i} \sum_Y \rho_{ij}^Y G_{XY},
\end{equation}
in which $G_{XY}$ is defined in \eqn{eq:G}. 
Now using \eqn{eq:Uexp} and retaining terms to first order \footnote{Retaining terms to first order in \eqn{eq:Uexp} is warranted in the event that the second-order term is small compared with the leading terms (as for example shown for Cu-Ni alloys
in \cite{najafabadi:wang:1991}). Further, retaining the second-order term leads to three-point correlation effects of the form $\langle s_i^X s_j^Y s_k^Z \rangle_{\cS}$ beyond the scope of the LOAA derivation.}, the average embedding energy in \eqn{eq:EAMexpectedE} follows as
\begin{align}
\langle E \rangle_{\cS}
&= \sum_i \sum_X \left\langle
\left[
U_X(\bar\rho_i)
+ U_X'(\bar\rho_i) (\rho_i - \bar\rho_i)
\right]
s_i^X \right\rangle_{\cS} \nonumber \\
&= \sum_i \sum_X 
\left[
U_X(\bar\rho_i) \langle s_i^X \rangle_{\cS}
+ U'_X(\bar\rho_i) 
\left( \langle \rho_i s_i^X \rangle_{\cS} 
- \bar\rho_i \langle s_i^X \rangle_{\cS} \right) 
\right] \nonumber \\
&= \sum_i \sum_X 
\Big[
U_X(\bar\rho_i) c_X
+ U'_X(\bar\rho_i) 
\Big( \sum_{j\ne i} \sum_Y \rho_{ij}^Y G_{XY} 
- \bar\rho_i c_X \Big) 
\Big] \nonumber \\
&= \sum_i \sum_X 
\Big[
U_X(\bar\rho_i) c_X
+ U'_X(\bar\rho_i) 
\Big( \sum_{j\ne i} \sum_Y \rho_{ij}^Y 
( G_{XY} - c_X c_Y ) \Big) \Big].
\label{eq:Uave}
\end{align}
We rewrite this in terms of an effective embedding energy and a pairwise first-order correction:
\begin{equation}
\langle E \rangle_{\cS} = \sum_i U_{\rm eff}(\bar\rho_i)
+ \sum_i \sum_{j\ne i} dU_{\rm eff}(\bar\rho_i, r_{ij}, R_{ij}),
\label{eq:Ueffsum}
\end{equation}
where
\begin{align}
U_{\rm eff}(\rho) 
&= \sum_X U_X(\rho) c_X \label{eq:Ueff} \\
dU_{\rm eff}(\rho, r, R)
&= \sum_X \sum_Y
U'_X(\rho)\, \rho_Y(r) 
\left( G_{XY}(R) - c_X c_Y \right).
\label{eq:dUeff}
\end{align}
We note that for the random alloy case where the species assignments are independent, $G_{XY}(R)=c_X c_Y$, so that the correction term $dU_{\rm eff}$ cancels, reducing \eqn{eq:Ueffsum} to the AA expression for the average embedding energy.

Computationally, the expression in \eqn{eq:Ueffsum} can be computed at little additional cost compared to the original embedding energy formulation, assuming that the functional forms of $U_X(\rho)$, $U'_X(\rho)$, and $\rho_Y(r)$ are stored on a discrete grid of points and interpolated on-the-fly at run time.  If so, the sum over species $Y$ in \eqn{eq:barrho} and $X$ in \eqn{eq:Ueff} can be pre-computed for a given set of species concentrations and stored on the same grids of $\rho$ and $r$ values. The calculation cost of \eqn{eq:dUeff} is comparable to that of $V_{\rm eff}$ in \eqn{eq:Veffg} and can be evaluated along with it saving time.

\section{Results}
\label{sec:results}

To explore the ability of the LOAA methodology to capture LO effects, we study two example problems: 1) a 2D binary alloy modeled using an LJ potential where the degree of LO can be tuned by the choice of LJ parameters to allow for systematic comparison with the AA approach; and 2) two 3D metallic alloy systems (FeNiCr and NiAl) modeled using EAM potentials where the effect of LO on mechanical properties and surface energies is explored.

In what follows, we refer to calculations using standard IPs with the actual atomic species as the ``true species'' (TS) case to differentiate with AA and LOAA simulations that employ an effective potential without species identities.

\subsection{2D Lennard-Jones Binary Alloy}
\label{sec:numerical}

As a first test case for LOAA, we consider a 2D binary alloy  modeled using an LJ potential. We study the effect of the LJ parameters on the short-range order seen within this alloy. The potential energy per atom and elastic constants of the 2D binary alloy are computed and compared for the TS, AA, and LOAA methods, examining the importance of LO effects.

\subsubsection{Problem Definition}
\label{sec:LJsetup}
Consider a 2D binary alloy of species $A$ and $B$,
with concentrations $c_A$ and $c_B$. We focus on  alloys that are majority $A$ with small amounts of $B$.
The atoms interact via a so-called Kob-Andersen potential that consists of a set of three LJ pair potentials with different parameters for each species combination \cite{kob:andersen:1994, kob:andersen:1995a, kob:andersen:1995b}:
\begin{equation}
V_{XY}(r) =
4\epsilon_{XY}\left[
\left(\frac{\sigma_{XY}}{r}\right)^{12} -
\left(\frac{\sigma_{XY}}{r}\right)^6
\right].
\label{eq:Vxy}
\end{equation}
Here $\sigma_{XY}$ and $\epsilon_{XY}$ are LJ parameters for interaction between species $X$ and $Y$ with units of length and energy, respectively, and $X, Y \in \{A,B\}$.
For simplicity, we set all $\sigma$ parameters to be the same, $\sigma_{AA}=\sigma_{BB}=\sigma_{AB}=\sigma$, and the potential is truncated at a cutoff radius $r_{\rm cut}$ for all interactions
($AA$, $BB$, $AB$) and shifted to have zero energy at the cutoff. Thus, we have
\begin{equation}
V_{XY}(r) = 4\epsilon_{XY} \varphi(r), \qquad
\varphi(r) = \hat\varphi(r) - \hat\varphi(r_{\rm cut}), \qquad
\hat\varphi(r)=
\left[
\left(\frac{\sigma}{r}\right)^{12} -
\left(\frac{\sigma}{r}\right)^6
\right].
\label{eq:Vxy:const:sig}
\end{equation}
In simulations, we take $\sigma=1$ and $r_{\rm cut}=2.0$. The ground state crystal structure is 2D hexagonal (triangular) with a nearest-neighbor distance $a_0$ that depends on the choice of the $\epsilon_{XY}$ values. For a crystal containing a single species, the result is $a_0=1.1159\sigma$.

\subsubsection{Local Ordering in a 2D LJ Binary Alloy}
\label{sec:LO_sec}
In general, the propensity to LO depends on the nature of bonding in the material. In the present case this is determined by the values of the LJ parameters defined in \sect{sec:LJsetup}. An analysis of the energetics of 2D binary hexagonal crystals for second-neighbor LJ interactions (given in \app{app:deriv_energy_phase_diag}) shows that the energy of a particular arrangement of $B$ atoms substituted into an $A$ lattice relative to the energy of a pure $A$ lattice  has the following general form:
\begin{equation}
\frac{\Delta E(a_0; \cS)}{\epsilon_{AA}} = 4C(\cS)\left[\left(\frac{\sigma}{a_0}\right)^{12} - \left(\frac{\sigma}{a_0}\right)^6\right] + 4D(\cS)\left[\left(\frac{\sigma}{\sqrt{3}a_0}\right)^{12} - \left(\frac{\sigma}{\sqrt{3}a_0}\right)^6\right].
\label{eq:ener_B_rel_allA}
\end{equation}
Here, $\cS$ is a species arrangement forming an LO pattern (defined by the number of $B$ atoms $N_B$, and the relative positions of the $B$ atoms on the $A$ lattice), $C(\cS)$ and $D(\cS)$ are constants that depend on the number of first- and second-neighbor bonds, respectively, for $AA$, $AB$, and $BB$ interactions, and on the LJ parameter ratios $\epsilon_{AB}/\epsilon_{AA}$ and $\epsilon_{BB}/\epsilon_{AA}$, and $a_0=1.1159$ as noted above (small relaxations that might occur in the vicinity of $B$ atoms are neglected).

\begin{algorithm}[t]
\centering
\caption{Calculation of the LO phase diagram for a 2D binary crystal with second-neighbor LJ interactions. (See text for discussion.)}\label{alg:bonding}
\begin{algorithmic}
\For{$\epsilon_{AB}/\epsilon_{AA} \in [0,5]$}
    \For{$\epsilon_{BB}/\epsilon_{AA} \in [0,5]$}
        \For{$N_B = 2, \dots, 7$}
            \State{Determine the set $\cB(N_B)$ of unique LO patterns for $N_B$}.
            \ForAll{$\cS \in \cB(N_B)$}
                \State Compute the coefficient $C(\cS)$ and $D(\cS)$ using \eqn{eq:C_and_D}.
                \State Compute the relative energy $\Delta E(a_0; \cS)/\epsilon_{AA}$ using \eqn{eq:ener_B_rel_allA}.
            \EndFor
        \EndFor
        \State Store the minimal relative energy and associated LO patterns $\cS$ identified $\forall \cS \in\ \cB(2)\dots\cB(7)$.
    \EndFor
\EndFor
\end{algorithmic}
\end{algorithm}

\begin{figure}
\centering
\begin{subfigure}{0.5\textwidth}
    \centering
    \includegraphics[width=\linewidth]{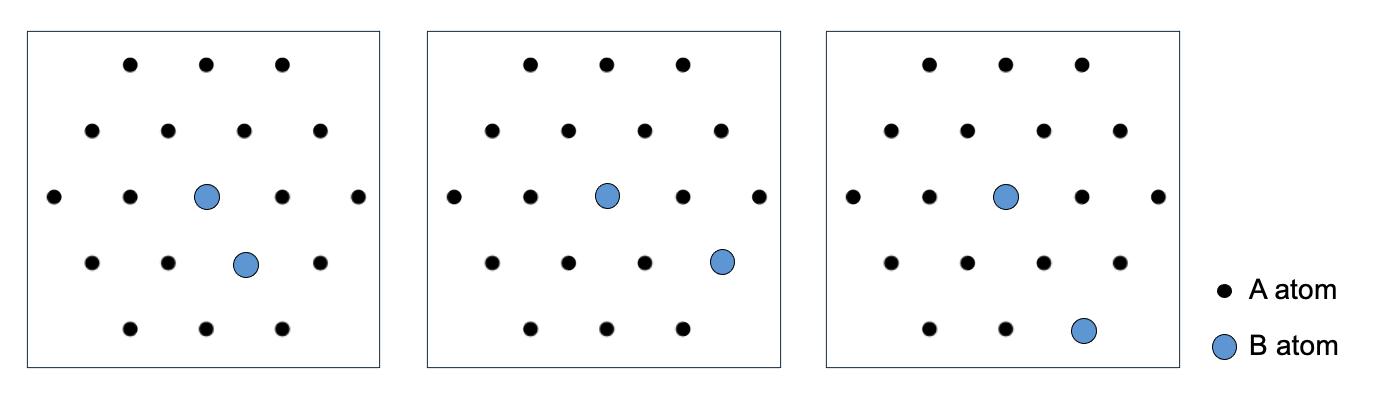}
    \caption{}
\end{subfigure}
\begin{subfigure}{0.7\textwidth}
    \centering
    \includegraphics[width=\linewidth]{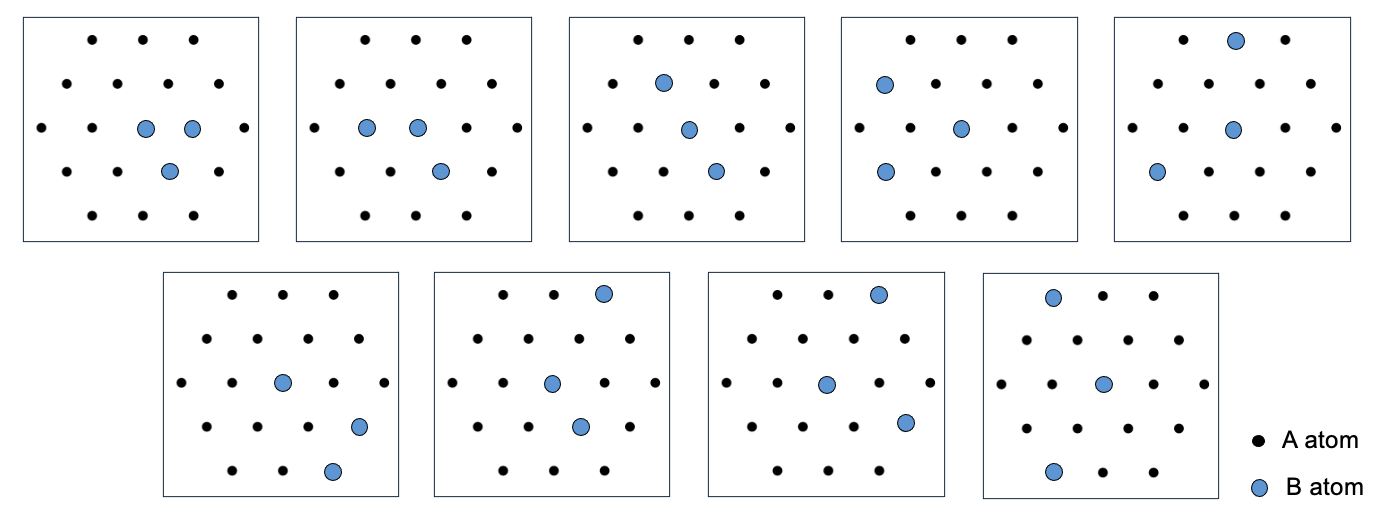}
    \caption{}
\end{subfigure}
\caption{Distinct LO patterns for (a) $N_B=2$, and (b) $N_B=3$ in a 2D hexgaonal lattice of type $A$ (black) atoms and type $B$ atoms (blue).}
\label{fig:permuations}
\end{figure}

To explore how the LJ parameters affect LO, we perform a systematic analysis of \eqn{eq:ener_B_rel_allA} identifying the ground state LO patterns across a range of LJ $\epsilon$ parameters. We follow the procedure in \alg{alg:bonding}. For each value $\epsilon_{AB}/\epsilon_{AA}$ and $\epsilon_{BB}/\epsilon_{AA}$, LO patterns are sought for $N_B=2$ to 7 $B$ atoms \footnote{The 2D LO patterns are constructed on a template of a hexagonal lattice centered on (0,0) with all points up to the third-neighbor distance from the origin, i.e.\ 18 neighboring sites plus the origin site for a total of 19. The upper limit of $N_B=7$ is the maximum number of $B$ atoms that can separate on the template beyond the second-neighbor distance (i.e.\ separate beyond interaction), which allows all possible configurations to be fully explored.}. For each value of $N_B$, all distinct LO pattern permutations (accounting for hexagonal symmetry) up to third-neighbors are generated \footnote{For a second-neighbor cutoff, it is necessary to consider atoms at third neighbor distance to include LO patterns in which one or more atoms are beyond the range of interaction.}. For $N_B=2$ there are three distinct LO patterns where the atoms are separated by the first-neighbor distance ($a_0$), second-neighbor distance ($\sqrt{3}a_0$), and third-neighbor distance ($2a_0$), for $N_B=3$ there are 9, and so on up to 70 configurations for $N_B=7$ (see \fig{fig:permuations}).
For each LO pattern $\cS$, the coefficients $C(\cS)$ and $D(\cS)$ in \eqn{eq:ener_B_rel_allA} are computed using \eqn{eq:C_and_D} based on the number of $AA$, $AB$, and $BB$ bonds in the configurations and specified values of $\epsilon_{AB}/\epsilon_{AA}$ and $\epsilon_{BB}/\epsilon_{AA}$.

\begin{figure}
\centering
\includegraphics[width=0.8\linewidth]{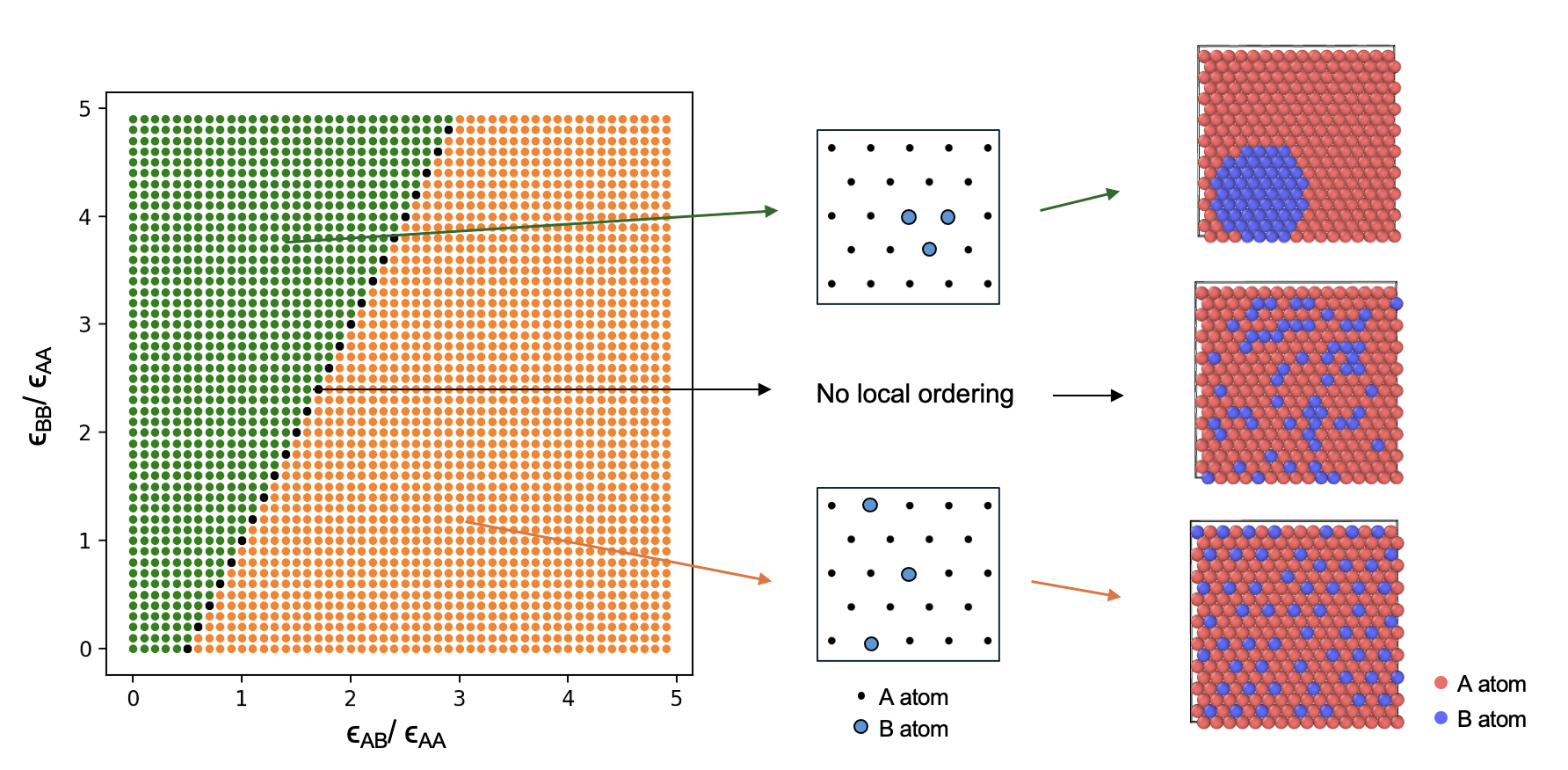}
\caption{Phase diagram and associated simulation results of 2D LJ binary alloy using different $\epsilon$ values. (Left) phase diagram with the green region corresponding to phase separation (PS), the orange region corresponding to a homogeneous solid solution (SS), and the black line corresponding to a random alloy; (middle) LO patterns corresponding to regions of the phase diagram; (right) MC simulation results for a 288 atom system with concentrations $c_A = 0.8$ and $c_B = 0.2$, corresponding to phase diagram regions. Visualized using OVITO \cite{ovito}.}
\label{fig:g:phase_diag_sim_results}
\end{figure}

\fig{fig:g:phase_diag_sim_results} (left) shows the resulting ``phase diagram'' where each point is colored according to the LO pattern corresponding to the ground state. Interestingly, there are only two distinct LO states in the ground state of this 2D binary system. Each point colored green favors $BB$ interactions leading to phase separation in the ground state in which $B$ atoms group together within the $A$ lattice (see \fig{fig:g:phase_diag_sim_results} (middle)). In the orange region, $BB$ interactions are unfavorable, and a solid solution is formed with $B$ atoms separated by a distance exceeding their range of interaction (see \fig{fig:g:phase_diag_sim_results} (middle)).
We will refer to these two regions as the phase separation (PS) and solid solution (SS) regions.
Along the black line separating the two regions, the energy of all LO patterns is the same and therefore the $B$ atoms arrange randomly in the ground state --- \emph{this is the ideal random alloy case}. We refer to this as the `no-LO line' --- see \app{app:noLO_line} for a more detailed discussion of the nature of this line.

\subsubsection{Monte Carlo Simulations of the Ground State}
\label{sec:ts:calcs}
To test the predictions of the phase diagram in \sect{sec:LO_sec}, we perform Monte Carlo (MC) simulations for the TS case (i.e.\ using the the species-dependent IP in \eqn{eq:Vxy}) to determine the ground state configuration for a 2D binary hexagonal crystal supercell. The simulations were performed for a crystal with concentrations $c_A = 0.8$ and $c_B = 0.2$. A convergence study  (see Section~I~A in the Supplementary Material (SM)) indicates that a supercell containing 2688 atoms is sufficient. We begin with a crystal of all $A$ atoms. The specified concentration of $B$ atoms is then distributed onto the lattice. Simulations are performed for three different choices of $\epsilon$ parameters corresponding to the two LO states and one no-LO state, and $\sigma=1$. MC swaps of atomic species are performed with a Metropolis acceptance criteria over a series of decreasing temperatures. 320,000 MC swaps are performed at each temperature, starting at $T=4.0$ (LJ temperature units) and halving each time until a final value of $T=1.953125\times 10^{-3}$ (see Section~I~B in the SM). No relaxation is performed following MC swaps, i.e.\ the ideal all-$A$ crystal structure is held fixed. Since relaxation effects are small, this procedure leads to a good approximation for the true ground state. All simulations are performed using the LAMMPS package \cite{lammps}.

The analysis in \sect{sec:LO_sec} identifies three distinct regions in $\epsilon$ space that control the resulting ground state crystal structure as seen in \fig{fig:g:phase_diag_sim_results} (right). $\epsilon$ values from the PS region of the phase diagram favor nearest-neighbor $B$ bonds, and as may be expected this leads to a ground state where the $B$ atoms clump together. Along the black line, no LO patterns are preferred, and the ground state shows a random pattern. Finally, in the SS region, bonding between $B$ atoms is unfavored, and in the ground state the $B$ atoms are spread apart in a solid solution. (Partial RDFs and Warren-Cowley parameters to further validate the distinct LO states are included in Section~I~C in the SM.)

\subsubsection{AA and LOAA Predictions for Ground State Energy}
\label{sec:energy:predictions}

As seen from the convergence analysis (Section~I~A in the SM), multiple realizations of large and computationally expensive MC simulations are necessary to obtain reliable statistical estimates for ground state properties like the cohesive energy (average energy per atom). In contrast, the AA and LOAA methods directly predict statistical averages with smaller supercells and without the need for repeated simulations \footnote{For LOAA a calculation of the partial RDFs is necessary, so no benefit is obtained unless multiple properties are computed.}.

The AA potential form is given in \eqns{eq:E_AA} and \eqnx{eq:Veff}. For the the special case of an LJ pair potential where the $\sigma$ parameter for all interactions are the same, the effective AA potential energy has the simple form
\begin{equation}
V_{\rm eff}(r) = 4\bar{\epsilon} \varphi(r),
\label{eq:veff:sh}
\end{equation}
where $\varphi(r)$ is defined in \eqn{eq:Vxy:const:sig} and
\begin{equation}
    \bar{\epsilon} = \epsilon_{AA} c_A^2 +2 \epsilon_{AB} c_A c_B + \epsilon_{BB} c_B^2.
\end{equation}
This can be used with the standard LJ implementation in LAMMPS.

The LOAA potential is defined in \eqns{eq:Veffg} and \eqnx{eq:G}. The partial RDFs used in \eqn{eq:G} are averaged over 10 different random seeds at 100 individual points across $\epsilon$ space using the TS simulations described in \sect{sec:ts:calcs}. To implement the LOAA formulation, we create a new pair style in LAMMPS. The LOAA potential is created by modifying the existing LJ pair style to include the additional $G_{XY}(R)$ term seen in \eqn{eq:Veffg}.

\begin{figure}
\centering
\includegraphics[width=0.5\linewidth]{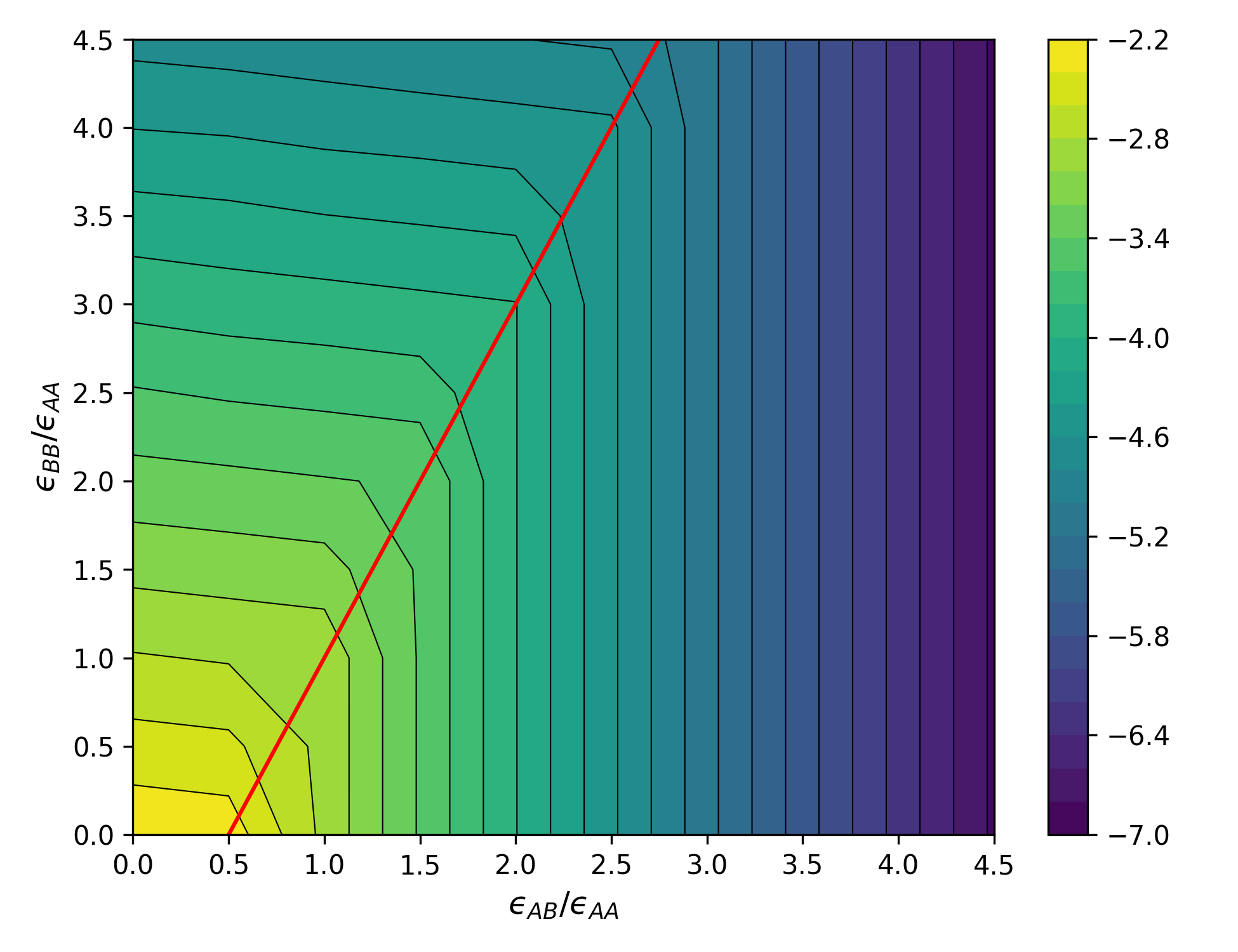}
\caption{TS cohesive energy contours for the 2D LJ binary alloy over $\epsilon$-space corresponding to the phase diagram. The no-LO line is shown in red.}
\label{fig:g:all_atom_energy_contour}
\end{figure}

The performance of the AA and LOAA methods is evaluated by comparing their predictions for the cohesive energy with the exact results obtained using TS simulations across $\epsilon$-space. The TS cohesive energy across $\epsilon$ space is visualized in the energy contour plot in \fig{fig:g:all_atom_energy_contour}. The energy is computed at 100 points across phase diagram space. At each point, results are averaged over 10 ground state realizations obtained via MC with different random seeds.
(See Section~I~D in the SM for an explanation of the form of this plot.)

\begin{figure}
\centering
\begin{subfigure}{0.48\textwidth}
\centering
\includegraphics[width=\linewidth]{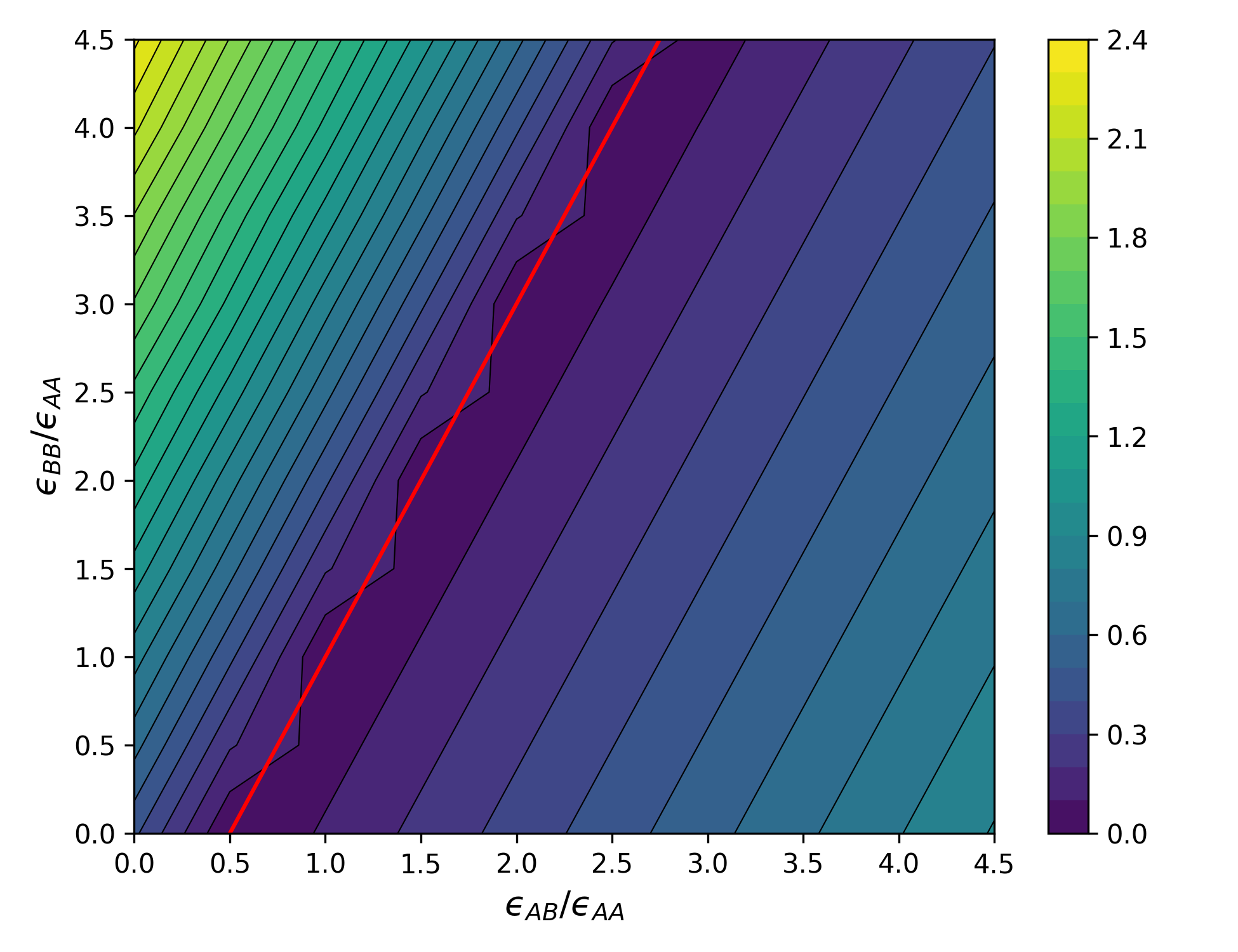}
\caption{}
\label{fig:g:AA_all_energy_dif_contour}
\end{subfigure}
\hspace{0.4cm}
\begin{subfigure}{0.48\textwidth}
\centering
\includegraphics[width=\linewidth]{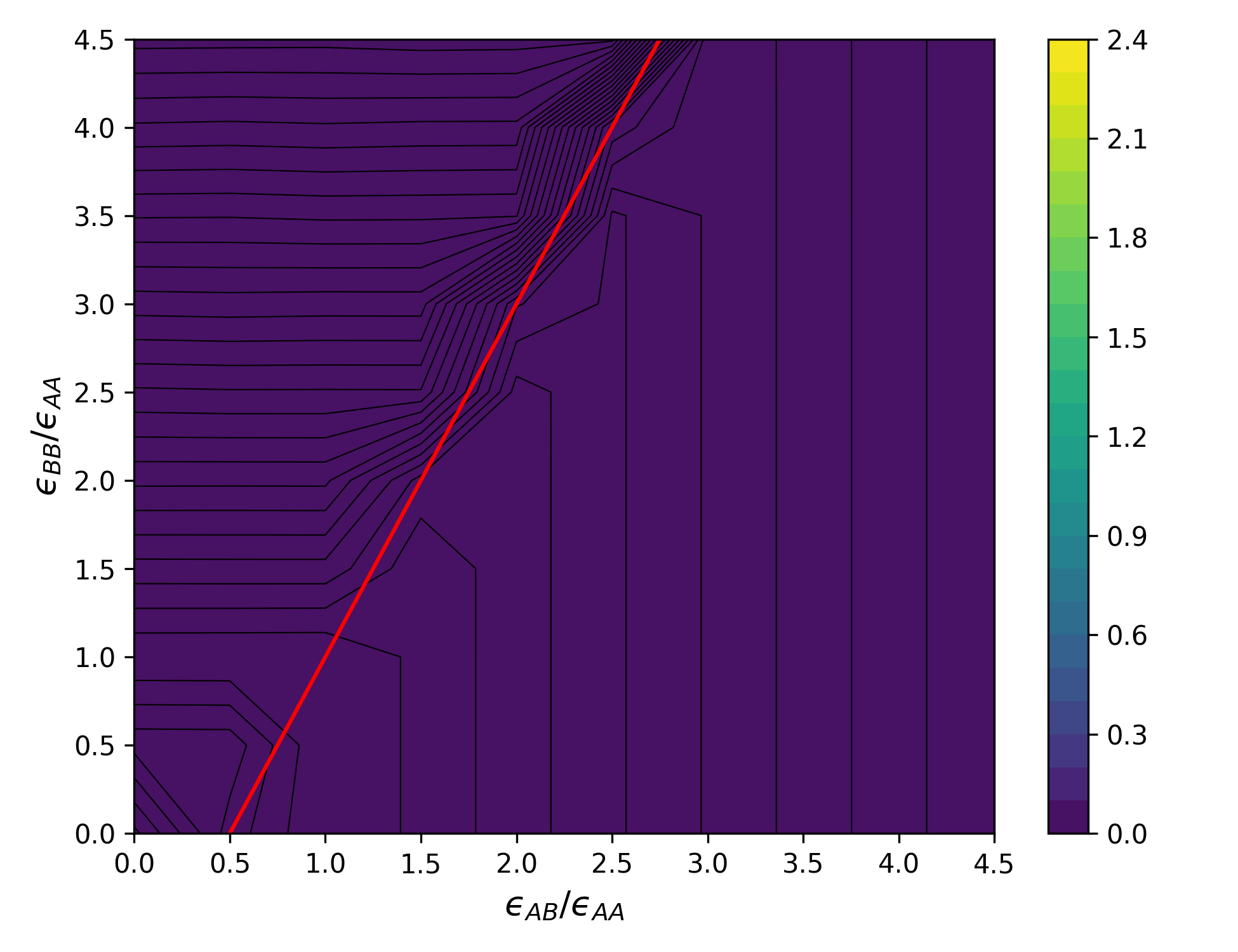}
\caption{}
\label{fig:g:LOAA_all_energy_dif_contour}
\end{subfigure}
\caption{Difference in the cohesive energy of the 2D LJ binary alloy between the (a) AA method, and (b) LOAA method, and the exact TS results plotted across $\epsilon$-space corresponding to the phase diagram. The no-LO line is shown in red.}
\label{fig:g:energy_dif_contour}
\end{figure}

The difference between the AA and TS results is presented in \fig{fig:g:AA_all_energy_dif_contour}. As expected, we see that the method is exact only in the case of no LO (shown as a red line in the plot). The error in the AA energy increases linearly with the distance from this line as shown by a theoretical analysis in \app{app:deltaE_LOAA_AA} (see also the discussion in Section~I~E in the SM). Comparing with the TS energy contour plot in \fig{fig:g:all_atom_energy_contour}, we find that the relative error in the AA energy prediction can be as large as 45\%.

The results for the LOAA method are presented in \fig{fig:g:LOAA_all_energy_dif_contour} using the same contour scale as for the AA method in \fig{fig:g:AA_all_energy_dif_contour}. The maximum error in energy for LOAA is about $6\times10^{-3}$, not discernible on this scale. This small error reflects finite size limitations of the TS simulation rather than the LOAA formalism, which is exact. Both the partial RDFs and the energy per atom computed from the TS simulations are obtained from 10 realizations of a 2688 atom system, both of which may not be fully converged (see Figs.~S2 and S3 in the SM).
As a result both the TS energy and the LOAA-predicted energy (which depends on the partial RDFs) have errors and do not match exactly.

\subsubsection{AA and LOAA Predictions for Elastic Constants}
To further validate the LOAA method, elastic constants are computed at various points in $\epsilon$-space using the TS, AA, and LOAA approaches. Similar to the ground state energy studies, we expect the AA method to be correct only for points on the no-LO line. In contrast, the LOAA method should yield accurate elastic constants for all $\epsilon$ values.

The elasticity tensor $\bc$ is computed using the expression for pair potentials from \cite[Section~11.5.2]{tadmor:miller:2011}:
\begin{equation}
c_{ijkl} = \frac{1}{2\Omega} \sum_{\substack{\alpha, \beta \\ \alpha \neq \beta}} \left[ V_{\text{eff}, \alpha \beta}''(r^{\alpha \mathring{\beta}}) -\frac{V_{\text{eff},\alpha \beta}'(r^{\alpha \mathring{\beta}})}{r^{\alpha \mathring{\beta}}} \right] \frac{ r_i^{\alpha \mathring{\beta}} r_j^{\alpha \mathring{\beta}} r_k^{\alpha \mathring{\beta}} r_l^{\alpha \mathring{\beta}} }{(r^{\alpha \mathring{\beta}})^2},
\label{eq:Cijkl}
\end{equation}
where $\Omega$ is the simulation cell volume, and the sum is over all atoms $\alpha$ and $\beta$ in the simulation cell, $\mathring\beta$ refers to the nearest periodic copy of atom $\beta$ assuming the minimum image convention, and $r^{\alpha \mathring{\beta}}$ is the distance between atoms $\alpha$ and $\mathring\beta$. Because the system is isotropic and 2D, the elasticity tensor has two independent elastic constants, which in Voigt notation are denoted $C_{11}$ and $C_{12}$.

\begin{figure}
\centering
\begin{subfigure}{0.45\textwidth}
\centering
\includegraphics[width=\linewidth]{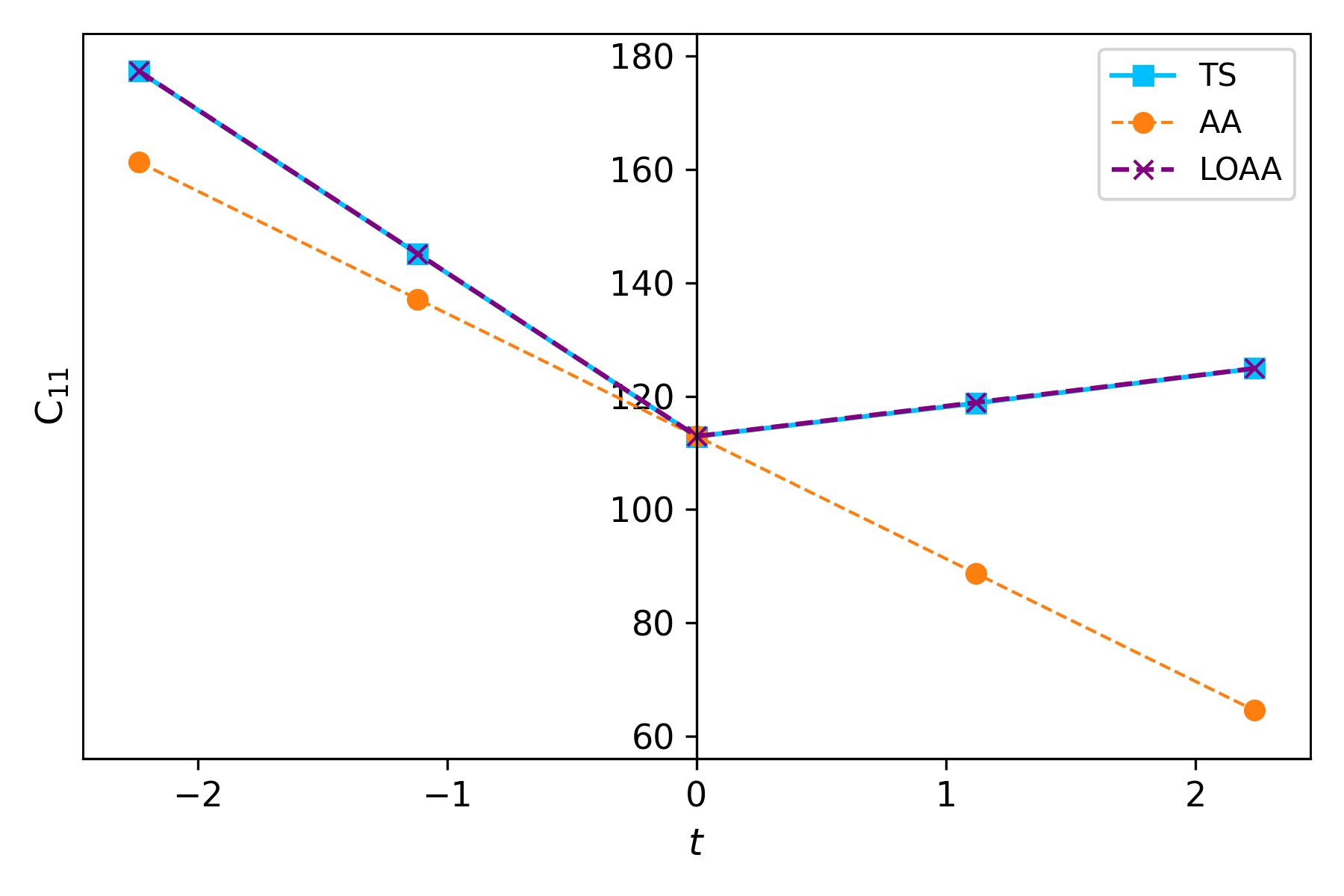}
\caption{}
\label{fig:g:C11}
\end{subfigure}
\hspace{0.5cm}
\begin{subfigure}{0.45\textwidth}
\centering
\includegraphics[width=\linewidth]{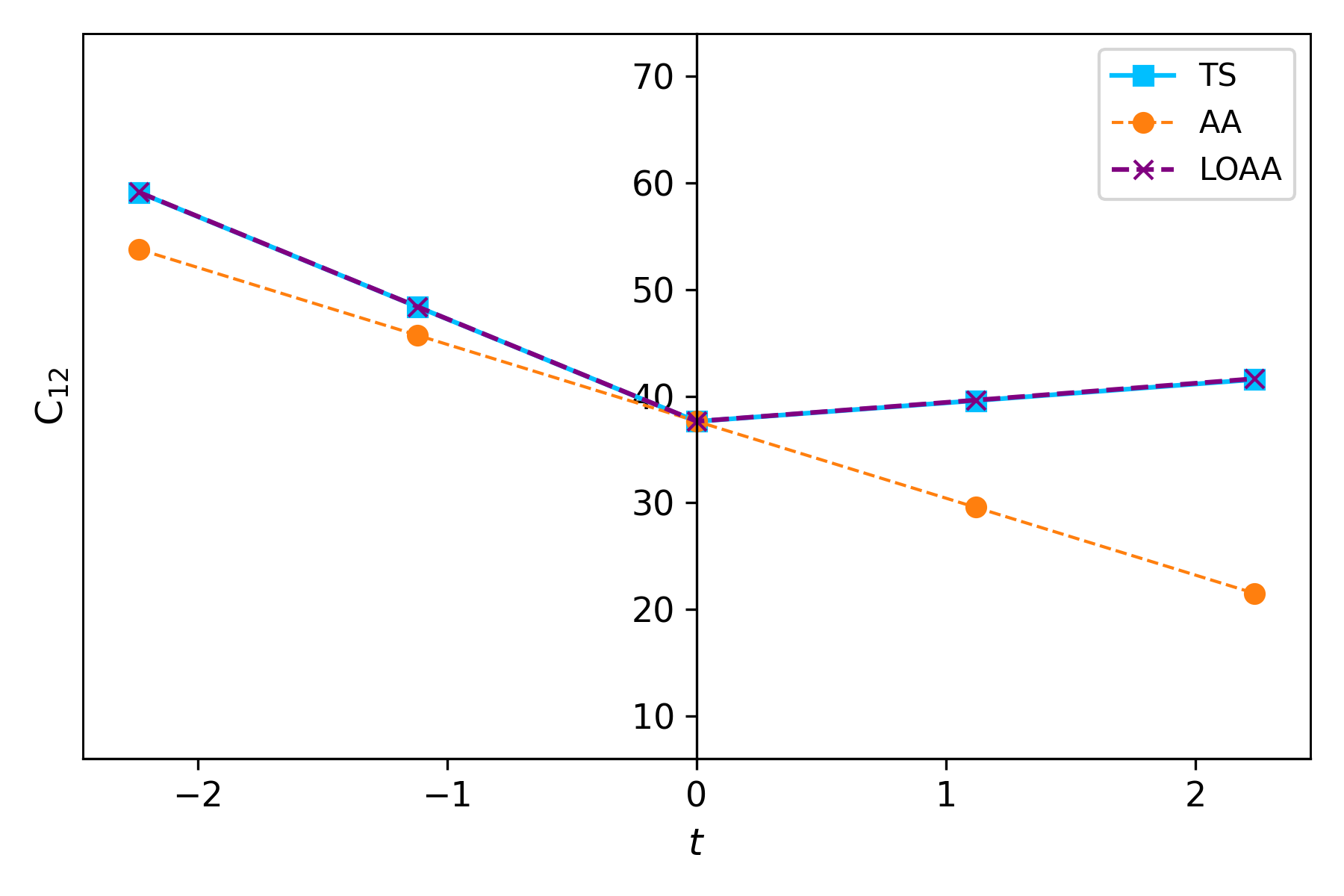}
\caption{}
\label{fig:g:C12}
\end{subfigure}
\caption{(a) $\text{C}_{11}$ and (b) $\text{C}_{12}$ as a function of perpendicular distance from the no-LO line, $t$, computed using the TS, AA, and LOAA methods for the 2D LJ binary system. Standard deviation error bars are plotted for TS.}
\label{fig:g:elastic_constants}
\end{figure}

Given the observation in \sect{sec:energy:predictions} that the error in AA prediction scales with the perpendicular distance $t$ from the no-LO line, we perform a similar analysis for the elastic constants. The results using each of the three methods (AA, LOAA, TS) are plotted in \figs{fig:g:C11} and \figx{fig:g:C12}. The plots are nearly identical at all points \textit{along} the no-LO line (i.e.\ at all values of $s$ (\fig{fig:change_of_var})). We see that LOAA predictions agree precisely with the TS ground truth, whereas the AA predictions diverge in proportion to the distance from the no-LO line. The effect is particularly noticeable to the right of the no-LO line ($t>0$) corresponding to the SS region where there is a change in slope that AA fails to capture. This change in behavior is explained by an analytical derivation of the AA and LOAA elastic constants expressions in \app{app:elastic_consts}. The slope of the elastic constants for the AA method depends only on the concentrations $c_A$ and $c_B$ and the crystal structure and so will be the same in the PS and SS regions as seen in \fig{fig:g:elastic_constants}. In contrast, the LOAA slope depends on the partial RDFs through the probability functions $\Phi_{AB}(r)$ and $\Phi_{BB}(r)$ (see \eqn{eq:phi}), which explains the change in slope across the no-LO line.

\subsection{3D EAM Alloys: FeNiCr and NiAl}
\label{sec:3deam}
As a further test of LOAA, we consider the more realistic case of 3D metallic alloys modeled via EAM potentials. First, we consider ordering effects on the lattice constant, cohesive energy, and elastic properties of \comp as described by the EAM potential of Bonny et al.\ \cite{Bonny_2011}. This  system was previously studied by Varvenne et al.\  \cite{varvenne:luque:2016} treating it as a random alloy and applying the AA method. We then consider surface energies of \nixal using the EAM potential of Purja Pun and Mishin \cite{pun:mishin:2009}. All 3D simulations are performed with LAMMPS and visualized using OVITO.

\subsubsection{Ordering Effects in FeNiCr}
In order to directly compare to the AA results of \cite{varvenne:luque:2016}, we consider alloys of the composition \linebreak \comp  for various Cr concentrations $x$.  As in the previous sections, we will denote the results of the full EAM description with atoms assigned their true species as ``TS''.  These can either be ``random'', in which case atoms are randomly assigned a species consistent with the concentrations, or ``ordered'', in which case the species of each lattice site is determined by MC as described below.  All TS results are shown for a system of $32{,}000$ atoms on $20\times20\times20$ units cells of an fcc crystal with periodic boundary conditions.

The ordered TS sample for each unique set of concentrations is determined by an on-lattice MC anneal of the $32{,}000$ atoms, initially assigned random species on an fcc lattice with the relaxed lattice constant determined by a separate minimization of the randomized species.  The MC annealing then proceeds to attempt $100{,}000$ species swaps at each of 6 progressively lower numerical temperatures ($1200K$, $1000K$, $800K$, $600K$, $400K$, $200K$).

\begin{figure}
\begin{subfigure}{0.45\textwidth}
\centering
\includegraphics[width=\linewidth]{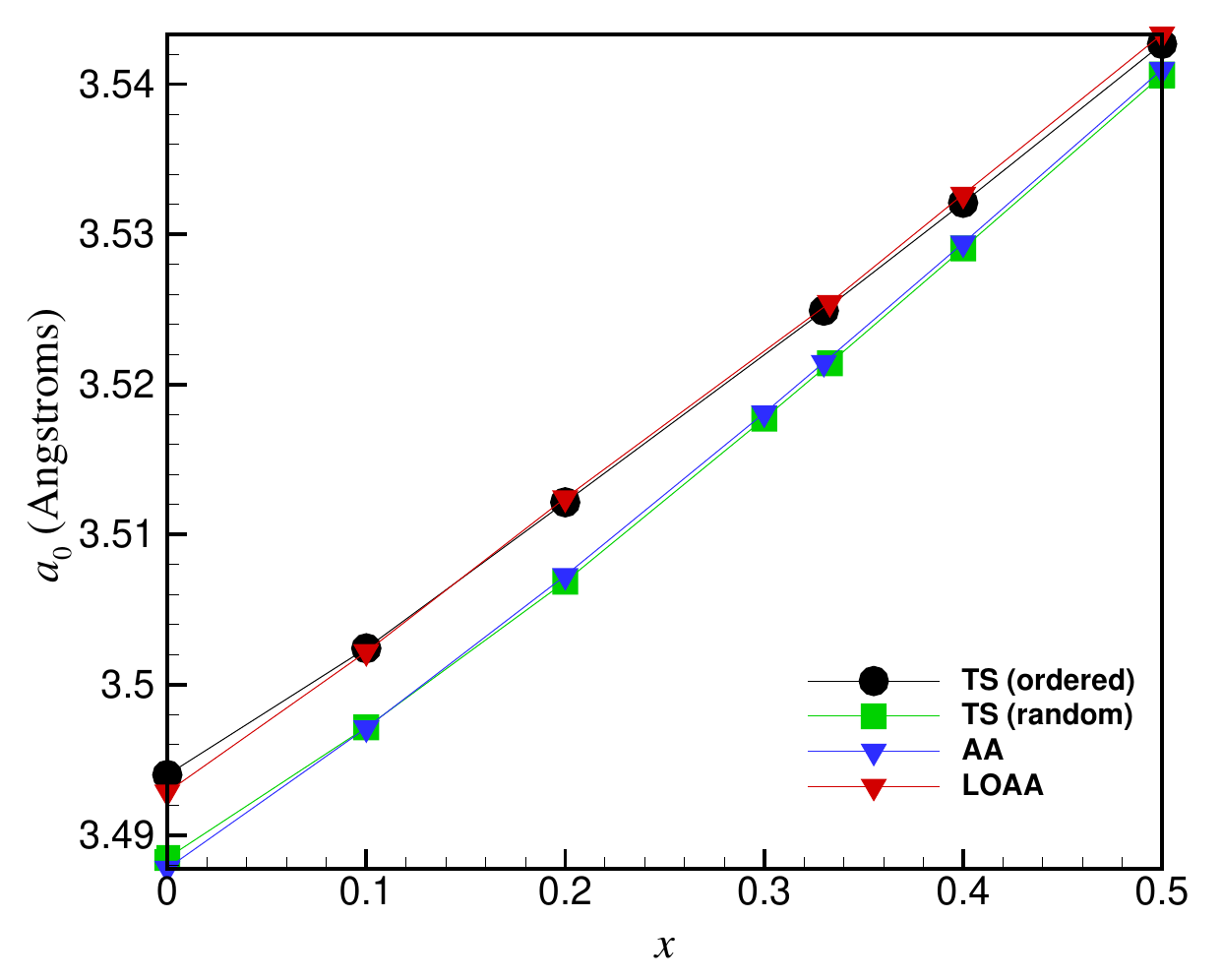}
\caption{}
\end{subfigure}
\hfill
\begin{subfigure}{0.45\textwidth}
\centering
\includegraphics[width=\linewidth]{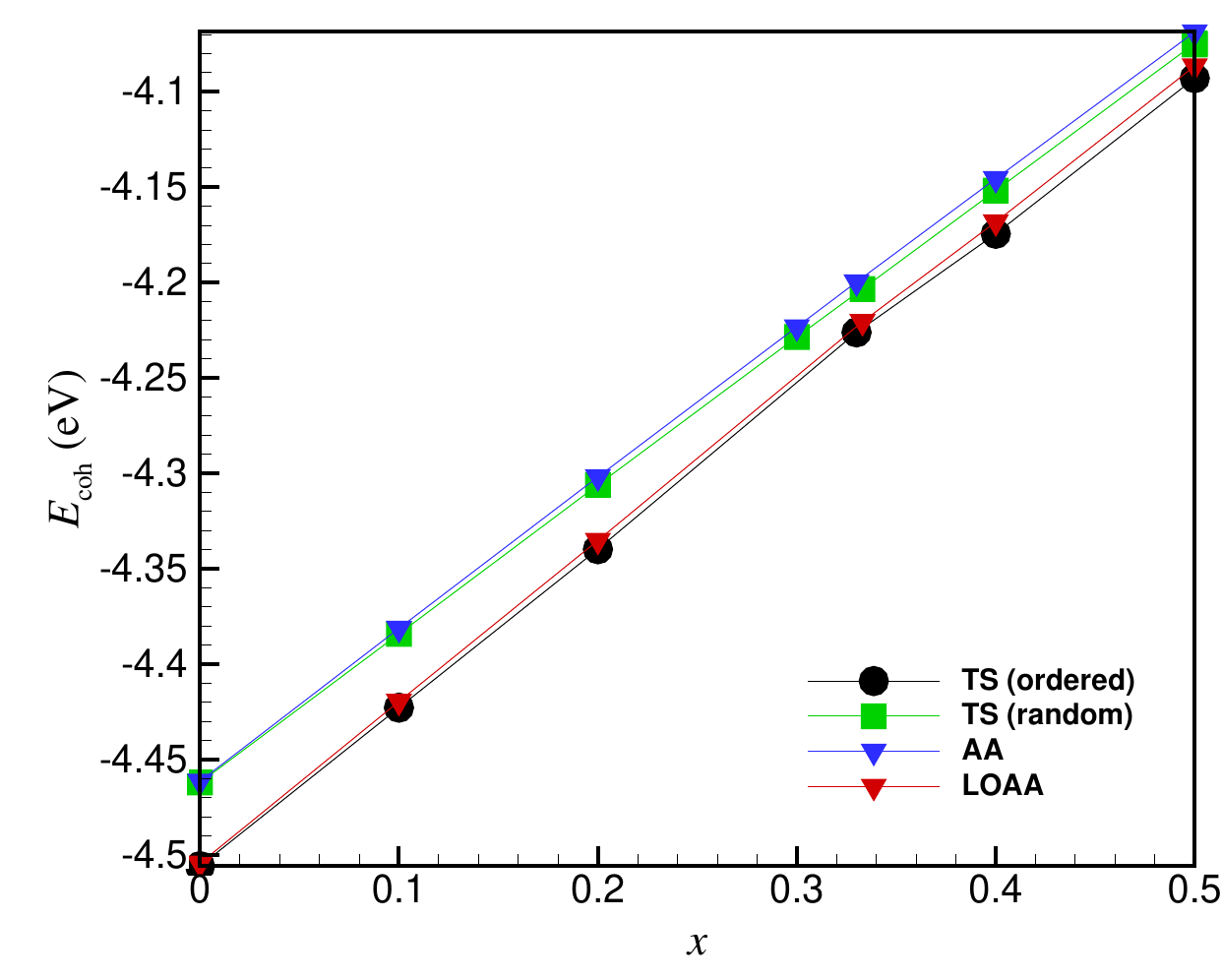}
\caption{}
\end{subfigure}
\vskip\baselineskip
\begin{subfigure}{0.45\textwidth}
\centering
\includegraphics[width=\linewidth]{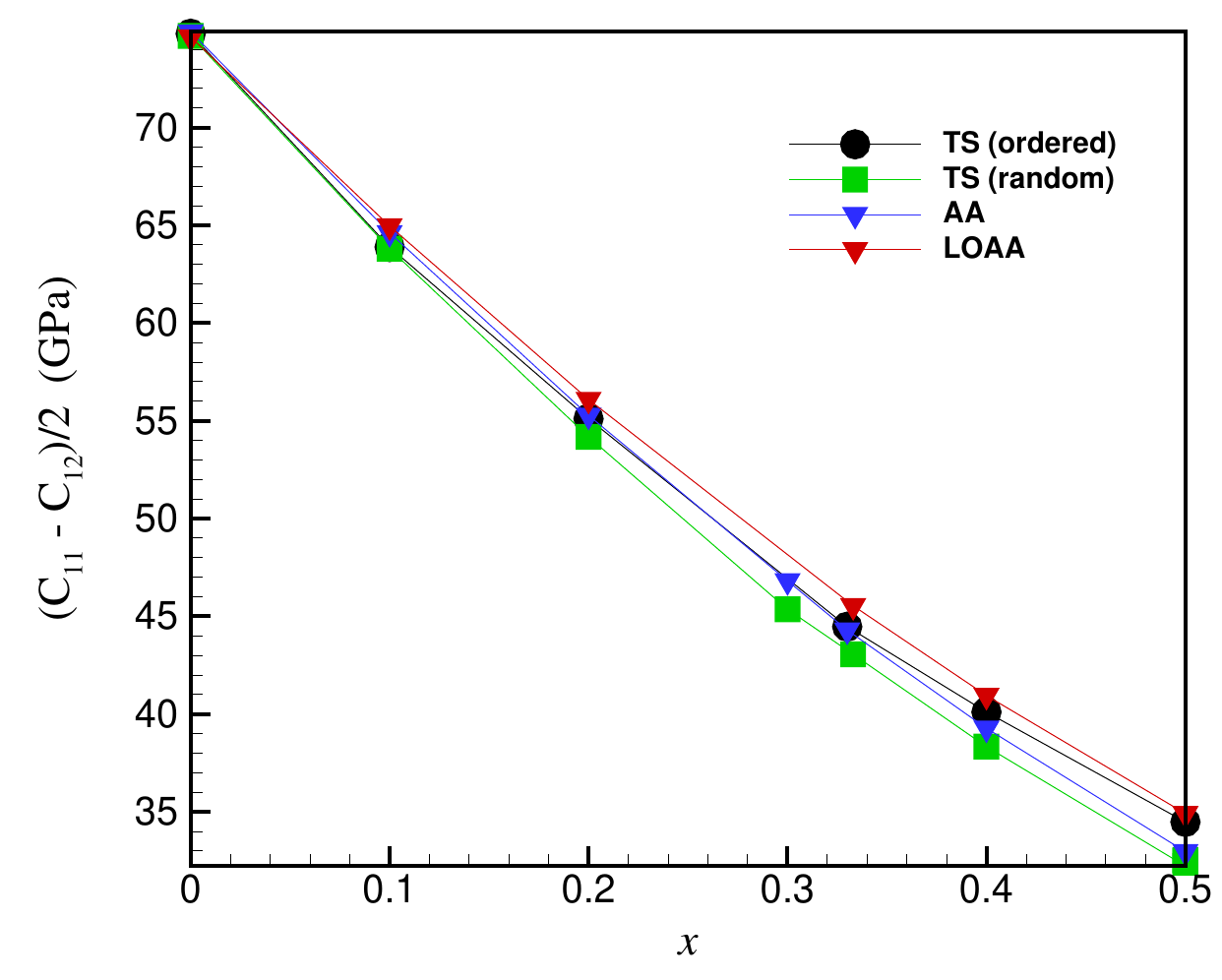}
\caption{}
\end{subfigure}
\hfill
\begin{subfigure}{0.45\textwidth}
\centering
\includegraphics[width=\linewidth]{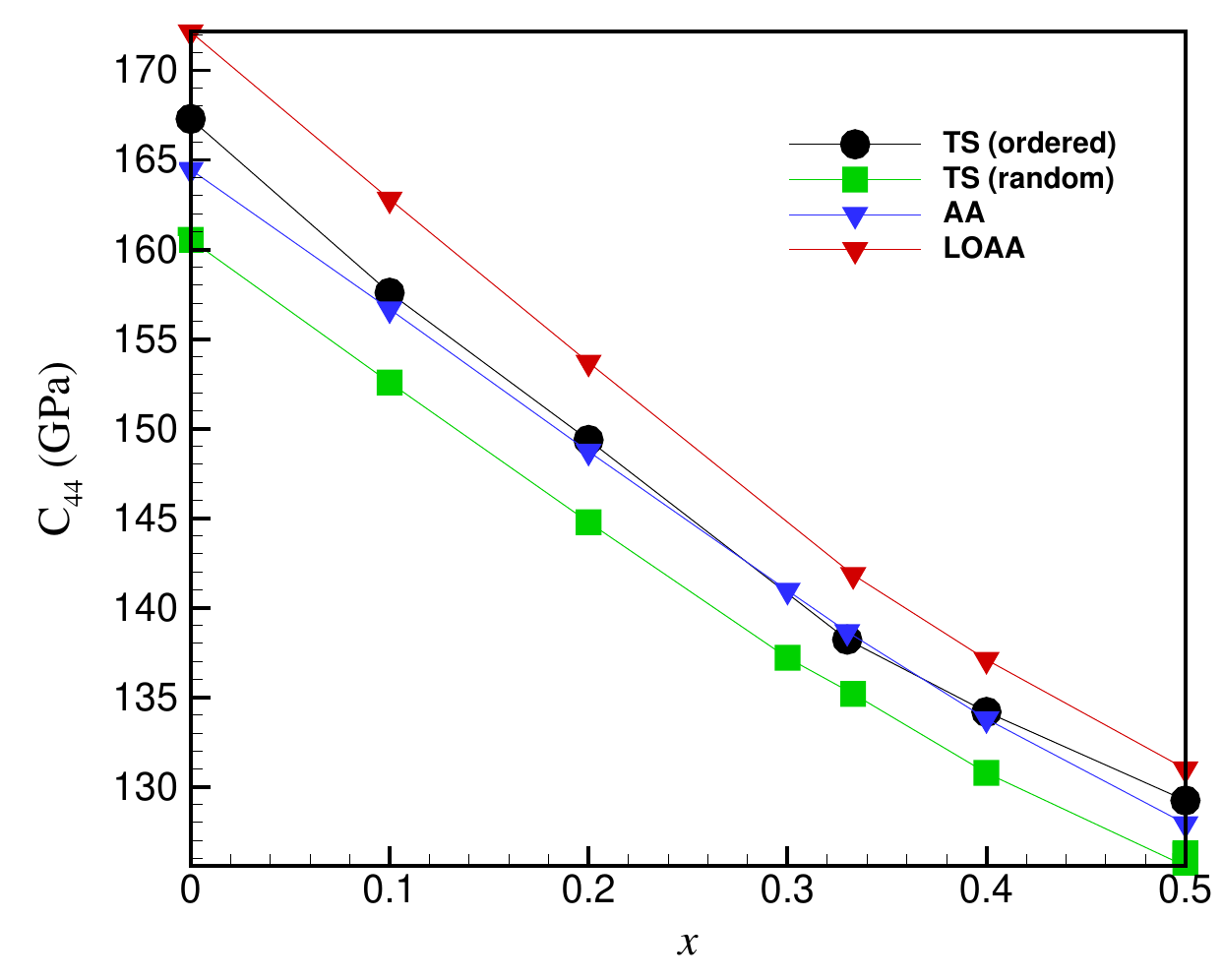}
\caption{}
\end{subfigure}
\caption{Comparison of TS, AA and LOAA calculations for 3D EAM \comp\ alloys. (a) Lattice constant, (b) cohesive energy, (c) and (d) elastic properties. TS quantities evaluated for relaxed structures.}
\label{fig:FeNiCr-properties}
\end{figure}

Results of the AA and LOAA formulations are compared to the TS results in \fig{fig:FeNiCr-properties}.  Note that both the AA and LOAA simulations can be performed on minimal periodic simulation cells (these were done on 32-atom fcc crystals).  Since the formulations homogenize the atoms to make them all equivalent, there is no longer a need to represent the complex ordered (or disordered) state or to have enough atoms to accurately match the concentrations of each species.  The additional requirement of the LOAA formulation --- the functions $G_{XY}(R)$ (\eqn{eq:G}) --- are determined from the partial RDFs of the final configuration of the same on-lattice MC anneal described above.

\begin{figure}
\begin{subfigure}{0.45\textwidth}
\centering
\includegraphics[width=\linewidth]{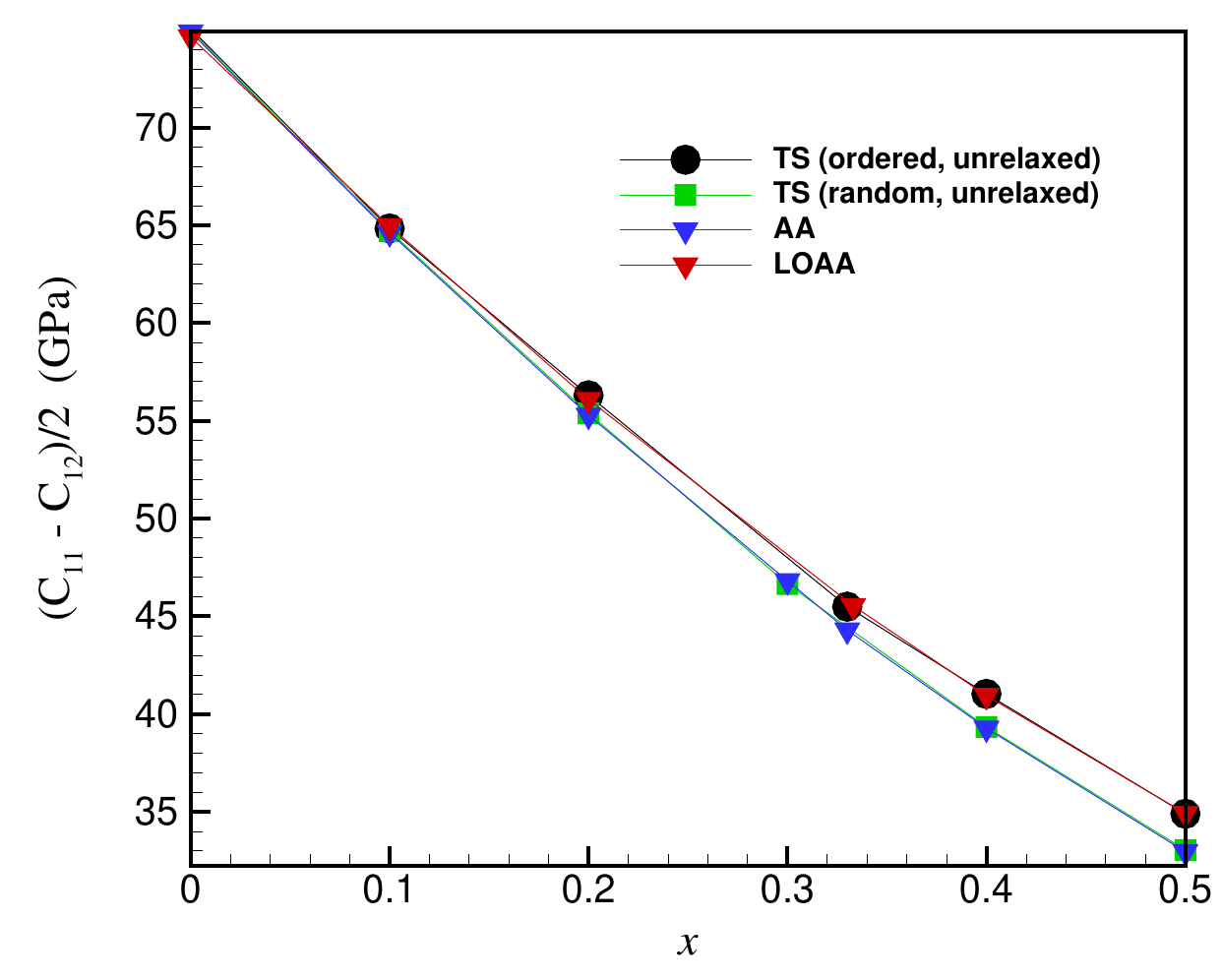}
\caption{}
\end{subfigure}
\hfill
\begin{subfigure}{0.45\textwidth}
\centering
\includegraphics[width=\linewidth]{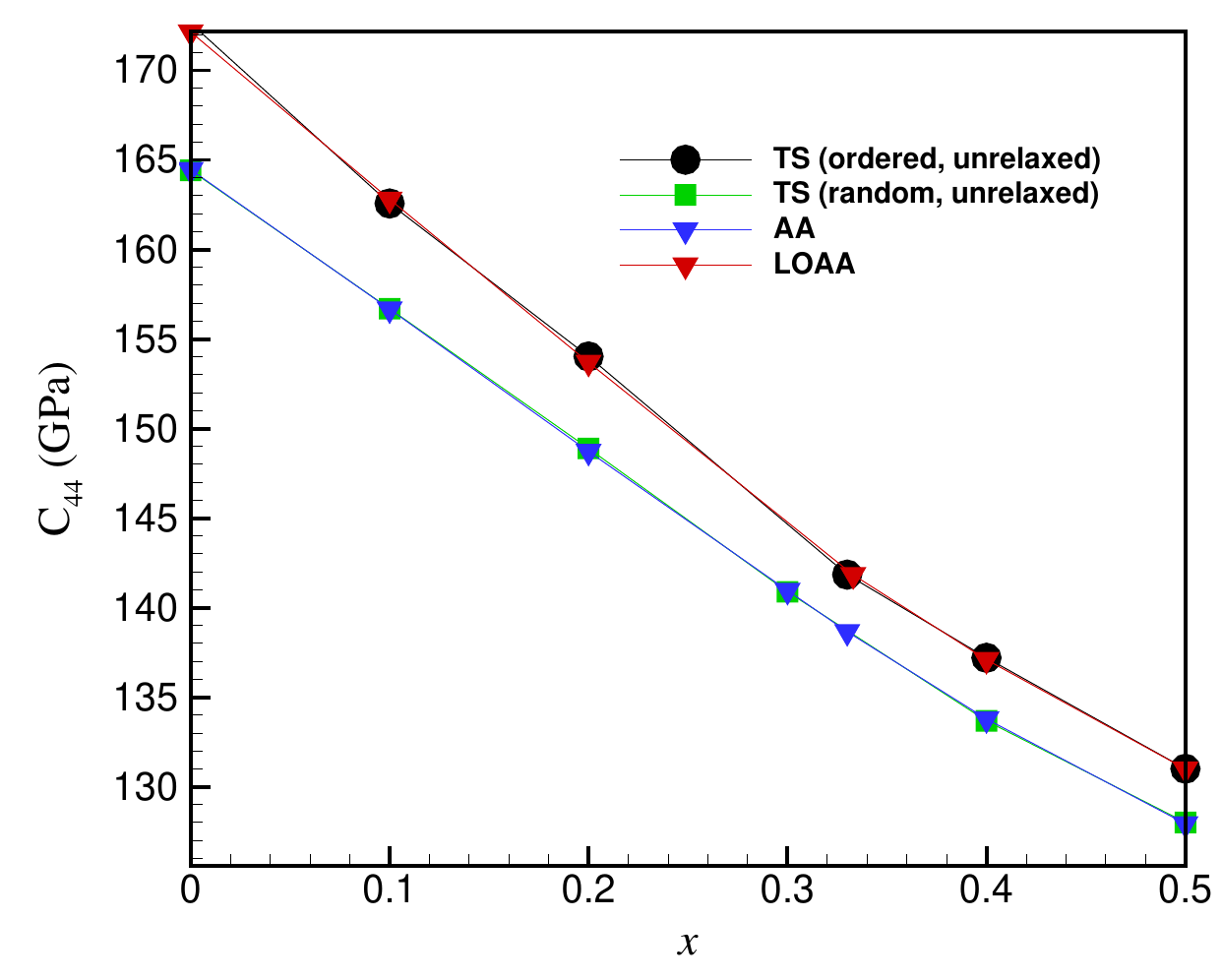}
\caption{}
\end{subfigure}
\caption{Comparison of TS, AA and LOAA elastic constants for 3D EAM \comp alloys when the TS results are not allowed to relax from perfect lattice positions.}
\label{fig:elastic-norelax-fenicr}
\end{figure}

First, we compare the two TS lines (random and ordered) in each frame of \fig{fig:FeNiCr-properties}.  The difference between these curves is the effect of LO, which is not insignificant in these alloys.  The AA formulation does an excellent job of capturing the lattice constant and cohesive energy for the random TS, and a reasonable job of the random TS elastic constants.  AA is of course unable to capture the ordered TS curves, since it knows nothing of the LO.  On the other hand, we see that the LOAA results are in excellent agreement for $a_0$ and $E_{\rm coh}$ case, and has comparable error to that of AA when considering the elastic constants.  The elastic constants are only in modest agreement with the TS results (for both AA and LOAA).  This is due to relaxation effects in the TS results, which arise from local variations of atomic environments. Relaxation effects are absent from the homogenized AA and LOAA results. Consequently, while AA and LOAA accurately reproduce average structural and energetic properties, they are not expected to accurately capture properties that depend strongly on local lattice distortions, such as elastic response or mechanisms involving defects or dislocation-solute interactions, like solid solution strengthening.
In \fig{fig:elastic-norelax-fenicr}, we compare AA and LOAA to TS results where the atoms are not allowed to relax from their perfect crystal positions, and we see the agreement largely restored.

\subsubsection{Surface Energies of NiAl}
\label{sec:nial:pt}
Next we consider the application of LOAA to predict the (100), (110), (111), and (112) surface energies of \nixal alloys for various Ni concentrations $x$ near $0.75$. In the previous section, we saw that AA and LOAA are unable to capture the relaxation effects of the TS results, so in this section we only include results for ideal unrelaxed surfaces in which the crystal is cleaved along a specific plane without allowing atoms to adjust their positions. Results for relaxed surface energies are included in the SM.

At $x=0.75$, Ni$_3$Al\xspace is stable in the ordered L1$_2$ phase at 0~K \cite{pun:mishin:2009, Shao:2023}. For $x\ne 0.75$, the alloy decomposes into a two-phase mixture of the ordered L1$_2$ phase and a disordered fcc solid solution. Here, we model \nixal on an fcc lattice without enforcing L1$_2$ ordering. Instead, local ground state configurations are found through the same MC annealing process described above for a range of Ni concentrations. The MC process allows the system to explore ordering, while restricting configurations to a single fcc phase for surface energy calculations. For $x$ between 0.65 and 1 at 0.05 intervals, atomic structures are generated on an fcc lattice containing approximately $10{,}000$ atoms with periodic boundary conditions, oriented such that the normal to the desired surface, (100), (110), (111), or (112), aligns with the $z$-direction. The functions $G_{XY}(R)$ (\eqn{eq:G}) required for LOAA are determined from the partial RDFs of the final TS configurations of the corresponding MC anneal. 

As above, we study ``random'' and ``ordered'' TS samples. For the random case, Al atoms are randomly distributed onto the lattice of Ni atoms consistent with the concentrations. For the ordered case, the species at each lattice site is determined by performing the MC annealing process on the random sample to find a locally ordered configuration.
Slab geometries are constructed by creating a free surface along $z$.
For TS ordered, five surface terminations are generated from a single ordered bulk configuration by sequentially removing one to five atomic layers. The surface energies of each termination are calculated and averaged to obtain a single value. For TS random, larger variations in surface energy are observed between different surface terminations due to the random local atomic configurations. To obtain a representative value, three different random bulk samples are generated, each providing five surface terminations. The surface energy is averaged over all fifteen calculations. AA and LOAA each only require a single sample.

\begin{figure}
\begin{subfigure}{0.45\textwidth}
\centering
\includegraphics[width=\linewidth]{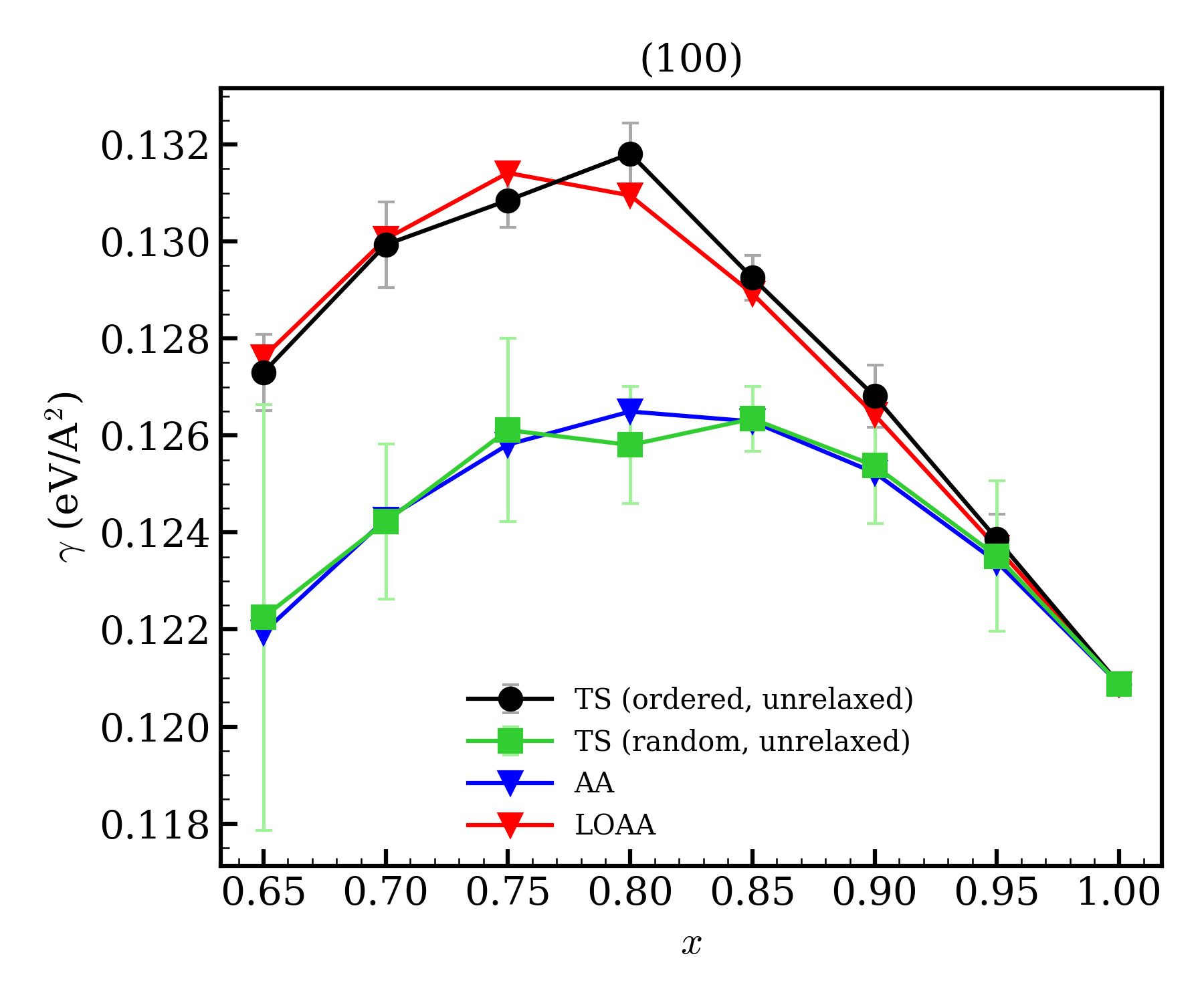}
\caption{}
\end{subfigure}
\hfill
\begin{subfigure}{0.45\textwidth}
\centering
\includegraphics[width=\linewidth]{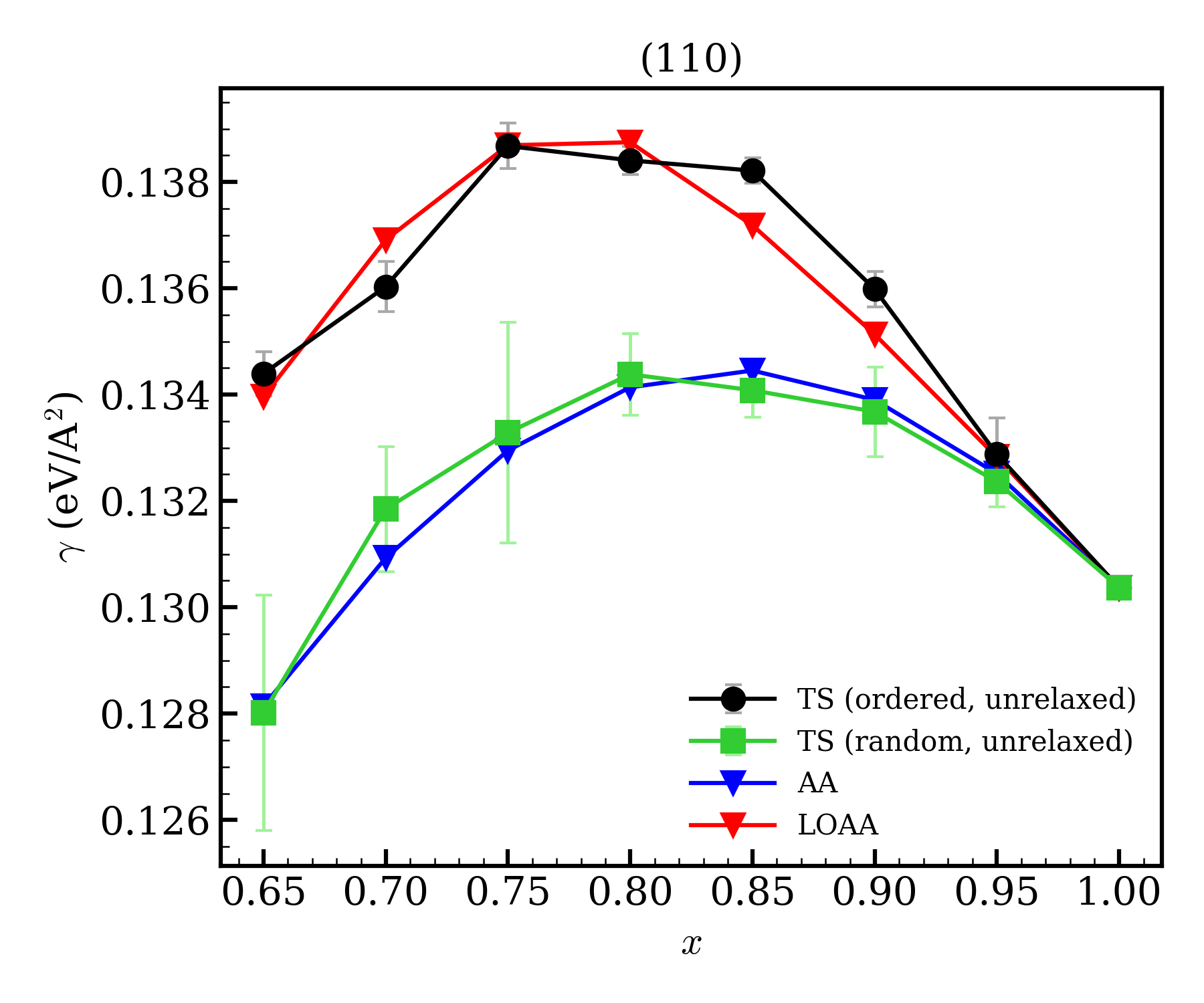}
\caption{}
\end{subfigure}
\vskip\baselineskip
\begin{subfigure}{0.45\textwidth}
\centering
\includegraphics[width=\linewidth]{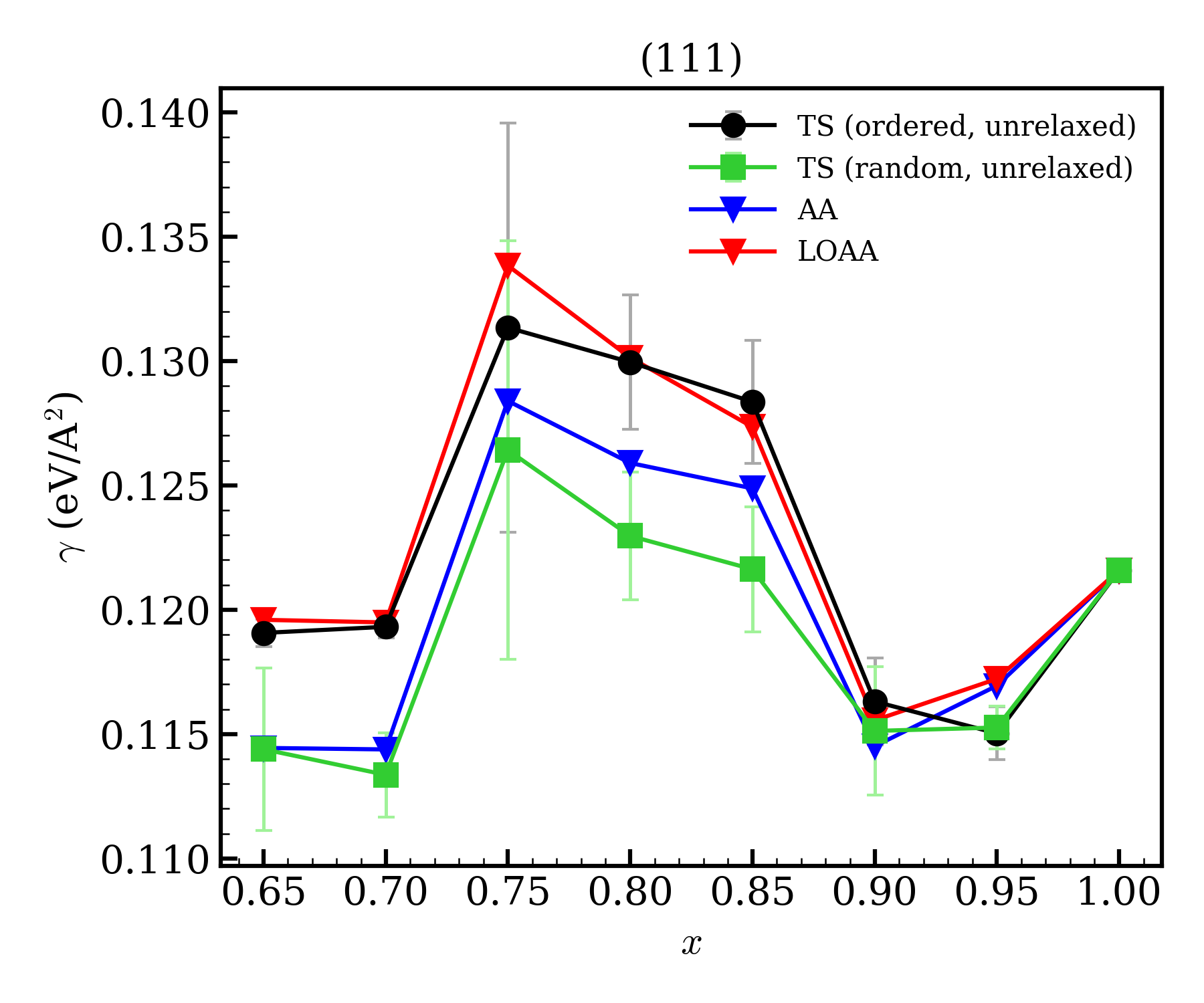}
\caption{}
\end{subfigure}
\hfill
\begin{subfigure}{0.45\textwidth}
\centering
\includegraphics[width=\linewidth]{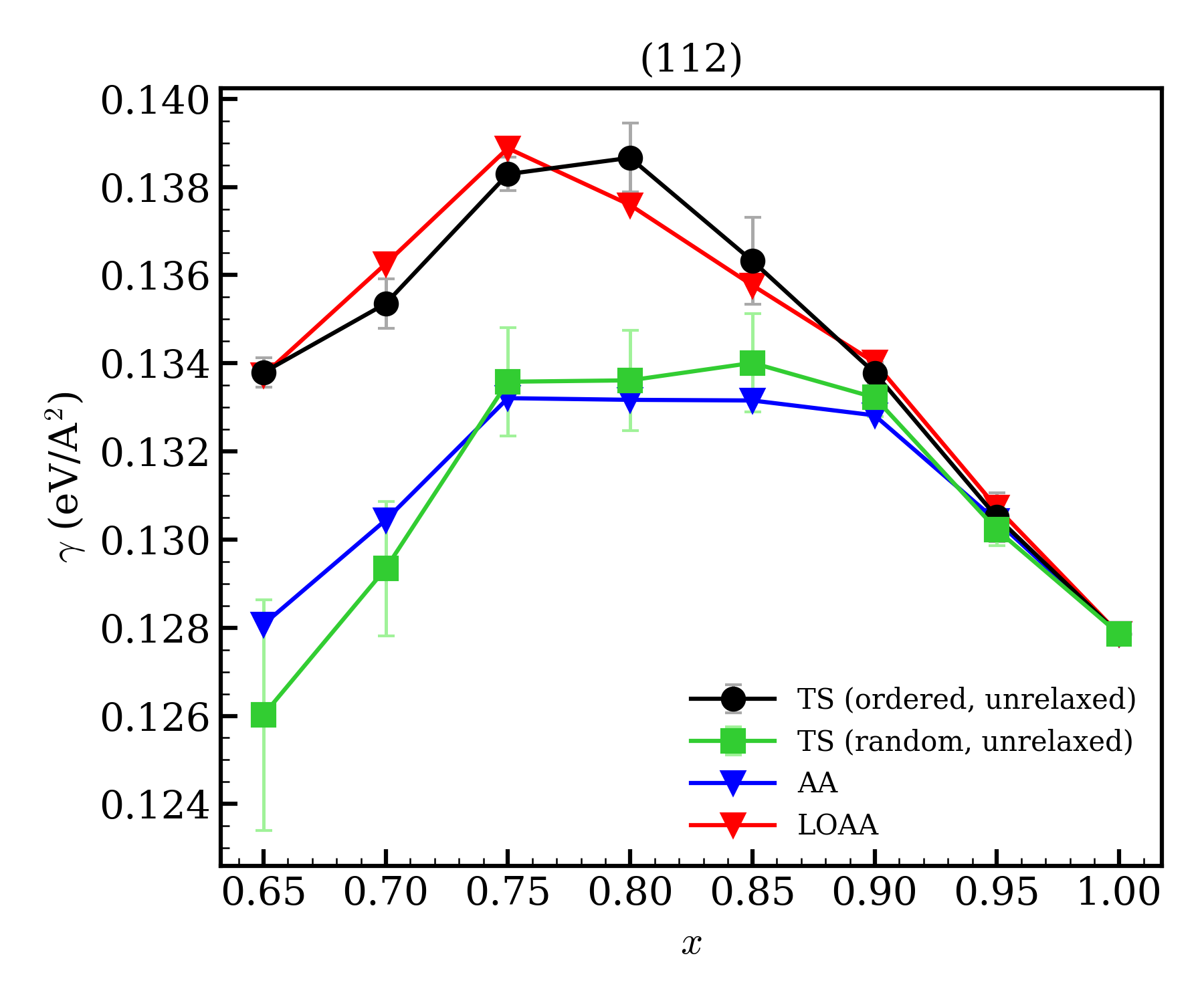}
\caption{}
\end{subfigure}
\caption{Comparison of TS, AA and LOAA surface energies for 3D EAM \nixal\ alloys. Surfaces (a) (100), (b) (110), (c) (111) and (d) (112). TS quantities evaluated for unrelaxed structures. Standard deviation error bars are plotted for TS.}
\label{fig:NiAl-properties}
\end{figure}

Results of the AA and LOAA formulations are compared with the TS results in \fig{fig:NiAl-properties}. We see that for all four surfaces there is a significant effect of LO, as the random and ordered TS curves are shifted from one another. LOAA generally agrees with the ordered TS curve within error. For the random case, TS and AA also largely agree within error. Note that the error is relatively high here because averaging was only performed for a few surface layers of a single, or over a few, relatively small sample(s). Better agreement could likely be reached by averaging over more and/or larger samples. AA and LOAA results are not size dependent, but LOAA does require that the partial RDF is fully representative. 

\section{Summary}
\label{sec:conclude}
LOAA, an extension to the AA approach accounting for LO (short-range order), is derived based on partial RDF information obtained on a reference configuration. The result is an effective IP defined for an atomistic system without species that in the limit of no structural relaxation provides exact results for an ensemble over all possible species arrangements on the reference configuration. Once the effective IP is constructed, it has a similar cost to the original ``true species'' IP upon which it is based, but enables simulations with far smaller supercells and without the need to run multiple times for good statistics, leading to significant computational savings. The LOAA method is attractive for simulations of complex materials, such as HEAs that require very large supercells to capture the off-stoichiometric structure, and where LO is known to play an important role.

The LOAA potential form is derived here for pair and EAM-style potentials, and can be extended to other IP functional forms in a similar fashion. For the EAM form, the effects of LO are approximated by expanding the embedding energy about the average electron charge density. The resulting first-order correction accounts for changes to the embedding energy due the deviation of the partial RDF from the ideal random alloy case.

LOAA is validated on several example problems. First, a 2D binary hexagonal crystal modeled via LJ interactions. A phase diagram is constructed showing the dependence of the nature and intensity of LO on the LJ parameters. It is shown that LOAA accurately reproduces the energy and elastic constants of the true species IP across this phase diagram, in contrast to AA that is only correct for the random alloy case where no LO exists. Second, LOAA is applied to two systems modeled via EAM potentials. In both cases it is shown to accurately account for LO effects: 1) Lattice constant, cohesive energy, and elastic properties of an ordered \comp\ alloy for $x$ in the range 0 to 0.5, and 2) Surface energies of ordered and disordered \nixal alloys for $x$ in the range of 0.65 to 1.

Code and example input suitable for running LOAA in LAMMPS is provided in the SM. Additional validation of results is also provided in the SM.

\section{Supplementary Material}
The supplementary material contains additional studies and results for the binary LJ system and the 3D EAM systems. For the LJ system, this includes figures showing supercell size convergence and distinct LO based on RDFs, details of the MC simulation setup, and an analysis of the energy contour plots. For the 3D EAM NiAl system, RDFs and effective pair potentials are included for Ni$_{67}$Al$_{33}$, along with surface energies where relaxation is allowed for \nixal. A tarball with the code and an example for running LOAA in LAMMPS is provided. The README explains in detail how to properly set up such simulations.

\begin{acknowledgments}
This work was supported by a NASA Space Technology Graduate Research Opportunity. CZ and ET acknowledge support by NSTGRO Award 21-NSTGRO22-0159. RM acknowledges the support of the Natural Sciences and Engineering Research Council of Canada (NSERC). The authors thank Ryan S. Elliott for helpful discussions, and Lauren Abbott and the computational materials group at NASA Ames Research Center for their valuable assistance.
\end{acknowledgments}

\appendix
\section{Radial Distribution Functions}
\label{app:pcf}
The radial distribution function for a 3D monoatomic system (or when not distinguishing atom species) is
\begin{equation}
g(r) = \frac{dn(r)}{4\pi r^2 dr (N/V)},
\label{eq:3dg}
\end{equation}
where $N/V$ is the number density (number of atoms $N$ per volume $V$),
and $dn(r)$ is the average number of atoms within a distance $[r,r+dr]$ of
any other atom,
\begin{equation}
dn(r) = \frac{1}{N}\sum_{i=1}^N \sumdd{j=1}{j\ne i}{N} \langle\dirac(r_{ij}-r)\rangle,
\end{equation}
where $\dirac()$ is the Dirac delta function, and $\langle\cdot\rangle$ indicates
the canonical phase average at a given temperature $T$.

Formally, the RDF in the canonical ensemble is
independent of structure. This is because phase averages are over all
configurations (subject to an overall volume constraint), which corresponds
to a liquid state. The dependence on structure comes in once the
ensemble is restricted in the sense described in Tadmor and Miller
\cite{tadmor:miller:2011} to a subset of configuration space,
e.g.\ a face-centered cubic (fcc) arrangement with perturbations about mean positions.
In this case, the RDFs will depend on the
restriction. In practice, the $g(r)$ computed in a molecular dynamics
simulation or measured in an experiment is inherently restricted by the
accessible time scale.

In the multiple species case, the partial RDF is
\begin{equation}
g_{XY}(r) = \frac{dn_{XY}(r)}{4\pi r^2 dr (N_Y/V)},
\label{eq:3dgxy}
\end{equation}
where $N_Y$ is the number of atoms of species $Y$, and $dn_{XY}(r)$ is the average number
of atoms of species $Y$ within a distance $[r,r+dr]$ of an atom of species $X$,
\begin{equation}
dn_{XY}(r) = \frac{1}{N_X}\sum_{i=1}^N \sumdd{j=1}{j \ne i}{N} s_i^X s_j^Y \langle\dirac(r_{ij}-r)\rangle.
\end{equation}
Here $s_i^X$ and $s_i^Y$ are the species indicator functions defined in \eqn{eq:siX}.

The probability of atoms $i$ and $j$ separated by distance $r_{ij}$ being of species $X$ and $Y$,
respectively, can be expressed in terms of the numbers $dn_{XY}(r)$ as follows:
\begin{equation}
p(s_i^X=1,s_j^Y=1 \,|\, T)=
\frac{c_X dn_{XY}(r)}{\sum_A \sum_B c_A dn_{AB}(r)}.
\end{equation}
The multiplication by the concentrations is required so that for all terms, the number
of neighbors is normalized by the total number of atoms, and not the number of atoms
of a specific species, e.g.\ for the $dn_{XY}$ term, $c_X/N_X=c_X/(c_X N)=1/N$.
Next using \eqn{eq:3dgxy} to express $dn_{XY}$ in terms of $g_{XY}$ and using
$N_Y=c_Y N$, we have
\begin{equation}
p(s_i^X=1,s_j^Y=1 \,|\, T)=
\frac{c_X c_Y g_{XY}(r)}{\sum_A \sum_B c_A c_B g_{AB}(r)}.
\label{eq:probxyT}
\end{equation}
This is the expression in \eqn{eq:sxsyg}.

For the special case where the species of the atoms are set independently, we have
\begin{equation}
dn_{XY}(r) = c_Y dn(r),
\label{eq:dnxyindep}
\end{equation}
since the number of neighbors of type $Y$ follows from the concentration $c_Y$.
Substituting \eqn{eq:dnxyindep} into \eqn{eq:3dgxy} gives
\begin{equation}
g_{XY}(r) = \frac{c_Y dn(r)}{4\pi r^2 dr (N_Y/V)} = g(r),
\label{eq:3dgxyindep}
\end{equation}
where we have used $N_Y=c_Y N$ to obtain the final equality. Thus when the
atom species are independent, all partial RDFs are
equal to the RDF, which does not account for species.
In this case, the probability in \eqn{eq:probxyT} is as expected
\begin{equation}
p(s_i^X=1,s_j^Y=1 \,|\, T)=
\frac{c_X c_Y g(r)}{\sum_A \sum_B c_A c_B g(r)}
= \frac{c_X c_Y}{\sum_A \sum_B c_A c_B}
= c_X c_Y,
\label{eq:probxyT:indep}
\end{equation}
where we have used $\sum_A \sum_B c_A c_B = \sum_A c_A (\sum_B c_B) = \sum_A c_A (1) = (1)(1) = 1$.

\section{Relative bonding energies in 2D Hexagonal Binary Crystals}
\label{app:deriv_energy_phase_diag}
We derive an expression for the energy of a 2D binary hexagonal crystal with $A$ and $B$ atoms relative to that of a crystal with all $A$ atoms. Assuming pairwise interactions, the energy of a system of $N$ atoms is given by
\begin{equation}
E = \frac{1}{2} \sum_i \sum_{j\ne i} \sum_X \sum_Y V_{XY}(r_{ij}) s_i^X s_j^Y,
\label{eq:E_norm}
\end{equation}
where $X$ and $Y$ are species, and we take an LJ form with different $\epsilon$ parameters for each pair of species, but the same $\sigma$, so that
\begin{equation}
V_{XY}(r) = \epsilon_{XY} \hat\varphi(r),
\label{eq:phi_pair}
\end{equation}
where $\hat\varphi(r)$ is the LJ function given in \eqn{eq:Vxy:const:sig} (since no relaxation takes place, there is no difference in using $\hat\varphi(r)$ rather than the truncated potential in \eqn{eq:Vxy:const:sig}).
Normalizing the energy by $\epsilon_{AA}$, the dimensionless energy for a periodic system of $N$ type $A$ atoms on a 2D hexagonal lattice with second neighbor interactions is
\begin{equation}
\frac{E(a; \cS_A)}{\epsilon_{AA}} = \frac{1}{2}(6N\varphi(a) + 6N\varphi(\sqrt{3}a)),
\label{eq:E_norm_allA}
\end{equation}
where $a$ is the lattice constant and $\cS_A$ refers to a species arrangement of all $A$ atoms. For the LJ functional form, this is
\begin{equation}
\frac{E(a; \cS_A)}{\epsilon_{AA}} = 12N\left[\left(\frac{\sigma}{a}\right)^{12} - \left(\frac{\sigma}{a}\right)^6 + \left(\frac{\sigma}{\sqrt{3}a}\right)^{12} - \left(\frac{\sigma}{\sqrt{3}a}\right)^6\right].
\label{eq:E_norm_allA_LJ}
\end{equation}
Setting $\frac{dE}{da}(a_0; S_A)=0$, we find that the equilibrium lattice constant (for second neighbor interactions) is $a_0 = 1.1159$.

\begin{figure}
\centering
\includegraphics[width=0.15\linewidth]{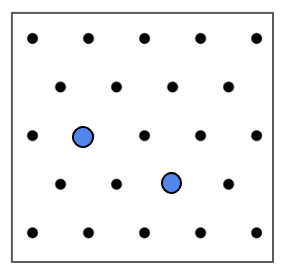}
\caption{2D hexagonal crystal of $A$ atoms (black) with two $B$ atoms (blue) at a second neighbor distance.}
\label{fig:g:hex_2B}
\end{figure}

Next, we consider the case of a crystal of $A$ atoms with two $B$ atoms interacting at a second neighbor distance (denoted $\cS_{AB}$) as shown in \fig{fig:g:hex_2B}. In order to obtain a closed-form expression, we neglect the effect of local relaxations about $B$ atoms (which are expected to be small). The energy for this case is derived by considering the number of $AA$, $AB$ and $BB$ bonds at first- and second-neighbor distances. The result is
\begin{equation}
\frac{E(a; \cS_{AB})}{\epsilon_{AA}} = \left(12\frac{\epsilon_{AB}}{\epsilon_{AA}} + 3N - 12\right)\varphi(a) + \left(10\frac{\epsilon_{AB}}{\epsilon_{AA}} + \frac{\epsilon_{BB}}{\epsilon_{AA}} + 3N - 11\right)\varphi(\sqrt{3}a).
\label{eq:E_norm_2B_LJ}
\end{equation}
In order to obtain an expression for an infinite system independent of $N$, we consider the energy \textit{relative} to $E(a; \cS_A)$, i.e.\ the difference between the energy of the system with $B$ atoms and the energy of the system with all $A$ atoms:
\begin{equation}
\frac{\Delta E(a; \cS_{AB})}{\epsilon_{AA}}\equiv\frac{E(a; \cS_{AB})-E(a; \cS_A)}{\epsilon_{AA}}  = \left(12\frac{\epsilon_{AB}}{\epsilon_{AA}} - 12\right)\varphi(a) + \left(10\frac{\epsilon_{AB}}{\epsilon_{AA}} + \frac{\epsilon_{BB}}{\epsilon_{AA}} - 11\right)\varphi(\sqrt{3}a).
\label{eq:E_norm_2B_rel}
\end{equation}
The general form of the relative energy for any species arrangement $\cS$ containing any number of $B$ atoms at any positions takes the form,
\begin{equation}
\frac{\Delta E(a; \cS)}{\epsilon_{AA}} = C(\cS)\varphi(a) + D(\cS)\varphi(\sqrt{3}a).
\label{eq:E_norm_gen_rel}
\end{equation}
where $C(\cS)$ and $D(\cS)$ are dependent on the number of first- and second-neighbor bonds, respectively, for $AA$, $AB$, and $BB$ interactions in a particular species arrangement $\cS$.

Finally, for the LJ form, we obtain
\begin{equation}
\frac{\Delta E(a; \cS)}{\epsilon_{AA}} = 4C(\cS)\left[\left(\frac{\sigma}{a}\right)^{12} - \left(\frac{\sigma}{a}\right)^6\right] + 4D(\cS)\left[\left(\frac{\sigma}{\sqrt{3}a}\right)^{12} - \left(\frac{\sigma}{\sqrt{3}a}\right)^6\right].
\label{eq:E_norm_gen_rel_KA}
\end{equation}
where $C(S)$ and $D(S)$ are defined by
\begin{subequations}
\begin{align}
    C(S) &= N^{(1)}_{AB}\frac{\epsilon_{AB}}{\epsilon_{AA}} + N^{(1)}_{BB}\frac{\epsilon_{BB}}{\epsilon_{AA}} - (N^{(1)}_{AB} + N^{(1)}_{BB}) \\
    D(S) &= N^{(2)}_{AB}\frac{\epsilon_{AB}}{\epsilon_{AA}} + N^{(2)}_{BB}\frac{\epsilon_{BB}}{\epsilon_{AA}} - (N^{(2)}_{AB} + N^{(2)}_{BB})
\end{align}
\label{eq:C_and_D}
\end{subequations}
where $N^{(1)}_{XY}$ and $N^{(2)}_{XY}$ are the number of $XY$ bonds at nearest neighbor and second neighbor distances, respectively.

By substituting $a = a_0$, we obtain a general expression for the minimum energy of a 2D hexagonal crystal with an arbitrary arrangement of $B$ atoms.

\section{Understanding the `no-LO' Line of the Phase Diagram}
\label{app:noLO_line}
The line in the phase diagram where no local ordering occurs (i.e.\ the `no-LO line') is defined by the equation:
\begin{equation}
    \frac{\epsilon_{BB}}{\epsilon_{AA}} = 2 \frac{\epsilon_{AB}}{\epsilon_{AA}} - 1.
\label{eq:nololine}
\end{equation}
Substituting this equation into the expressions for $C(\cS)$ and $D(\cS)$ in \eqn{eq:C_and_D}, the result is:
\begin{subequations}
\begin{align}
C(\cS) &= \left(\frac{\epsilon_{AB}}{\epsilon_{AA}}-1\right)
\left(N_{AB}^{(1)} + 2 N_{BB}^{(1)}\right), \\
D(\cS) &= \left(\frac{\epsilon_{AB}}{\epsilon_{AA}}-1\right)
\left(N_{AB}^{(2)} + 2 N_{BB}^{(2)}\right).
\end{align}
\end{subequations}
In addition, due to the geometry of the lattice \footnote{In a hexagonal lattice, each $B$ atom has six first and six second neighbors; summed over all $B$ atoms, this total equals the number of $A$-$B$ bonds plus twice the number of $B$-$B$ bonds, since each $B$-$B$ bond is shared between two $B$ atoms.}, we have $N_{AB}^{(1)} + 2 N_{BB}^{(1)} = N_{AB}^{(2)} + 2 N_{BB}^{(2)} = 6 N_B$, so that
\begin{equation}
C(S)=D(S)= 6 N_B \left(\frac{\epsilon_{AB}}{\epsilon_{AA}}-1\right).
\end{equation}
This means that at any point on the no-LO line for a given number of $B$ atoms, the energy of every LO pattern is the same. Physically this happens because, as seen in \eqn{eq:nololine}, along this line the energy of a $BB$ bond is equal to the energy of two $AB$ bonds, so that neither LO pattern is favorable over the other (the remaining ``$-1$'' term is simply accounting for the \textit{relative} energy, as described in \app{app:deriv_energy_phase_diag}.)

\section{Theoretical Analysis of the Error in Energy Predictions of the AA Method}
\label{app:deltaE_LOAA_AA}
\fig{fig:g:AA_all_energy_dif_contour} is a plot of the difference between the energy per atom predicted by the AA method and the exact TS case. The plot shows that the energy difference increases as a function of the vertical distance from the no-LO line (shown in red). In this section we derive an analytical result for the energy difference in order to explain this observation. Since an analytical expression for the TS case does not exist (as it involves numerical MC simulations), we use the LOAA solution as a surrogate since it is in close agreement with the TS results.

The effective potentials for AA and LOAA are given in \eqns{eq:Veff} and \eqnx{eq:Veffg}, respectively. The difference between them $\Delta V_{\rm eff} = V_{\rm AA} - V_{\rm LOAA}$ is
\begin{align}
\Delta V_{\rm eff}(r,R)
&= \sum_X \sum_Y V_{XY}(r) \left(c_X c_Y - G_{XY}(R)\right) \nonumber \\
&= \sum_X \sum_Y V_{XY}(r) \left(c_X c_Y - \frac{c_X c_Y g_{XY}(R)}{\sum_A \sum_B c_A c_B g_{AB}(R)}\right) \nonumber \\
&= \sum_X \sum_Y V_{XY}(r) c_X c_Y \left(1- \frac{g_{XY}(R)}{\sum_A \sum_B c_A c_B g_{AB}(R)}\right) \nonumber \\
&= \sum_X \sum_Y V_{XY}(r) c_X c_Y \left(1 - \Phi_{XY}(R)\right),
\end{align}
where $\Phi_{XY}(R)$ is defined in \eqn{eq:phi}. The difference in the energy per atom between the prediction of the AA method and LOAA is then
\begin{align}
\Delta E_{\rm atom}
&= \frac{1}{2}\sum_n N_n \Delta V_{\rm eff}(R_n) \nonumber \\
&= \frac{1}{2}\sum_n N_n \sum_X \sum_Y V_{XY}(R_n) c_X c_Y \left(1 - \Phi_{XY}(R_n)\right),
\label{eq:deltaV_tot}
\end{align}
where $N_n$ is the number of atoms in neighbor shell $n$, at distance $R_n$ in the reference configuration from an atom (all atoms in a hexagonal lattice have the same environment, so are equivalent). The results in \fig{fig:g:AA_all_energy_dif_contour} are for a binary hexagonal lattice with LJ interactions up to second neighbors. To simplify the expressions, we neglect second-neighbor interactions (which are small) so that \eqn{eq:deltaV_tot} is
\begin{equation}
    \Delta E_{\rm atom} = 3 \left[V_{AA}(a) c_A^2 (1-\Phi_{AA}(a))
    + 2V_{AB}(a) c_A c_B (1-\Phi_{AB}(a))
    + V_{BB}(a) c_B^2 (1-\Phi_{BB}(a))\right].
\end{equation}
Substituting in $V_{XY}(r)$ from \eqn{eq:Vxy:const:sig} and normalizing by $\epsilon_{AA}$, we have
\begin{equation}
    \frac{\Delta E_{\rm atom}}{\epsilon_{AA}} = 12\varphi(a)
    \left[
    \eta + \gamma \frac{\epsilon_{AB}}{\epsilon_{AA}} +
    \zeta \frac{\epsilon_{BB}}{\epsilon_{AA}}
    \right].
    \label{deltaV_tot_abg}
\end{equation}
where
\begin{equation}
\eta = c_A^2 (1-\Phi_{AA}(a)), \qquad
\gamma = 2 c_A c_B (1-\Phi_{AB}(a)), \qquad
\zeta = c_B^2 (1-\Phi_{BB}(a)).
\end{equation}
Note that parameters $\eta$, $\gamma$, $\zeta$ do not depend on the potential parameters ($\epsilon_{XY}$), since the same partial RDF is observed for all $\epsilon$ values within a given LO region (as explained in \sect{sec:ts:calcs}) and hence $\Phi_{XY}(a)$ are the same as well.

\begin{figure}
    \centering
    \includegraphics[width=0.5\linewidth]{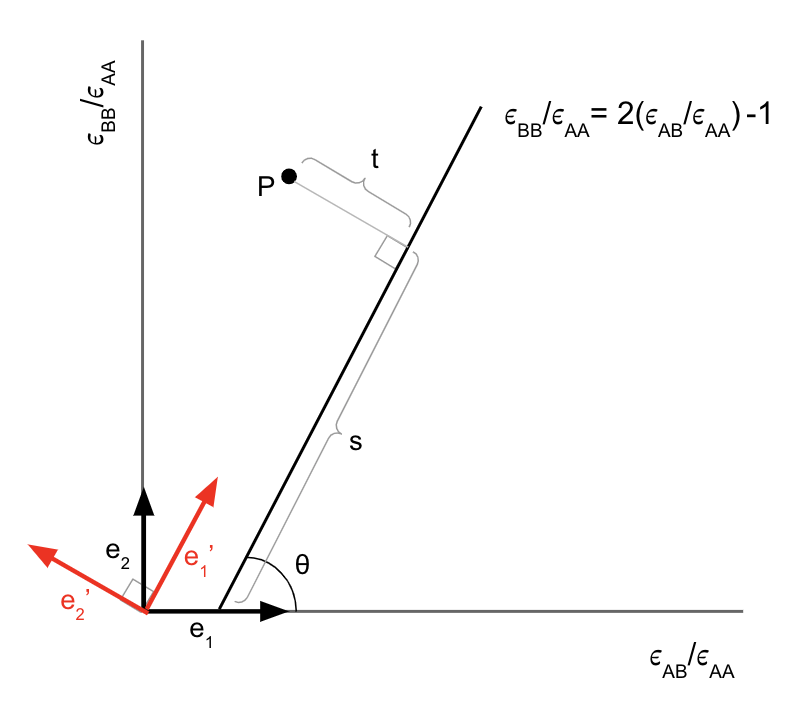}
    \caption{Change of variables in $\epsilon$-space to coordinates $s$ and $t$ that are parallel and perpendicular to the no-LO line in the LO phase diagram.}
   \label{fig:change_of_var}
\end{figure}

We wish to study how the energy difference in \eqn{deltaV_tot_abg} changes with vertical distance from the no-LO line to understand the observations in \fig{fig:g:AA_all_energy_dif_contour}. To this end, we perform a change of variables in $\epsilon$-space from $\epsilon_{AB}/\epsilon_{AA}$ and $\epsilon_{BB}/\epsilon_{AA}$ to the coordinates $s$ and $t$ that are parallel and perpendicular to the no-LO line as shown in \fig{fig:change_of_var}.
The equation of the no-LO line is
\begin{equation}
    \frac{\epsilon_{BB}}{\epsilon_{AA}} = 2 \frac{\epsilon_{AB}}{\epsilon_{AA}} - 1.
\end{equation}
The transformation from basis vectors $\be_1$ and $\be_2$ aligned with the horizontal and vertical axes in \fig{fig:change_of_var} to a new basis where $\be_1'$ is aligned with the no-LO line is
\begin{equation}
   \begin{bmatrix}
   \be_1' \\ \be_2'
   \end{bmatrix}
=
   \begin{bmatrix}
        \cos\theta &
        \sin\theta \\
        -\sin\theta &
        \cos\theta
   \end{bmatrix}
   \begin{bmatrix}
   \be_1 \\ \be_2
   \end{bmatrix}
\end{equation}
where $\theta=\tan^{-1} 2 = 63.43^\circ$ is the angle between $\be_1$ and the no-LO line, for which $\cos\theta=1/\sqrt{5}$ and $\sin\theta=2/\sqrt{5}$. Referring to \fig{fig:change_of_var}, the position of an arbitrary point P is
\begin{align}
    P
    & = 0.5 \be_1 + s \be_1' + t \be_2' \nonumber \\
    & = (0.5 + s \cos\theta - t \sin\theta) \be_1 + (s \sin \theta + t
    \cos\theta) \be_2.
    \label{eq:P1}
\end{align}
We also have that
\begin{equation}
    P = \frac{\epsilon_{AB}}{\epsilon_{AA}} \be_1 + \frac{\epsilon_{BB}}{\epsilon_{AA}} \be_2.
    \label{eq:P2}
\end{equation}
Equating \eqns{eq:P1} and \eqnx{eq:P2}, we find
\begin{equation}
    \frac{\epsilon_{AB}}{\epsilon_{AA}} = \frac{1}{2} + \frac{1}{\sqrt{5}}(s-2t), \qquad
    \frac{\epsilon_{BB}}{\epsilon_{AA}} = \frac{1}{\sqrt{5}}(2s+t).
\end{equation}
Substituting these expressions into \eqn{deltaV_tot_abg} and rearranging gives
\begin{equation}
    \frac{\Delta E_{\rm atom}}{\epsilon_{AA}}
    = 12 \varphi(a) \left[
    \eta + \frac{1}{2} \gamma + \frac{1}{\sqrt{5}}(2\zeta + \gamma)s
    + \frac{1}{\sqrt{5}}(\zeta - 2\gamma)t
    \right].
    \label{eq:AA:energy:error}
\end{equation}
Thus, considering only near-neighbor interactions and using the LOAA solution as an approximation for the exact TS solution, we expect a linear variation of the per-atom-energy difference between the AA and exact result moving parallel and perpendicular to the no-LO line. However, \fig{fig:g:AA_all_energy_dif_contour} shows contour lines parallel to the no-LO lines, which implies no dependence on $s$. This requires that
\begin{equation}
    2\zeta + \gamma = 0.
\end{equation}
Computing this value across $\epsilon$-space, it is found to be of order  $10^{-3}$--$10^{-4}$ in both LO regions. In contrast the term multiplying the $t$ coordinate ($\zeta - 2\gamma$) is of order $0.1$. This explains the observations in \fig{fig:g:AA_all_energy_dif_contour}.

\section{Elastic Constant Equations for AA and LOAA}
\label{app:elastic_consts}
In this section, we derive analytical expressions for the elastic constant predictions by the AA and LOAA approaches and demonstrate their linear dependence on distance from the no-LO line. We begin with the AA approach. Replacing $V_{\rm eff}$ in \eqn{eq:Cijkl} for AA, we have
\begin{equation}
c_{ijkl}^{\rm AA} = \frac{1}{2\Omega} \sum_{\substack{\alpha, \beta \\ \alpha \neq \beta}} Q^{\alpha \beta}_{ijkl} \left[ \sum_X \sum_Y \left( V_{XY}''(r^{\alpha \mathring{\beta}}) -\frac{V_{XY}'(r^{\alpha \mathring{\beta}})}{r^{\alpha \mathring{\beta}}} \right) c_x c_y \right],
\end{equation}
where for simplicity we denote,
\begin{equation}
Q^{\alpha\beta}_{ijkl} = \frac{ r_i^{\alpha \mathring{\beta}} r_j^{\alpha \mathring{\beta}} r_k^{\alpha \mathring{\beta}} r_l^{\alpha \mathring{\beta}} }{(r^{\alpha \mathring{\beta}})^2}.
\label{eq:Cijkl_Q}
\end{equation}
Simplifying further by expanding and pulling terms out of the summations, we have
\begin{equation}
c_{ijkl}^{\rm AA} = \frac{2}{\Omega} \left[ c_A^2 \epsilon_{AA} + 2 c_A c_B \epsilon_{AB} + c_B^2 \epsilon_{BB} \right] \sum_{\substack{\alpha, \beta \\ \alpha \neq \beta}} Q^{\alpha\beta}_{ijkl} \left[ \varphi''(r^{\alpha \mathring{\beta}}) -\frac{\varphi'(r^{\alpha \mathring{\beta}})}{r^{\alpha \mathring{\beta}}} \right].
\end{equation}
We rewrite this as
\begin{equation}
\frac{c_{ijkl}^{\rm AA}}{\epsilon_{AA}} = \frac{2}{\Omega} \left[ \eta^{\rm AA} + \gamma^{\rm AA} \frac{\epsilon_{AB}}{\epsilon_{AA}} + \zeta^{\rm AA} \frac{\epsilon_{BB}}{\epsilon_{AA}} \right] \sum_{\substack{\alpha, \beta \\ \alpha \neq \beta}} Q^{\alpha\beta}_{ijkl} \left[ \varphi''(r^{\alpha \mathring{\beta}}) -\frac{\varphi'(r^{\alpha \mathring{\beta}})}{r^{\alpha \mathring{\beta}}} \right],
\end{equation}
where
\begin{equation}
\eta^{\rm AA} = c_A^2, \qquad
\gamma^{\rm AA} = 2 c_A c_B, \qquad
\zeta^{\rm AA} = c_B^2.
\end{equation}
The motivation for this substitution will be more evident in the LOAA formulation below. Following the same change of variables procedure in \app{app:deltaE_LOAA_AA}, this can be re-written as
\begin{equation}
\frac{c_{ijkl}^{\rm AA}}{\epsilon_{AA}} = \frac{2}{\Omega} \left[ \eta^{\rm AA} + \frac{1}{2}\gamma^{\rm AA} + \frac{1}{\sqrt{5}}\left( \gamma^{\rm AA} + 2\zeta^{\rm AA} \right)s
+ \frac{1}{\sqrt{5}} \left( \zeta^{\rm AA} - 2\gamma^{\rm AA} \right)t \right] \sum_{\substack{\alpha, \beta \\ \alpha \neq \beta}} Q^{\alpha\beta}_{ijkl} \left[ \varphi''(r^{\alpha \mathring{\beta}}) -\frac{\varphi'(r^{\alpha \mathring{\beta}})}{r^{\alpha \mathring{\beta}}} \right],
\label{eq:cAA}
\end{equation}
where $s$ and $t$ are distances parallel and perpendicular to the no-LO line as shown in \fig{fig:change_of_var}. A similar analysis for LOAA yields,
\begin{equation}
\frac{c_{ijkl}^{\rm LOAA}}{\epsilon_{AA}} = \frac{2}{\Omega} \sum_{\substack{\alpha, \beta \\ \alpha \neq \beta}}
Q^{\alpha\beta}_{ijkl}
\left( \eta^{\rm LOAA} + \gamma^{\rm LOAA} \frac{\epsilon_{AB}}{\epsilon_{AA}} + \zeta^{\rm LOAA} \frac{\epsilon_{BB}}{\epsilon_{AA}} \right)
\left[ \varphi''(r^{\alpha \mathring{\beta}}) -\frac{\varphi'(r^{\alpha \mathring{\beta}})}{r^{\alpha \mathring{\beta}}} \right],
\end{equation}
where
\begin{equation}
\eta^{\rm LOAA} = c_A^2 \Phi_{AA}(r^{\alpha \mathring{\beta}}), \qquad
\gamma^{\rm LOAA} = 2 c_A c_B \Phi_{AB}(r^{\alpha \mathring{\beta}}), \qquad
\zeta^{\rm LOAA} = c_B^2 \Phi_{BB}(r^{\alpha \mathring{\beta}}).
\label{eq:LOAA_vars}
\end{equation}
Then after the change of variables to $s$ and $t$ coordinates,
\begin{multline}
\frac{c_{ijkl}^{\rm LOAA}}{\epsilon_{AA}} = \frac{2}{\Omega} \sum_{\substack{\alpha, \beta \\ \alpha \neq \beta}}
Q^{\alpha\beta}_{ijkl} \left[
\varphi''(r^{\alpha \mathring{\beta}}) -\frac{\varphi'(r^{\alpha \mathring{\beta}})}{r^{\alpha \mathring{\beta}}} \right] \times \\
\left( \eta^{\rm LOAA} + \frac{1}{2}\gamma^{\rm LOAA} +
\frac{1}{\sqrt{5}}\left( \gamma^{\rm LOAA} + 2\zeta^{\rm LOAA} \right)s + \frac{1}{\sqrt{5}} \left( \zeta^{\rm LOAA} - 2\gamma^{\rm LOAA} \right)t
\right).
\end{multline}
We can derive expressions for the slopes of the elastic constants when plotted against $t$ by separating out the terms multiplying $t$,
\begin{subequations}
\begin{align}
m^{\rm AA}_{11} &= \frac{2\epsilon_{AA}}{\sqrt{5}\Omega} \left( c_B^2 - 4 c_A c_B \right) \sum_{\substack{\alpha, \beta \\ \alpha \neq \beta}}
\left[ \varphi''(r^{\alpha \mathring{\beta}}) -\frac{\varphi'(r^{\alpha \mathring{\beta}})}{r^{\alpha \mathring{\beta}}} \right]
\frac{ \left(r_1^{\alpha \mathring{\beta}}\right)^4}{(r^{\alpha \mathring{\beta}})^2},
\\
m^{\rm LOAA}_{11} &= \frac{2\epsilon_{AA}}{\sqrt{5}\Omega} \sum_{\substack{\alpha, \beta \\ \alpha \neq \beta}}
\left(c_B^2 \Phi_{BB}(r^{\alpha \mathring{\beta}}) - 4 c_A c_B \Phi_{AB}(r^{\alpha \mathring{\beta}})\right)
\left[ \varphi''(r^{\alpha \mathring{\beta}}) -\frac{\varphi'(r^{\alpha \mathring{\beta}})}{r^{\alpha \mathring{\beta}}} \right]
\frac{ \left(r_1^{\alpha \mathring{\beta}}\right)^4}{(r^{\alpha \mathring{\beta}})^2}.
\end{align}
\end{subequations}

\clearpage % to put images in immediate pages

\bibliography{loaa}

\end{document}

\title{\textbf{Supplementary Material for \protect\\
``Local Order Average-Atom Interatomic Potentials''}
}%

\author{Chloe A. Zeller}
\affiliation{
Department of Aerospace Engineering and Mechanics,
University of Minnesota, Minneapolis, MN 55455, USA}
\author{Ronald E. Miller}
\affiliation{
Department of Mechanical and Aerospace Engineering,
Carleton University, Ottawa, Canada}
\author{Ellad B. Tadmor}
\email{Contact author: tadmor@umn.edu}
\affiliation{
Department of Aerospace Engineering and Mechanics,
University of Minnesota, Minneapolis, MN 55455, USA}

% figures path
\graphicspath{{figures/}{./}}

\maketitle

\section{Verification of Calculations for the 2D LJ Binary System}

\subsection{Supercell Size Convergence for RDF and Energy Calculations}

Local-order average atom (LOAA) calculations require an evaluation of the partial radial distribution functions (RDFs). In this section we determine the minimal supercell size required for sufficiently accurate RDF calculations.

We begin by examining the RDF for the pure $A$ system. Adapting Eq.~(A1) from Appendix~A in the main manuscript (MS) to a discrete 2D system, the RDF is
\begin{equation}
g(r) = \frac{\Delta n(r)}{2\pi r \Delta r (N/\Omega)},
\label{eq:2dg}
\end{equation}
where $N/\Omega$ is the number density (number of atoms $N$ per periodic area $\Omega$),
$\Delta r=r_{\rm max}/N_{\rm bins}$, $r_{\rm max}$ is the maximum
radius computed, $N_{\rm bins}$ is the number of bins in the histogram used in the computation,
and $\Delta n(r)$ is the average number of atoms within a distance $[r,r+\Delta r]$ of
any other atom,
\begin{equation}
\Delta n(r) = \frac{1}{N}\sum_{i=1}^N \sumdd{j=1}{j\ne i}{N} \cH((r_{ij}-r)(r+\Delta r-r_{ij})),
\end{equation}
where $\cH(\cdot)$ is the Heaviside step function. It is important to note that the results of
the calculation in \eqn{eq:2dg} depend on the choice of $\Delta r$
(the smaller $\Delta r$, the larger $g(r)$).

\begin{figure}
\centering
\includegraphics[width=0.6\linewidth]
{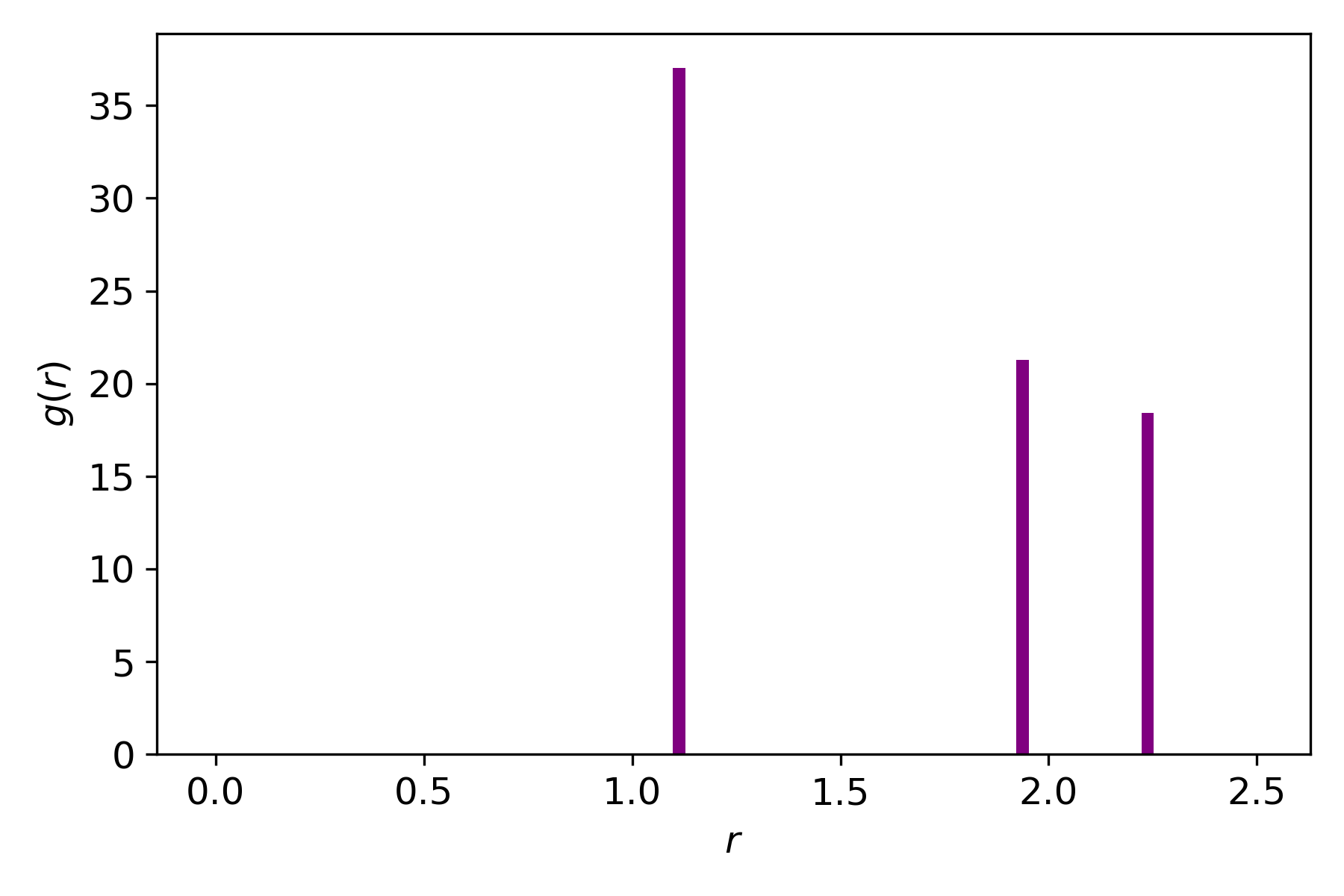}
\caption{RDF for a pure $A$ 2D hexagonal crystal structure. \newline
\textit{Alt text}: RDF plot showing three histogram peaks at $r=1.1159$, $1.9328$, and $2.2318$ (first, second, and third neighbor shells), with zero elsewhere, demonstrating the perfect 2D hexagonal crystal lattice.}
\label{fig:g:pureA}
\end{figure}

In the results below, $r_{\rm max}=r_{\rm cut}$ and
$N_{\rm bins}=100$, so that $\Delta r = 2.0/100 = 0.02$.
For a hexagonal crystal, $N/\Omega=(3 \times 1/6)/(\sqrt{3}a^2/4)=2/(\sqrt{3}a^2)$,
which for $a=1.1159\sigma$ (where $\sigma=1$) is $N/\Omega=0.9273$.
The RDF for the pure $A$ crystal is show in \fig{fig:g:pureA}.
There are three peaks in the range $[0,r_{\rm cut}]$ associated with the first,
second and third neighbor shells at $r=a=1.1159$, $r=\sqrt{3}a=1.9328$ and $r=2a=2.2318$,
which each have 6 neighbors.

Next we consider a binary alloy with concentrations $c_A=0.8$ and $c_B=0.2$.
We begin with the random alloy case where the species of any atom is independent of its
environment and is set randomly based on the
probabilities defined by the concentration values. Since there are now multiple species, we compute partial RDFs,
\begin{equation}
g_{XY}(r) = \frac{\Delta n_{XY}(r)}{2\pi r \Delta r (N_Y/\Omega)},
\label{eq:2dgxy}
\end{equation}
where $N_Y$ is the number of atoms of species $Y$, and $\Delta n_{XY}(r)$ is the average number
of atoms of species $Y$ within a distance $[r,r+\Delta r]$ of an atom of species $X$,
\begin{equation}
\Delta n_{XY}(r) = \frac{1}{N_X}\sum_{i=1}^N \sumdd{j=1}{j \ne i}{N} s_i^X s_j^Y \cH((r_{ij}-r)(r+\Delta r-r_{ij})).
\end{equation}
Here $s_i^X$ and $s_i^Y$ are the species indicator functions defined in Eq.~(2) in MS.

As shown in Appendix~A,
for the case where the species of the atoms are set independently,
we have $g_{XY}(r)=g(r)$. This requirement is helpful in determining a minimal system size
to ensure that a single snapshot provides good statistics as assumed in Eq.~(14) in MS,
which leads to the effective potential definition in Eq.~(16).

\begin{figure}
\centering
\begin{subfigure}{0.44\textwidth}
\centering
\includegraphics[width=\linewidth]{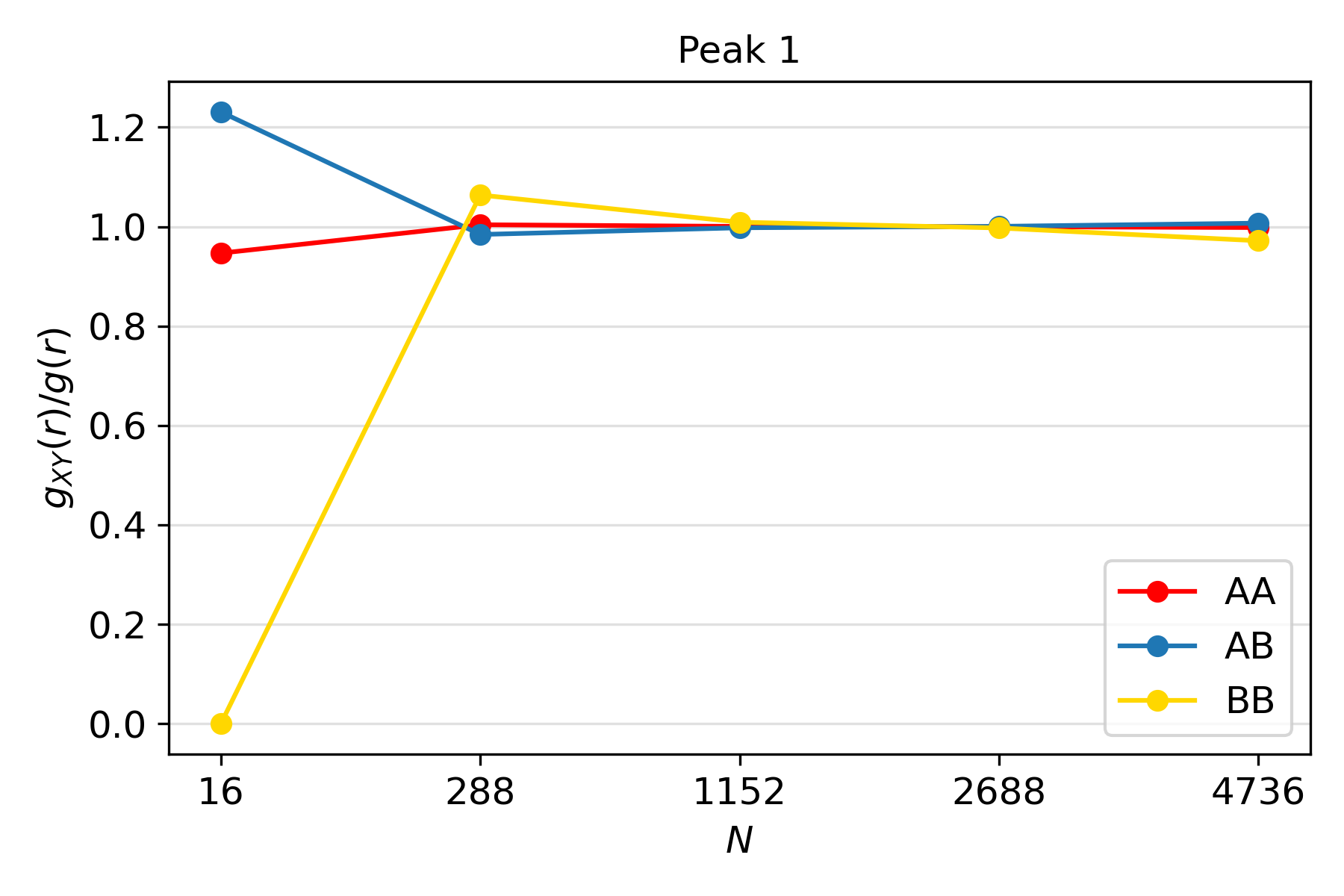}
\caption{}
\end{subfigure}
\hspace{0.5cm}
\begin{subfigure}{0.44\textwidth}
\centering
\includegraphics[width=\linewidth]{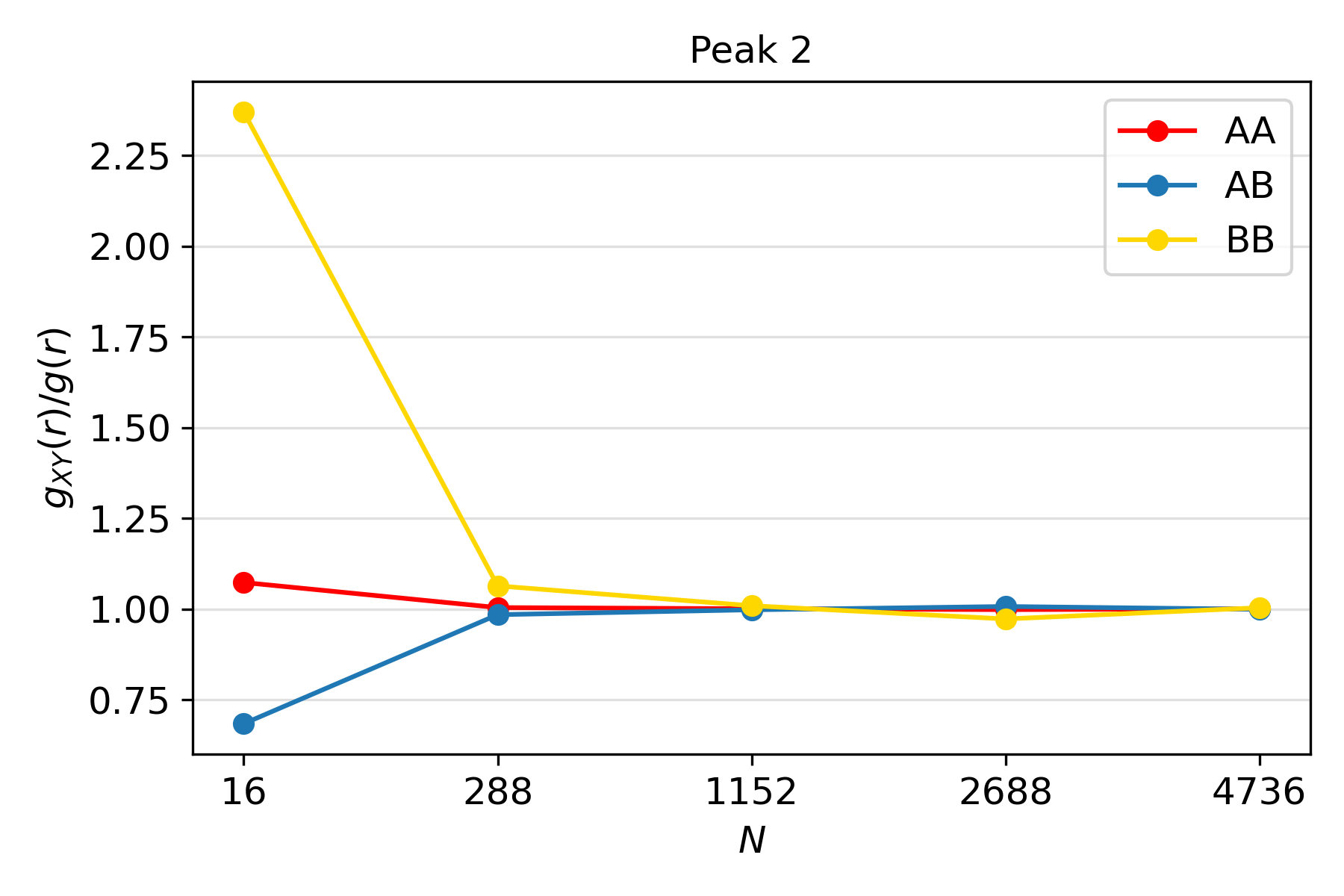}
\caption{}
\end{subfigure} \\[1ex]
\begin{subfigure}{0.44\textwidth}
\centering
\includegraphics[width=\linewidth]{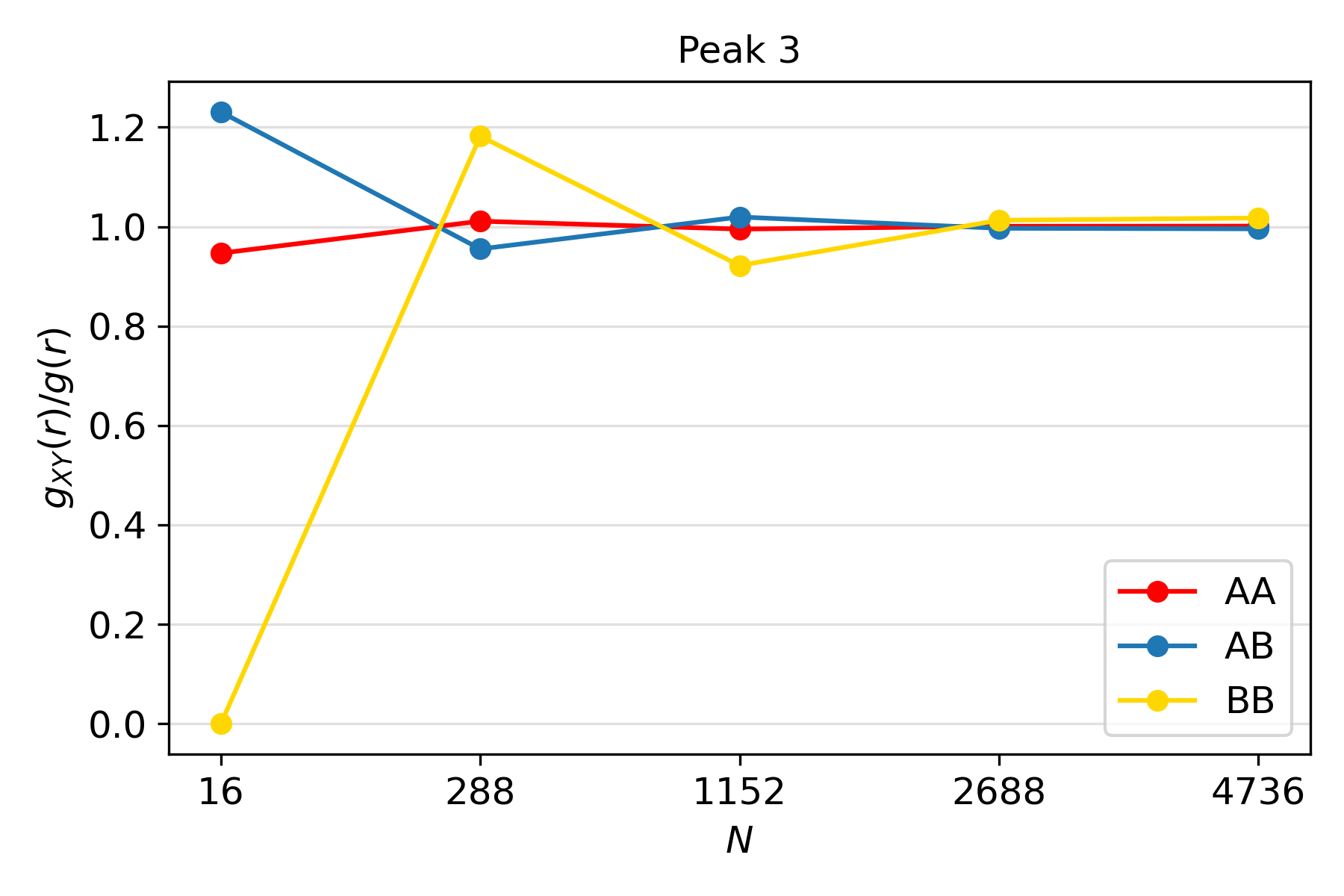}
\caption{}
\end{subfigure}
\caption{Normalized partial RDF versus number
of atoms in 2D binary hexagonal lattice system at the (a) first, (b) second and (c) third
peaks. Calculations performed for system sizes of
 $4\times2$ ($N=16$),
 $16\times9$ ($N=288$),
 $32\times18$ ($N=1152$),
 $48\times28$ ($N=2688$), and
 $64\times37$ unit cells ($N=4736$), with 2 atoms per cell.
 \newline
\textit{Alt text}: Line plots showing the first three peaks for $AA$, $AB$, and $BB$ normalized partial RDF values converging to $1.0$ as N is increased from $16$ to $4736$ atoms.}
\label{fig:g:syssize}
\end{figure}
\fig{fig:g:syssize} presents $g_{XY}(r)/g(r)$ at the three peaks of the RDF for the 2D LJ binary system for different system
sizes. We see that the results are largely converged at $N=2688$ atoms, although the $BB$ interactions remain noisier due to the lower concentration of $B$ atoms.

\begin{figure}
\centering
\includegraphics[width=0.7\linewidth]{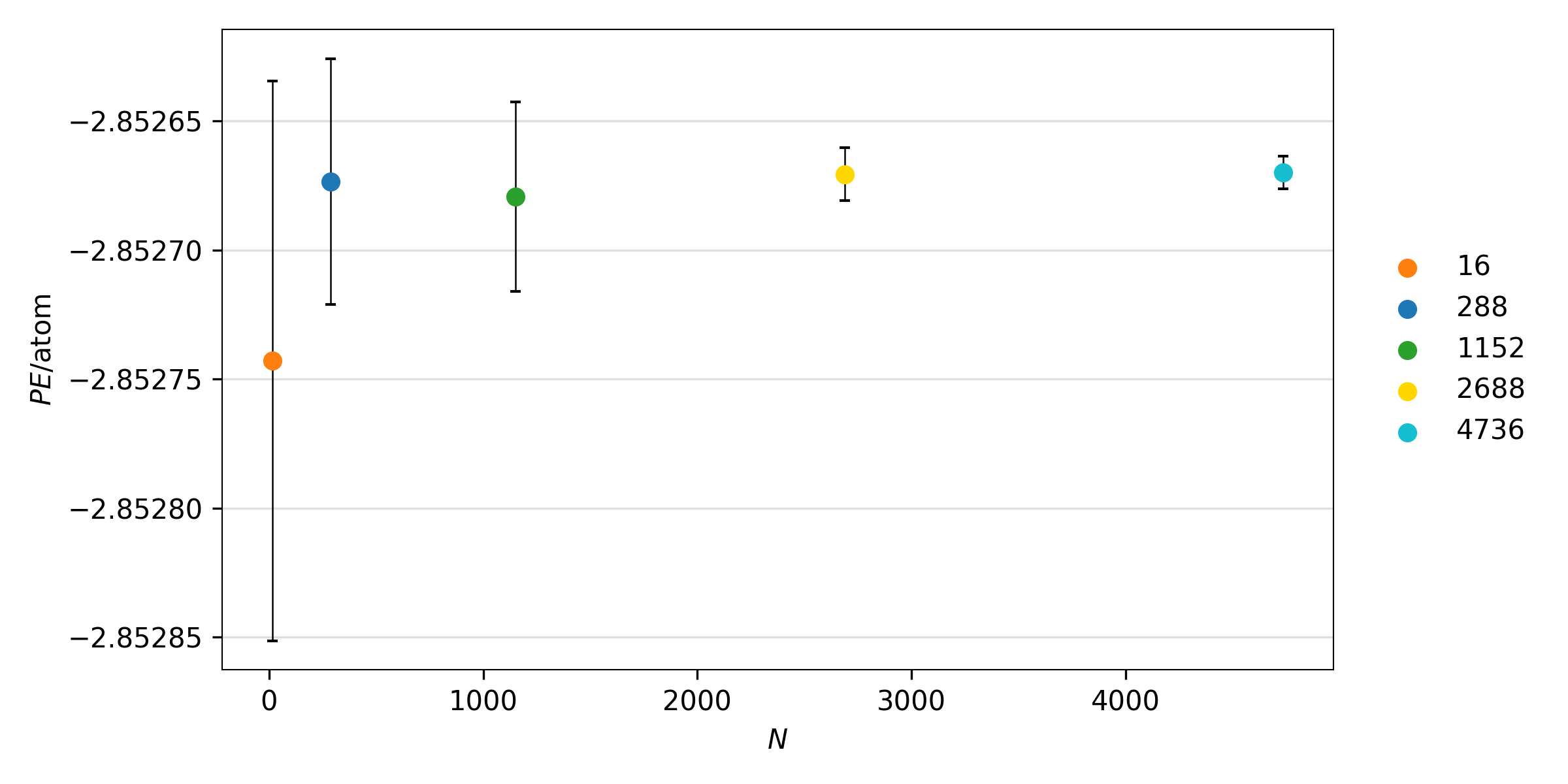}
\caption{Potential energy per atom versus system size of 2D LJ binary alloy. Each point is an average over 10 computations with standard deviation error bars.  Computed at $(\epsilon_{AB}, \epsilon_{BB})=(1.0, 0.8)$.
\newline
\textit{Alt text}: Scatter plot with one averaged point per system size N, each with vertical error bars. The error bars shrink as N increases from $16$ to $4736$ atoms, and the PE/atom converges around $-2.85267$ between N=$2688$ and N=$4736$.}
\label{fig:g:PE_RVE}
\end{figure}

\fig{fig:g:PE_RVE} presents an energy convergence study to determine the necessary minimum system size. Increasing from $N=16$ to $N=4736$ atoms, the potential energy per atom is computed for 10 realizations (different random seeds) and standard deviation error bars are plotted in \fig{fig:g:PE_RVE}. We again see that the results converge around $N=2688$ atoms. To balance simulation costs with accuracy, TS calculations are performed with this system size (2688 atoms).

\subsection{Monte Carlo Simulations of Ground State LJ Structures}
To find the ground state configuration of $A$ and $B$ atoms in the 2D hexagonal lattice, Monte Carlo (MC) swaps of atomic species are performed. The MC swaps are performed with a Metropolis acceptance criteria over a series of decreasing temperatures. A hyperparameter search was performed to determine the optimal sequence of temperatures and number of MC swaps.

We begin by looking at a small system of 16 atoms. At this system size, we can visualize the atom swapping performed at each temperature. At higher temperatures, atoms swap rapidly, and as the temperature is lowered, the atoms settle into the ground state configuration. It is found that it is best to start at a relatively high temperature of $T = 4.0$ (LJ temperature units), and halve each time until a final value of $T = 1.953125\times 10^{-3}$ to ensure that atoms have fully explored the configuration space.

The system is then scaled up to 2688 atoms. For the fixed number of temperature steps determined previously, the final energy of the system for increasing numbers of MC swaps is recorded. Larger systems require more swaps to reach a ground state energy, so this is increased from $40{,}000$ swaps for the 16 atom system to $320{,}000$ swaps for the 2688 atom system.

\begin{figure}
\centering
\includegraphics[width=0.6\linewidth]{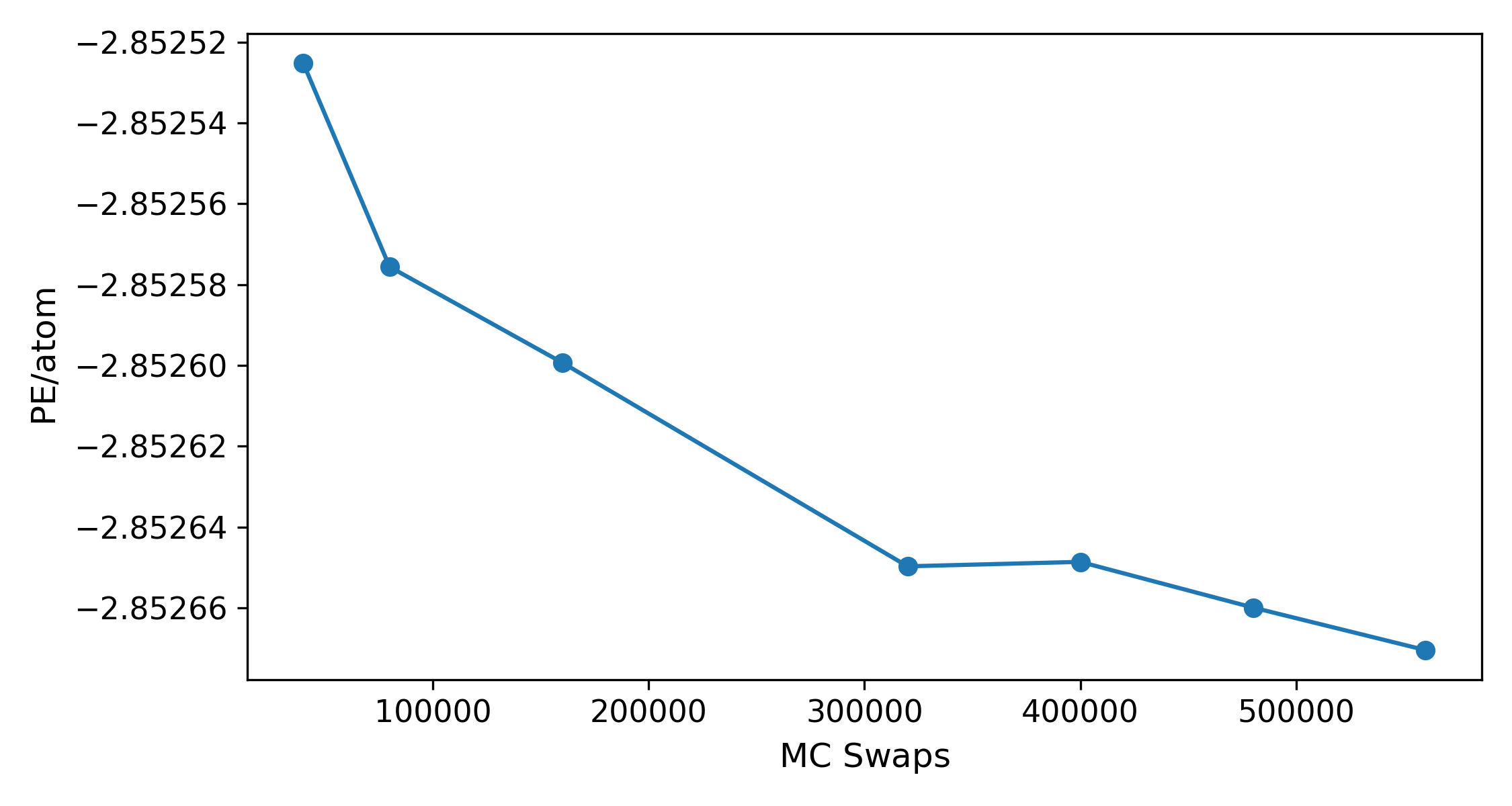}
\caption{Final potential energy per atom vs number of Monte Carlo swaps in the 4736 atom 2D binary system. $\epsilon$ values chosen from the SS region of the phase diagram (Fig.~2 in MS).
\newline
\textit{Alt text}: Line plot of potential energy per atom decreasing monotonically from $-2.85252$ to $-2.85266$ as MC swaps increase from $40{,}000$ to $560{,}000$, with the change in energy leveling off towards convergence after around $320{,}000$ swaps.}
\label{fig:MC_convergence}
\end{figure}

The necessary number of MC swaps is determined through a convergence study, weighing simulation cost with final energy. For a system of 4736 atoms it is seen from \fig{fig:MC_convergence} that as the number of MC swaps is increased from $40{,}000$ to $320{,}000$, the final energy continues to drop rapidly. However, as this is increased beyond $320{,}000$ to $560{,}000$ swaps, the energy difference is smaller. While the time required to perform the simulation consistently increases, the energy does not decrease at the same rate. Therefore, $320{,}000$ MC swaps was chosen as a reasonably converged number to achieve a ground-state structure. The minimal lower energy that can be achieved through further MC swaps is not worth the cost of the increased simulation time, especially given that we need to perform $1{,}000$ simulations at minimum to achieve properly averaged results across $\epsilon$ space.

\subsection{RDFs and Warren-Cowley Parameters of Binary LJ Structures}
\label{sec:2d_rdfs}
\begin{figure}
\centering
\begin{subfigure}{0.44\textwidth}
\centering
\includegraphics[width=\linewidth]{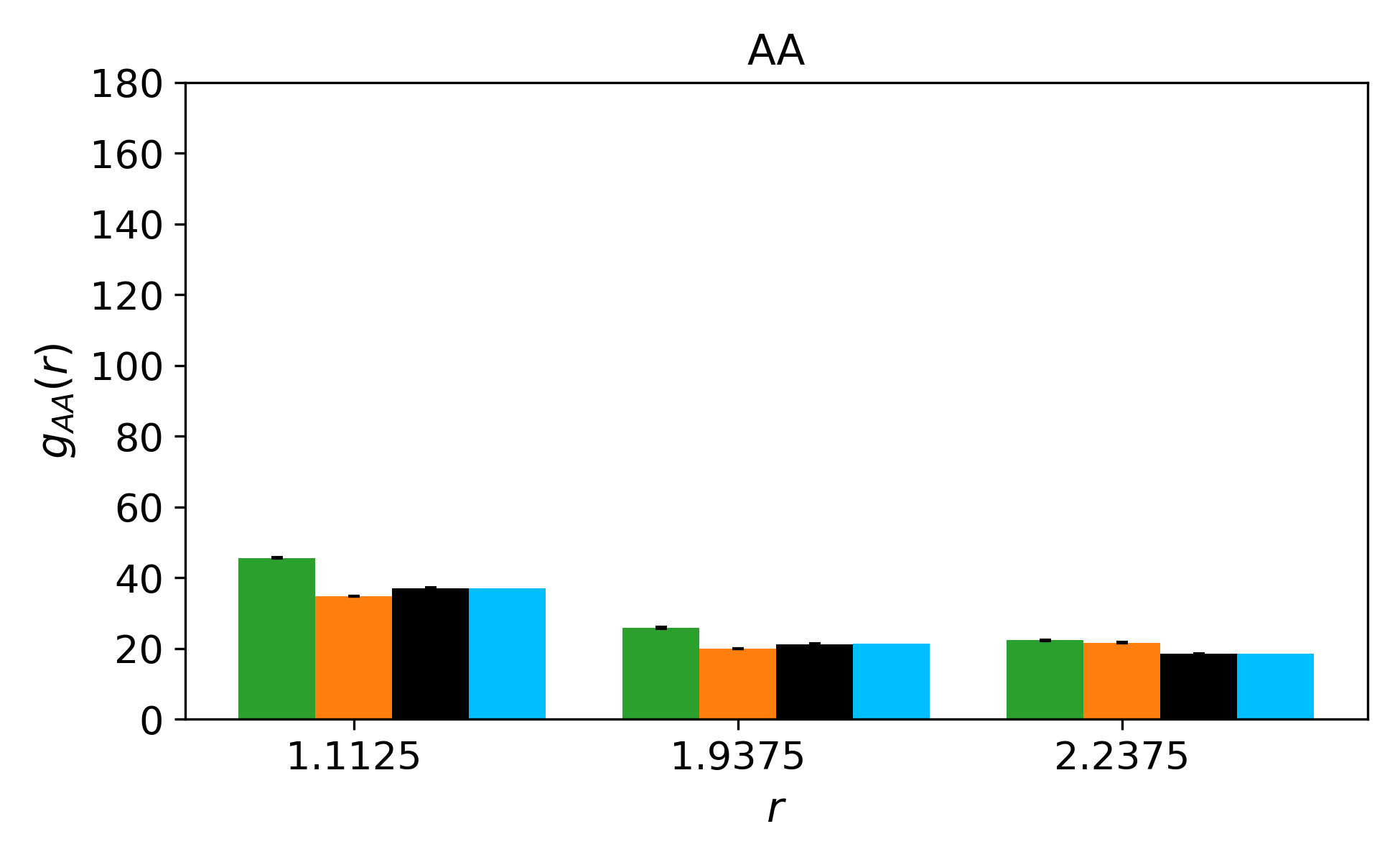}
\caption{}
\end{subfigure}
\begin{subfigure}{0.44\textwidth}
\centering
\includegraphics[width=\linewidth]{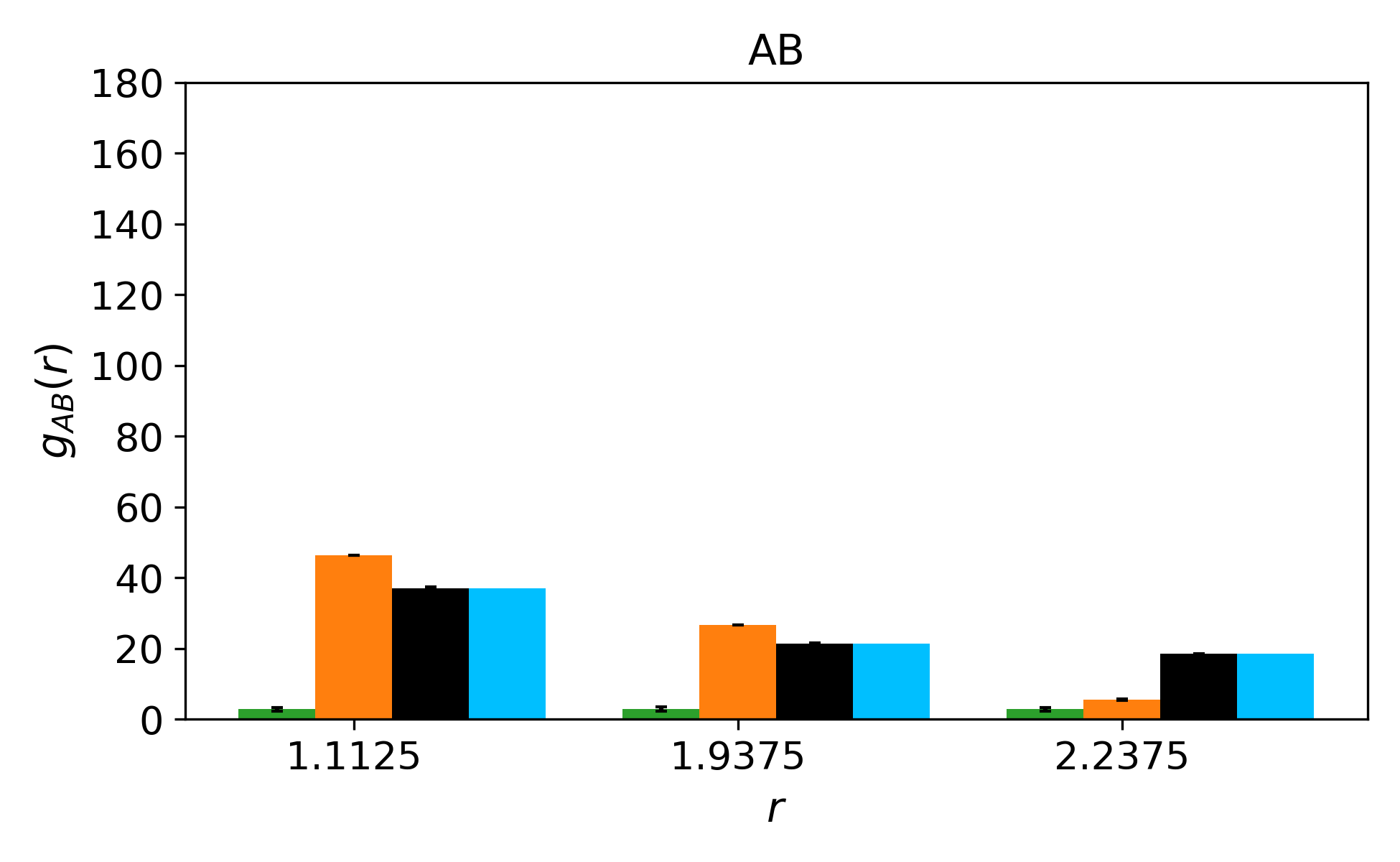}
\caption{}
\end{subfigure} \\[1ex]
\begin{subfigure}{0.44\textwidth}
\centering
\includegraphics[width=\linewidth]{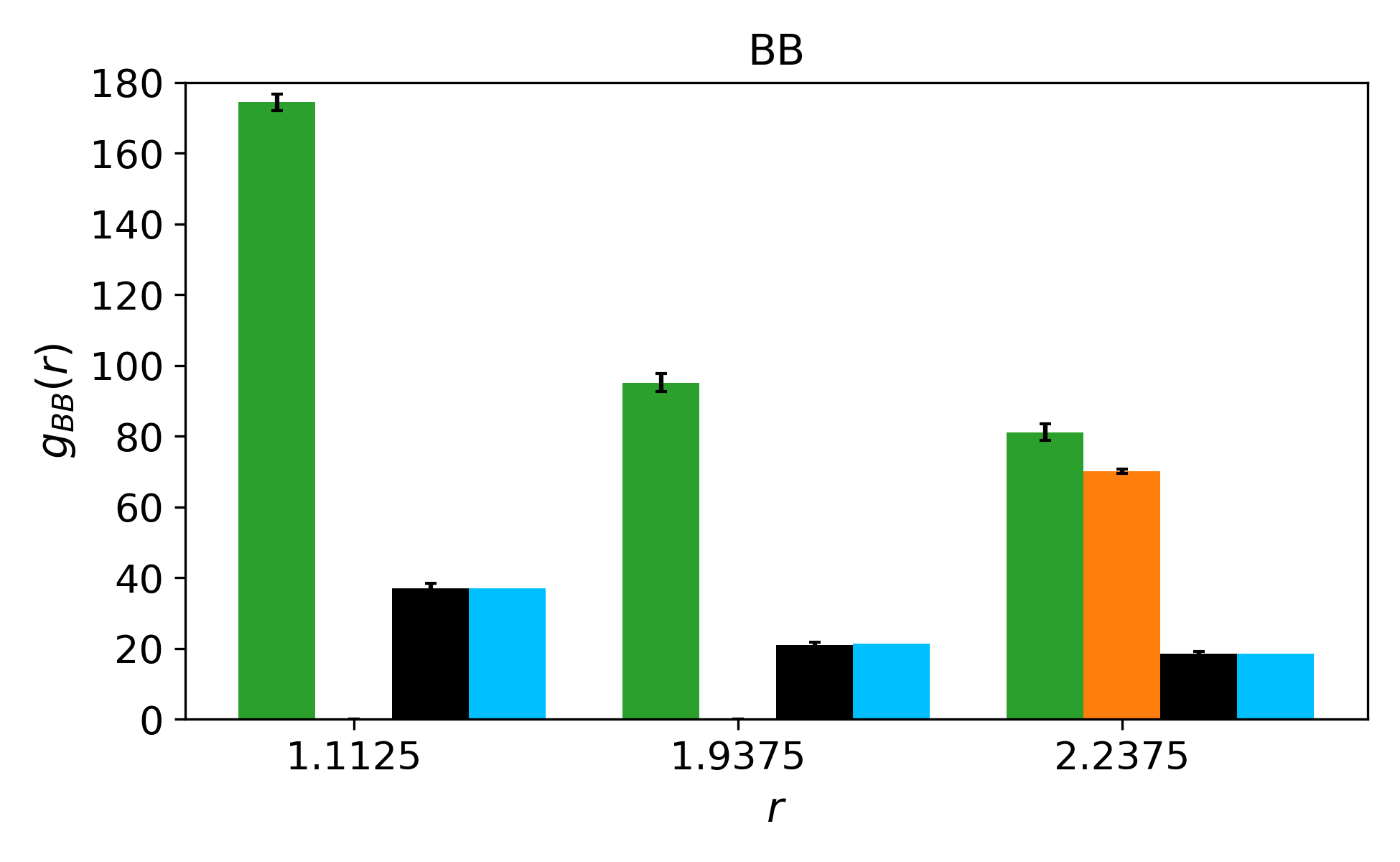}
\caption{}
\end{subfigure}
\caption{
Partial radial distribution functions at first, second, and third neighbors for (a) $AA$, (b) $AB$ (c) $BB$ interactions in the 2D binary LJ system. Calculations performed for system size of
 $48\times28$ unit cells ($N=2688$). Bar chart colors correspond to $\epsilon$ values from different regions of the phase diagram (Fig.~2 in MS): green=PS, orange=SS, and black=no-LO. The cyan bars correspond to the total (non-partial) RDF.  Partial RDFs shown are for $(\epsilon_{AB}/\epsilon_{AA}, \epsilon_{BB}/\epsilon_{AA}) = (1,3), (3,2), (1,1).$
 \newline
 \textit{Alt text}: Bar charts of partial RDFs at three peak r values ($1.1125$, $1.9375$, and $2.2375$), with four bars at each peak. In (a), all bars are nearly equal. In (b), PS is close to zero. In (c), PS is high at all three peaks, while SS is zero at the first two peaks and nonzero at the third. In all three panels, the no-LO bars match the total RDF. Error bars are very small.}
\label{fig:g:partials_g_LO}
\end{figure}

Beyond visually validating the LO states, the results from the TS simulations are also examined in terms of their radial distribution functions. Partial RDFs for $AA$, $AB$, and $BB$ interactions from each region in the phase diagram (Fig.~2 in MS) are computed for first-, second-, and third-neighbor distances and shown in \fig{fig:g:partials_g_LO}. The colors correspond to those in the phase diagram (see figure caption). The blue bars correspond to the total RDF of the system. The partial RDF results are averaged over 10 realizations with error bars shown.
As expected from the derivation in Appendix A in MS, the black bars of the randomly distributed system are equal to the total RDF.
It can be seen that the three LO states exhibit  distinct RDF profiles in accordance with their LO states.

Interestingly, the partial RDFs for all points within the same region of the phase diagram  are found to be equal, therefore \fig{fig:g:partials_g_LO} represents all partial RDFs for the ground state configurations of this system. This observation is explained by the no-LO line relation in Eq.~(C1) in MS. To the left of the no-LO line, $\epsilon_{BB}/\epsilon_{AA} > 2\epsilon_{AB}/\epsilon_{AA}-1$, which means that there is a preference for $BB$ bonds since these are lower in energy and hence the same clumping of $B$ atoms is observed at all points in the PS region. In the SS region, the effect is opposite with $AB$ bonds preferred over $BB$ and hence the spreading of $B$ atoms.

\begin{figure}
\centering
\includegraphics[width=0.5\linewidth]{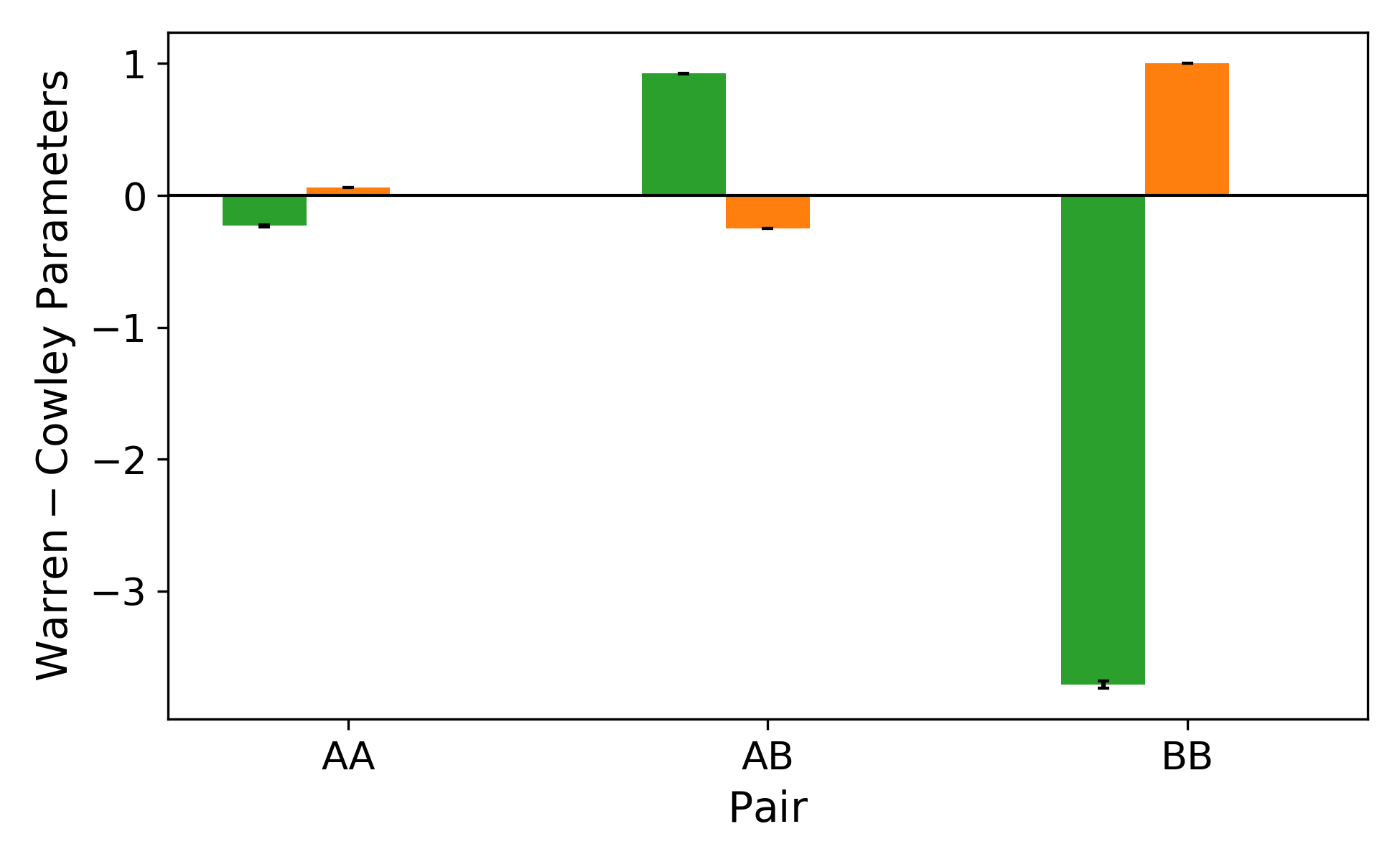}
\caption{Warren-Cowley order parameters for nearest neighbor interactions of the 2D LJ binary system. Calculations performed for system size of $48\times28$ unit cells ($N=2688$). Colors correspond to those from phase diagram regions. $(\epsilon_{AB}/\epsilon_{AA}, \epsilon_{BB}/\epsilon_{AA}) = (1,3), (3,2), (1,1).$
\newline
\textit{Alt text}: Bar chart comparing Warren-Cowley order parameters for $AA$, $AB$, and $BB$ pairs. The $AA$ pair remains close to zero for all cases. The $AB$ pair is around $+1.0$ for PS and weakly negative for SS. The $BB$ pair is strongly negative for PS (around $-3.5$) and $+1.0$ for SS. Error bars are very small. }
\label{fig:warren_cowley}
\end{figure}

We can further characterize the ordering seen in each LO state through the Warren-Cowley (WC) order parameters \cite{cowley:1965, norman:warren:1951}, defined by
\begin{equation}
    \alpha_{ij}^{XY} = 1 - \frac{p(s_i^X=1, s_j^Y=1 \,|\, \cC)}{c_X c_Y}
\end{equation}
for atoms $i$ and $j$ of species $X$ and $Y$, where the probability is defined in Eq.~(14) in MS. Note that the standard Warren-Cowley parameter formula involves a conditional probability, but it can trivially be shown that the given formulation with a joint probability is also correct \cite{rao:curtin:2022}. \fig{fig:warren_cowley} displays the WC parameters for each atom type combination at nearest neighbor distance. The three LO states are plotted in accordance with their coloring in Fig.~2 in MS. A WC parameter that is $<0$ indicates an attractive interaction between atom types, while $>0$ indicates repulsion. A WC parameter equal to $0$ suggests that there is no LO, which explains why the no-LO state is not visible in \fig{fig:warren_cowley}. Only first neighbor WC parameters were plotted as the second neighbor parameters are nearly identical.

\subsection{TS Energy Contour Plot of LJ Structures}
The TS energy contour plot (Fig.~3 in MS) can be understood by considering the different LO states on the phase diagram. In the SS region, $B$ atoms are not (or are minimally) interacting, which explains the vertical energy contour lines in this region since there is little dependence on $\epsilon_{BB}$. In the PS region of the phase diagram $B$ atoms form clumps, and here we see that the energy has little dependence on $\epsilon_{AB}$. This  makes sense since there are fewer $AB$ interactions as these are only present along the borders of the clumps of $B$ atoms.

\subsection{AA Energy Contour Plot of LJ Structures}
The AA energy contour plot (Fig.~4a in MS) can be also be understood.
Since, as explained in \sect{sec:2d_rdfs}, the partial RDFs are the same at all points in a given region of the phase diagram, the increasing error in AA method predictions is due to an increase in the $BB$ bond strength rather than an increase in the degree of LO. We note that errors are larger to the left of the no-LO line (corresponding to the PS region in Fig.~2 in MS) relative to the right (SS region). Referring to the energy difference expression in Eq.~(D11) in MS, we see that the error in the direction $t$ perpendicular to the no-LO line scales with $\zeta - 2\gamma$. Due to the nature of the partial RDF profiles in the PS and SS regions, this factor is considerably larger in magnitude (around four times) in the PS region than in the SS region. A systematic increase in error occurs on both sides of the no-LO line, but this explains why the error grows more rapidly with distance to the left of the no-LO line.

\section{3D EAM N\MakeLowercase{i}A\MakeLowercase{l}}

\subsection{Monte Carlo Simulation Convergence of NiAl}

\begin{figure}
\centering
\includegraphics[width=0.6\linewidth]{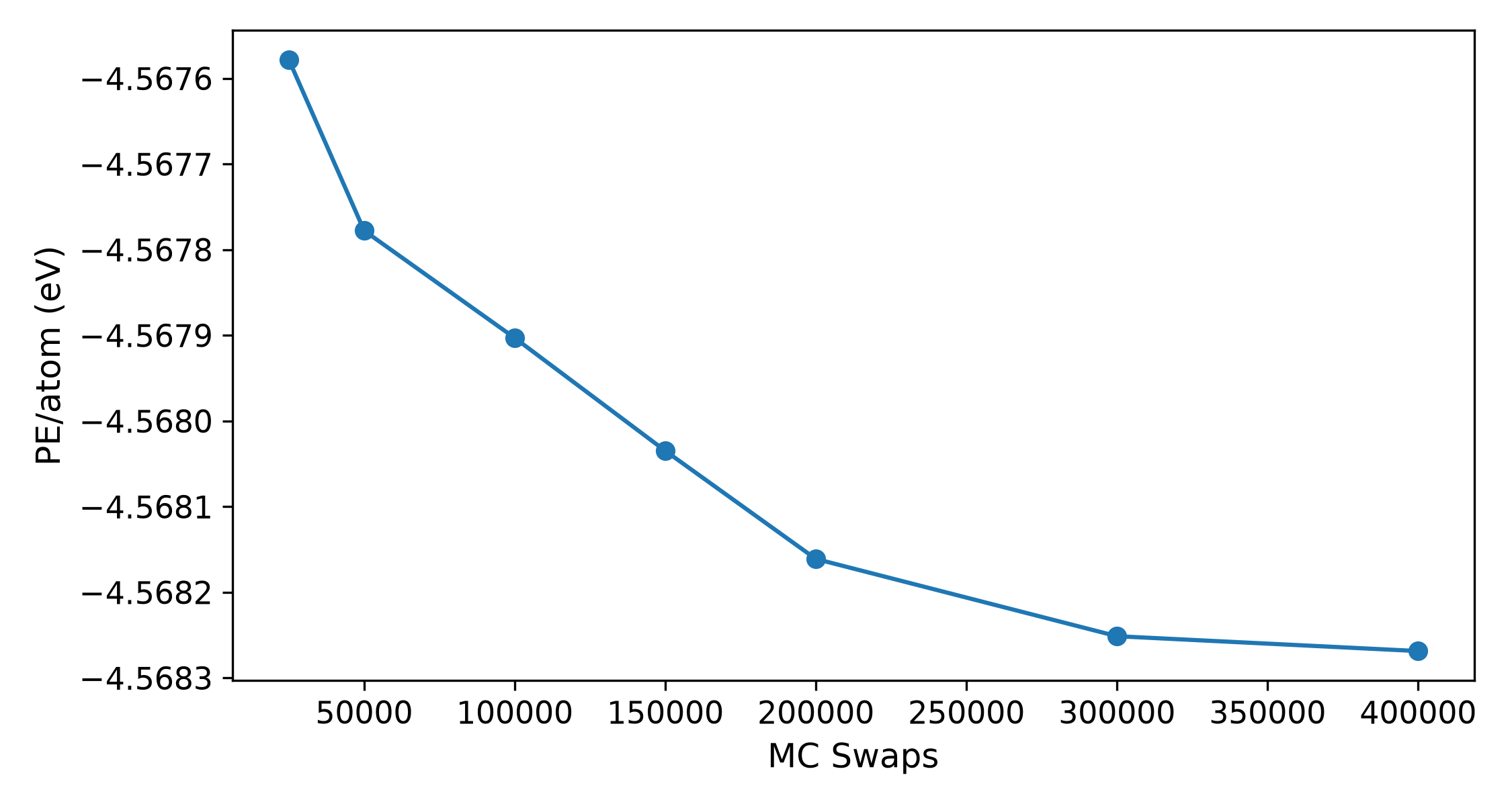}
\caption{Final potential energy per atom vs number of Monte Carlo swaps in the 3D fcc Ni$_{x}$Al$_{1-x}$ system for $x=0.85$. 
\newline
\textit{Alt text}: Line plot of potential energy per atom decreasing monotonically from $-4.5676$ to $-4.5682$ as MC swaps increase from $25{,}000$ to $400{,}000$, with the change in energy leveling off around $300{,}000$ swaps.}
\label{fig:MC_convergence_nial}
\end{figure}

Similarly to Sec.\ I.B, we perform a convergence study to determine the necessary number of MC swaps per temperature step to achieve a sufficiently converged, low-energy ordered 3D NiAl structure. For this, we take 3D fcc Ni$_{x}$Al$_{1-x}$ with $x=0.85$ as a representative system with an off-stoichiometric composition. This system was used in the surface energy calculations of Sec.\ III.B.2 in MS. For this system of $10{,}976$ atoms, \fig{fig:MC_convergence_nial} demonstrates that the final energy decreases with increasing number of swaps, with the largest changes occurring at lower swap counts. 
Increasing the number of MC swaps from $25{,}000$ to $300{,}000$ results in a continued decrease in energy, while further increasing up to $400{,}000$ swaps produces little additional change. However, the total difference in energy across the entire range of swap counts is only $0.015\%$, indicating that the system's energy is relatively insensitive to the precise number of swaps once sufficient species rearrangement has occurred. The initially random configuration has an energy of $-4.5179$ eV, corresponding to a $1.09321\%$ and $1.10832\%$ difference from the configurations at the first and last plotted points, respectively. However, increasing from $100{,}000$ to $300{,}000$ swaps changes the energy by only $0.007\%$. Thus, while additional MC swaps continue to lower the energy, the magnitude of this improvement quickly becomes negligible. We take $100{,}000$ MC swaps as a reasonable level of convergence to avoid unnecessary computational expense, and use this for all NiAl and FeCrNi systems.

\subsection{Radial Distribution Functions for BCC N\MakeLowercase{i}A\MakeLowercase{l}}

\begin{figure}
\begin{subfigure}{0.49\textwidth}
\centering
\includegraphics[width=\linewidth]{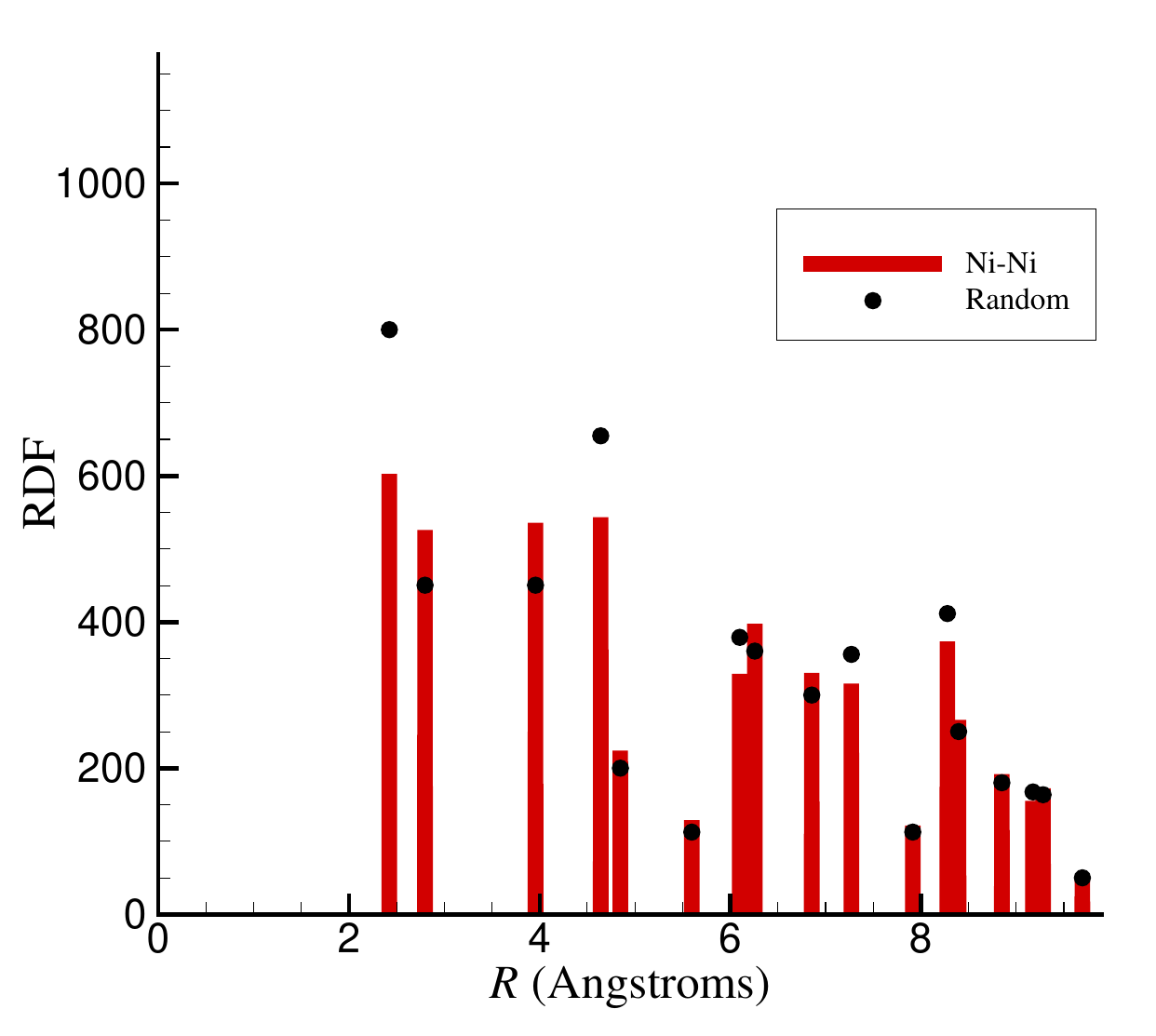}
\caption{}
\end{subfigure}
\hfill
\begin{subfigure}{0.49\textwidth}
\centering
\includegraphics[width=\linewidth]{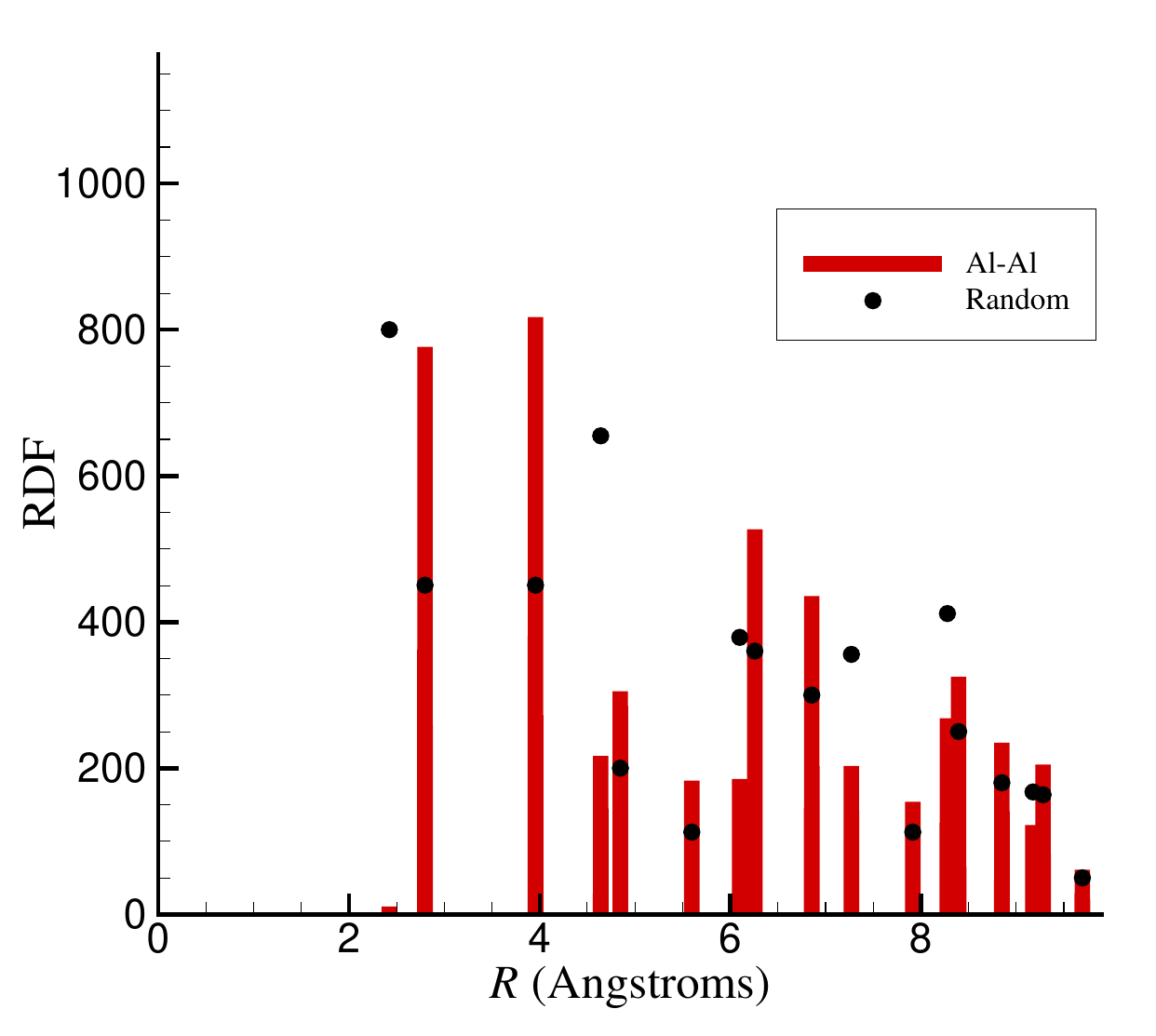}
\caption{}
\end{subfigure}
 \vskip\baselineskip
\begin{subfigure}{0.49\textwidth}
\centering
\includegraphics[width=\linewidth]{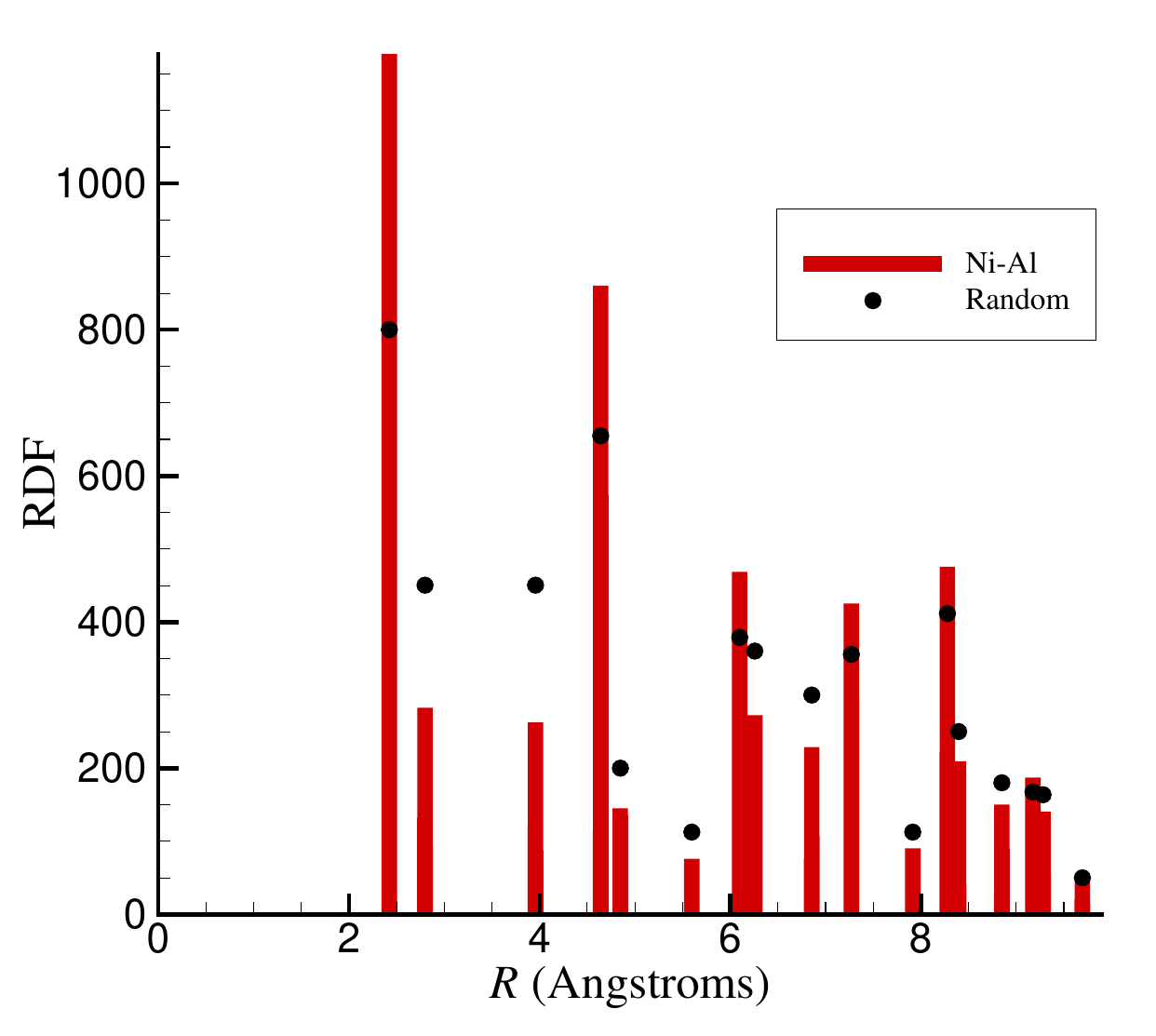}
\caption{}
\end{subfigure}
\hfill
\begin{subfigure}{0.49\textwidth}
\centering
\includegraphics[width=\linewidth]{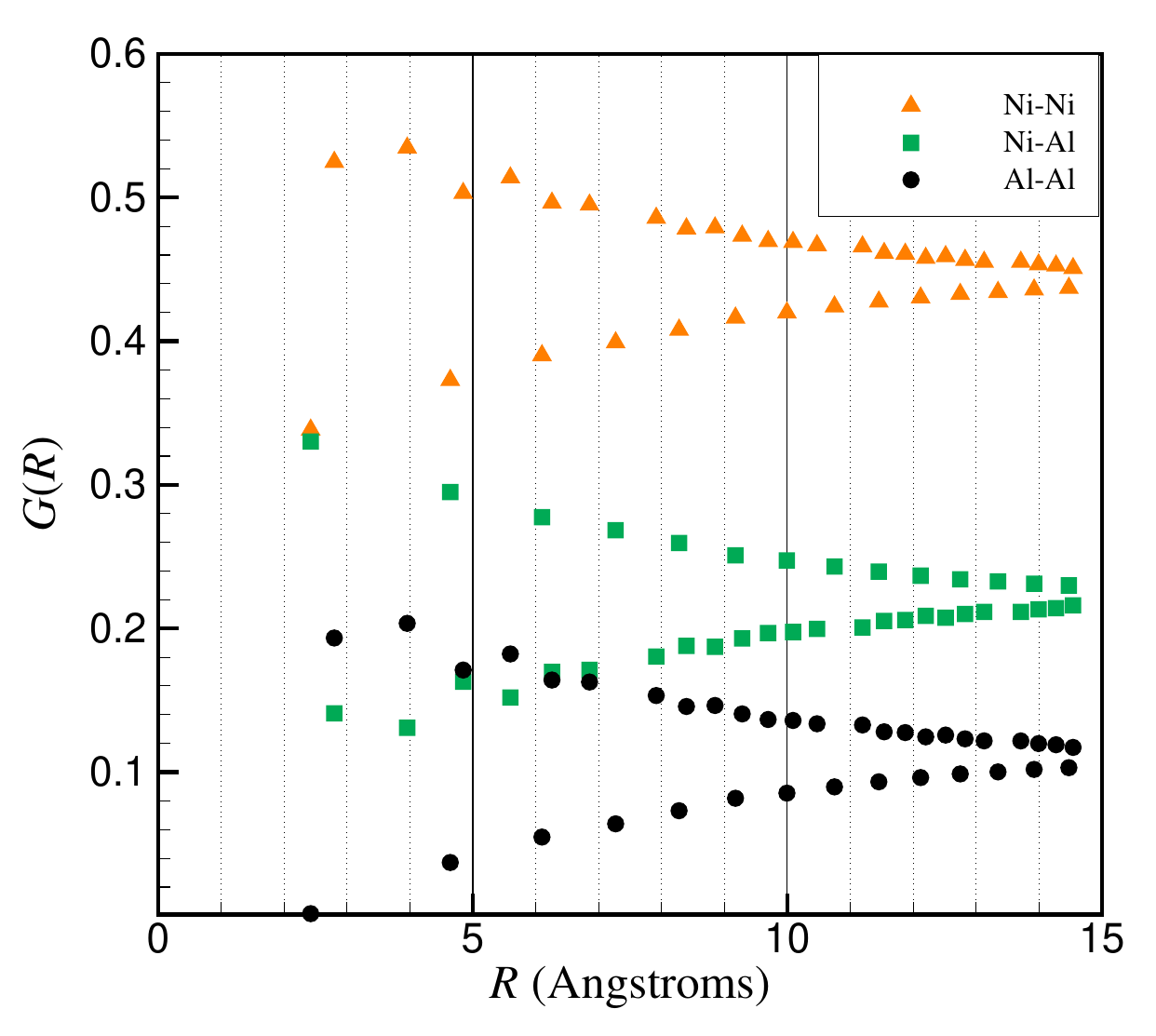}
\caption{}
\end{subfigure}
\caption{(a)-(c) Partial RDFs of NiAl for the ordered TS sample, compared with the expected RDF for a random distribution of the Ni and Al atoms. (d) resulting $G(R_0)$ for the \nial\ alloy.
\newline
\textit{Alt text}: Partial RDFs of (a) Ni-Ni, (b) Al-Al, and (c) Ni-Al for the ordered bcc \nial TS sample compared with a random sample, showing strong ordering through large deviations between some peaks and agreement on others. (d) Scatter plot of $G(R_0)$ versus interatomic separation, with Ni-Ni values generally highest, followed by Ni-Al and then Al-Al.}
\label{fig:nial-rdfs}
\end{figure}

Here, as an example system, we take the bcc phase of 3D \nial, and we examine the partial RDFs, the resulting $G(R)$, and the effect these have on the effective pair potential.

In \fig{fig:nial-rdfs}, we compare the partial RDFs for the Ni--Ni, Ni--Al and Al--Al pairs in the ordered TS sample after the same MC annealing protocol as described in the MS (recall that this configuration is used to generate $G(R)$ for the LOAA effective pair potentials).  In each case, we compare to the bcc RDF for the case where the atomic species are randomly assigned, which is just the RDF for ideal bcc ignoring species.  These plots emphasize the effect of the ordering.  For example, we see virtually no Al--Al bonds at the nearest neighbor distance, compared to the number of Ni--Ni and Ni--Al bonds.  The LOAA formulation takes this into account through the $G(R)$ functions, as shown in \fig{fig:nial-rdfs}(d).  When considering a bond that is at some distance $R_0$ in the reference configuration, its contribution is a weighted sum of the three types of interactions, not only by average overall concentration (as in AA), but also by the relative probability of finding those bonding pairs in the ordered structure.

\begin{figure}
\begin{subfigure}{0.49\textwidth}
\centering
\includegraphics[width=\linewidth]{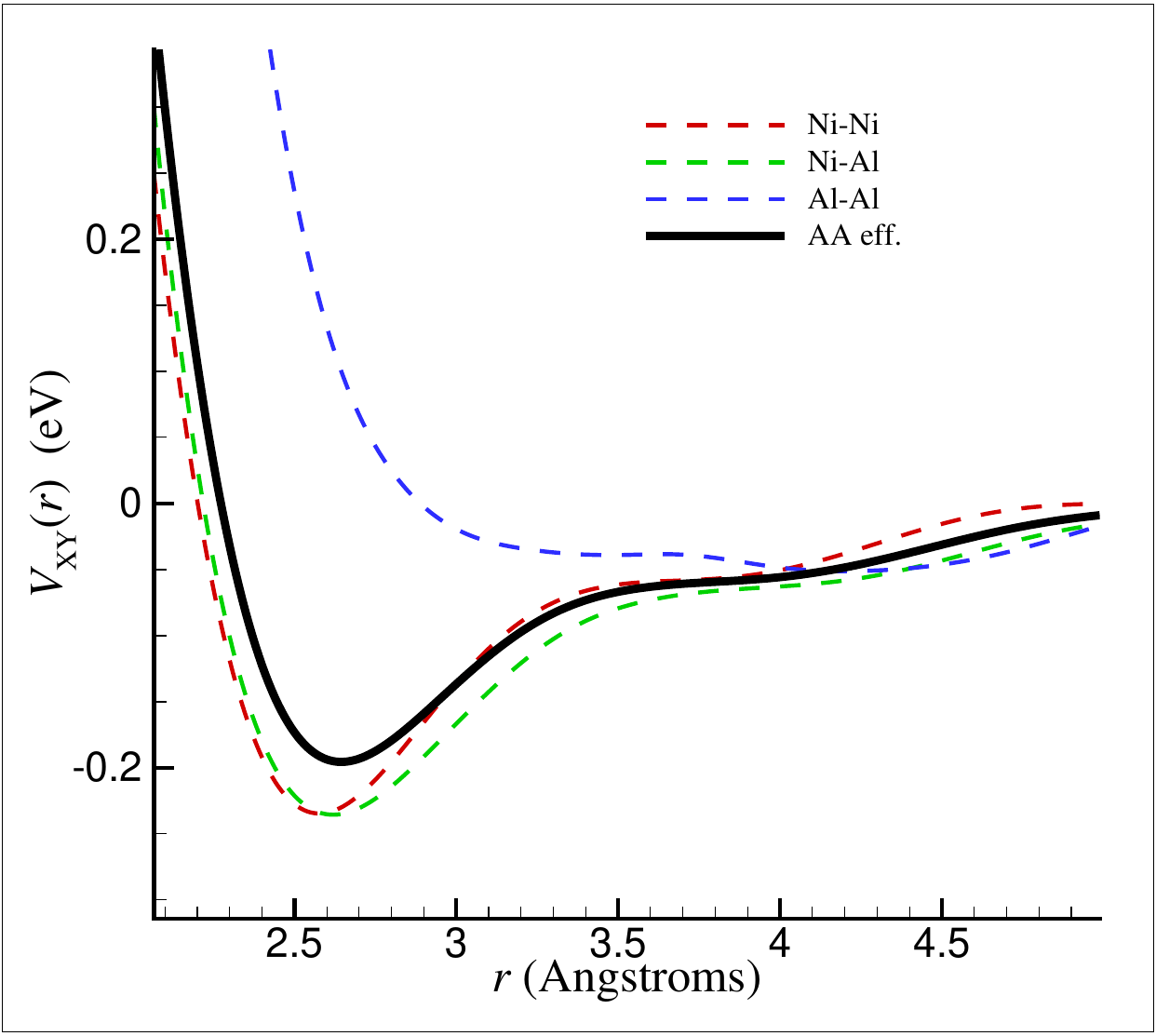}
\caption{}
\end{subfigure}
\begin{subfigure}{0.49\textwidth}
\centering
\includegraphics[width=\linewidth]{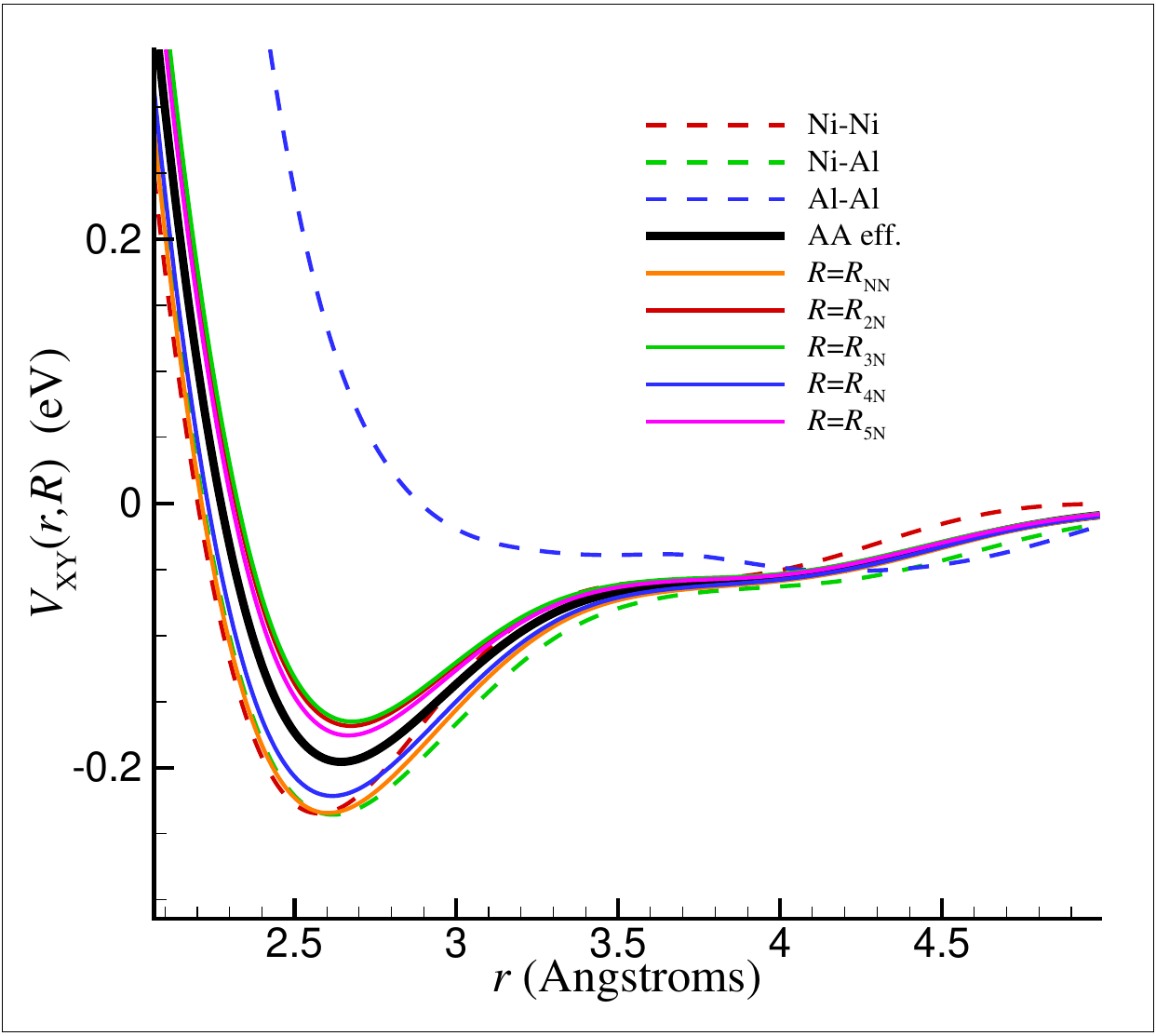}
\caption{}
\end{subfigure}
\caption{(a) Comparison of the TS pair potentials with the effective pair potential for the AA formulation, for the bcc \nial\ structure.  (b) Further comparison to the LOAA effective pair potentials, one for each of the first 5 non-zero values in the RDF functions.
\newline
\textit{Alt text}: Line plots of the effective pair potentials vs $r$ for (a) AA, showing the AA curve is an average of the three inter-species pair potentials, and (b) LOAA, where each peak's pair potential curve closely surrounds the AA curve.}
\label{fig:veff}
\end{figure}

The effective pair potentials for AA and LOAA are presented in \fig{fig:veff}.  In AA, $V_{\rm eff}(r)$ is simply a blend of the three inter-species pair potentials, weighted by their respective compositions.  In LOAA, each peak in the RDF curves, corresponding to a different reference configuration pair distance, $R_0$, produces a different effective pair potential, through the effect of $G(R)$.  This is seen in \fig{fig:veff}(b), where the LOAA potential for the first five neighbor shell distances in the reference configuration are compared to the original EAM potentials and the blended result from AA.  

%This highlights the pronounced differences between the AA and LOAA formulation,  allowing LOAA to predict the stabilization of the B2 phase at high temperature.

\subsection{Surface Energies of FCC N\MakeLowercase{i}A\MakeLowercase{l}}

\begin{figure}
\begin{subfigure}{0.45\textwidth}
\centering
\includegraphics[width=\linewidth]{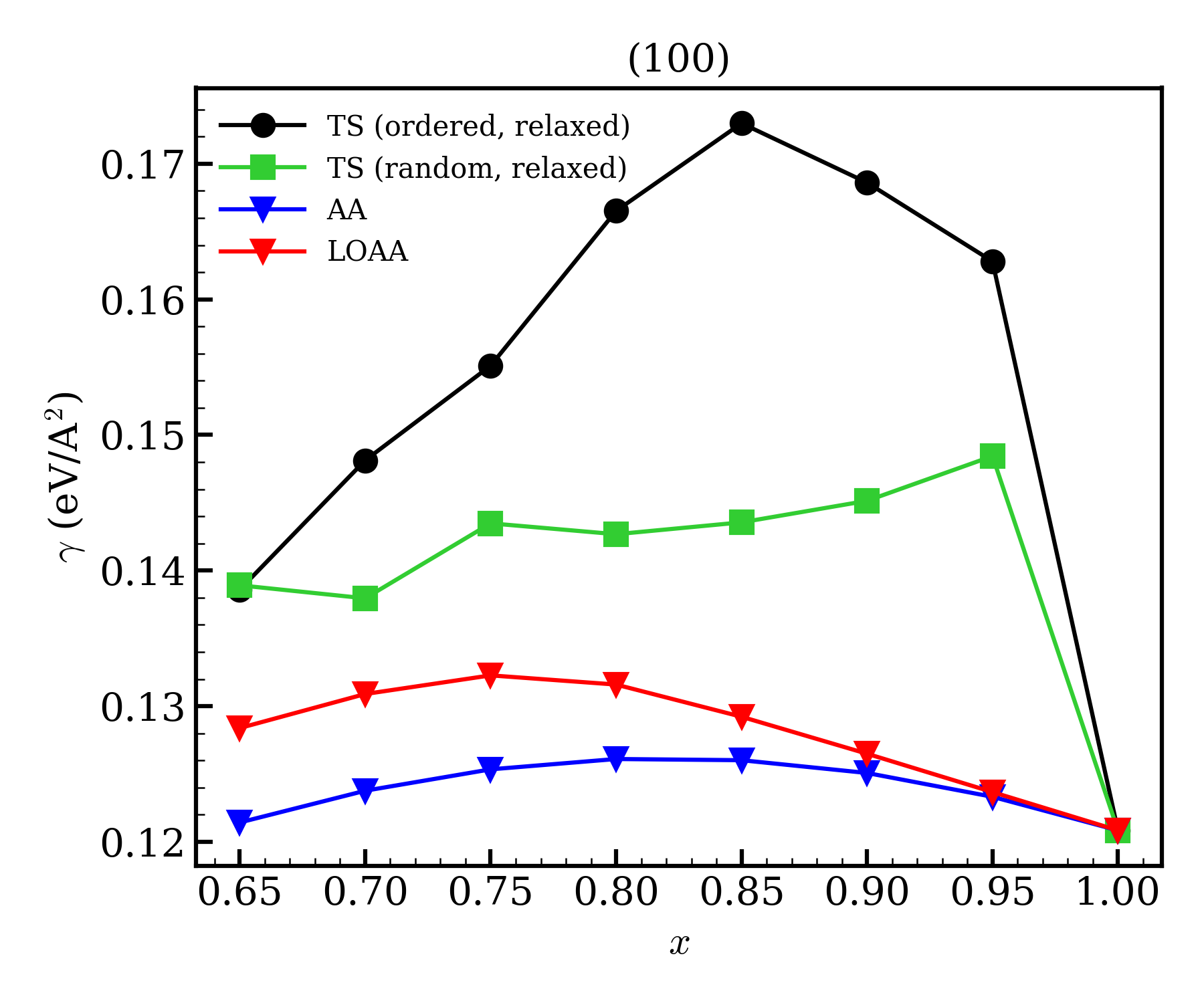}
\caption{}
\end{subfigure}
\hfill
\begin{subfigure}{0.45\textwidth}
\centering
\includegraphics[width=\linewidth]{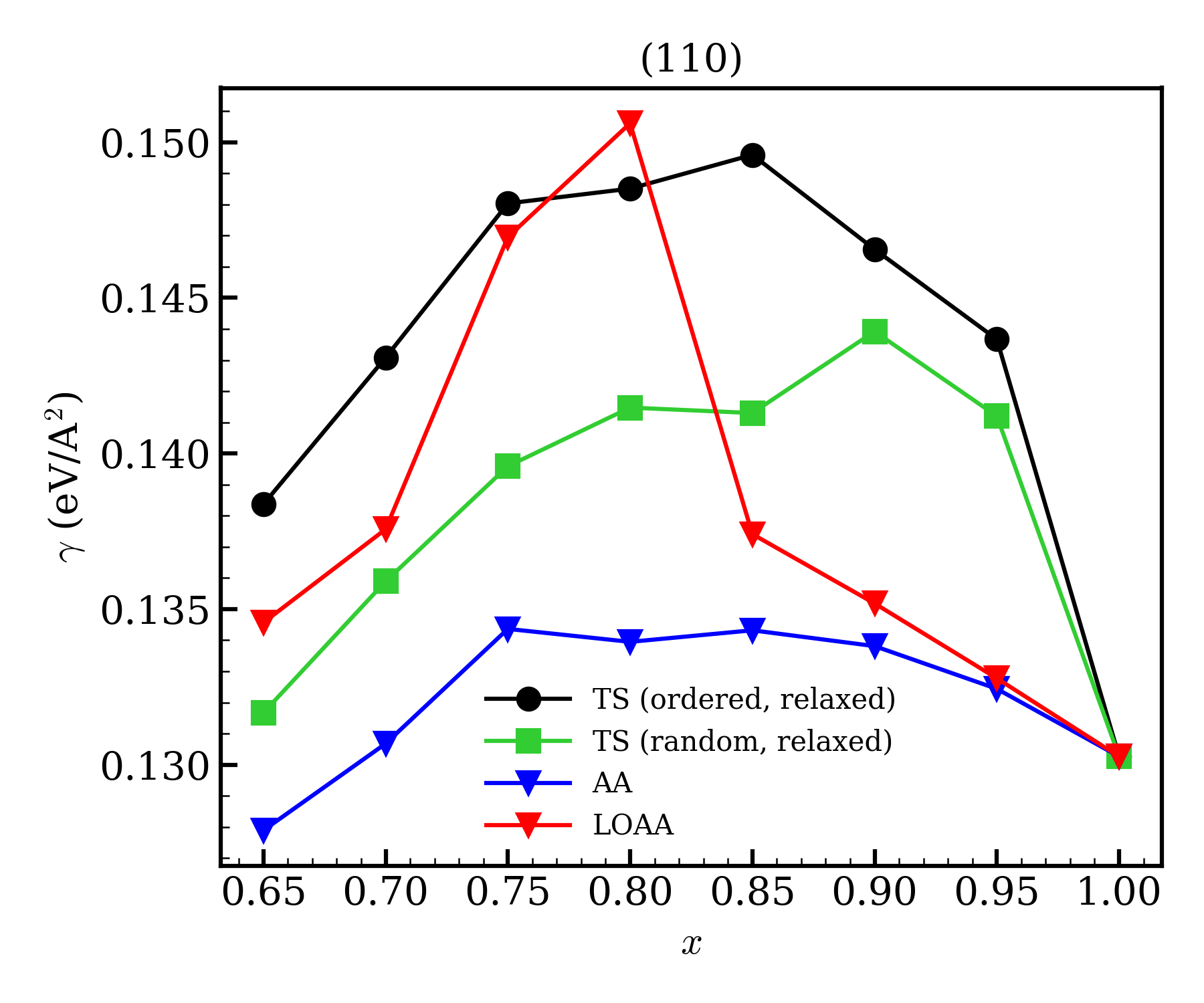}
\caption{}
\end{subfigure}
\vskip\baselineskip
\begin{subfigure}{0.45\textwidth}
\centering
\includegraphics[width=\linewidth]{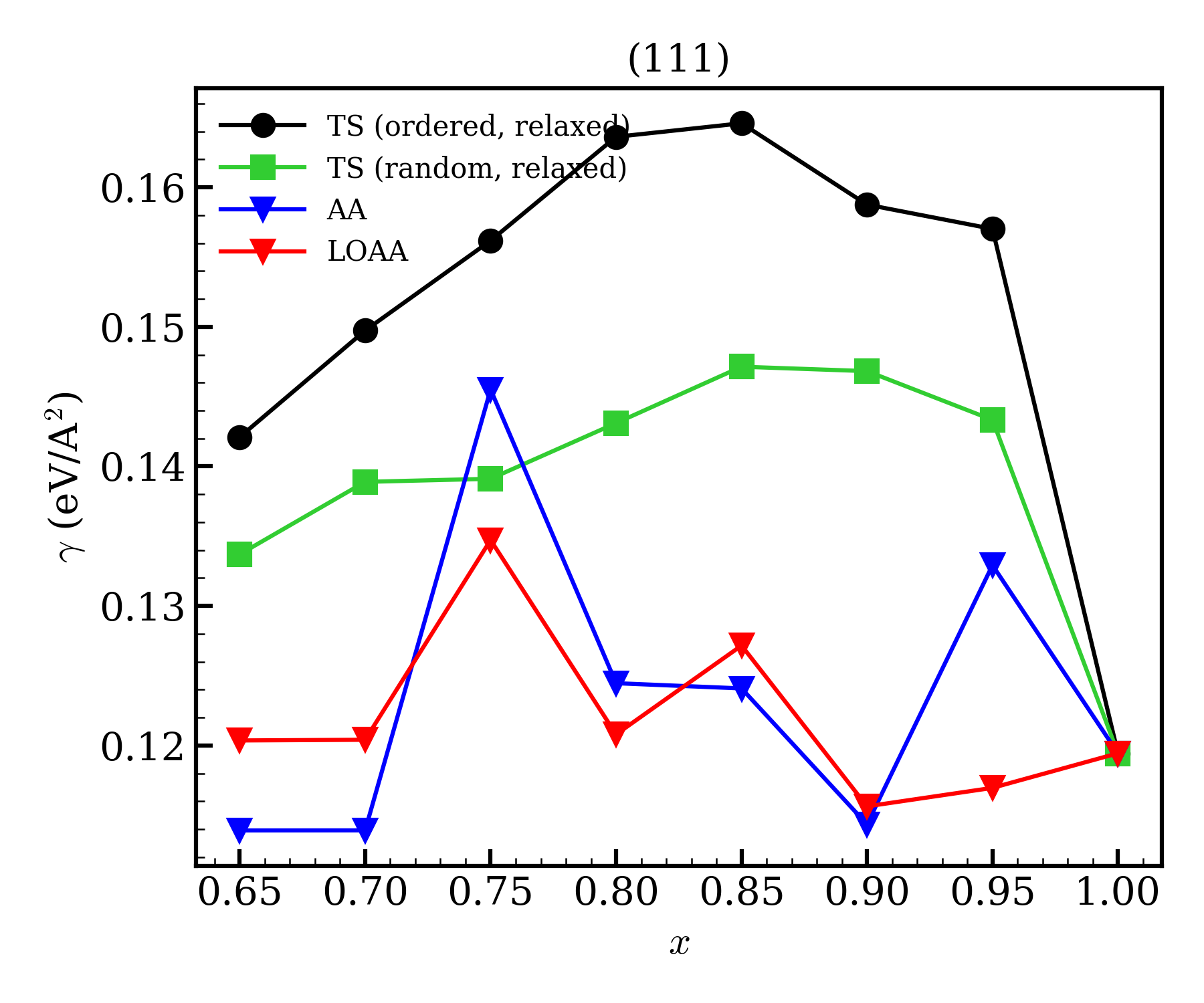}
\caption{}
\end{subfigure}
\hfill
\begin{subfigure}{0.45\textwidth}
\centering
\includegraphics[width=\linewidth]{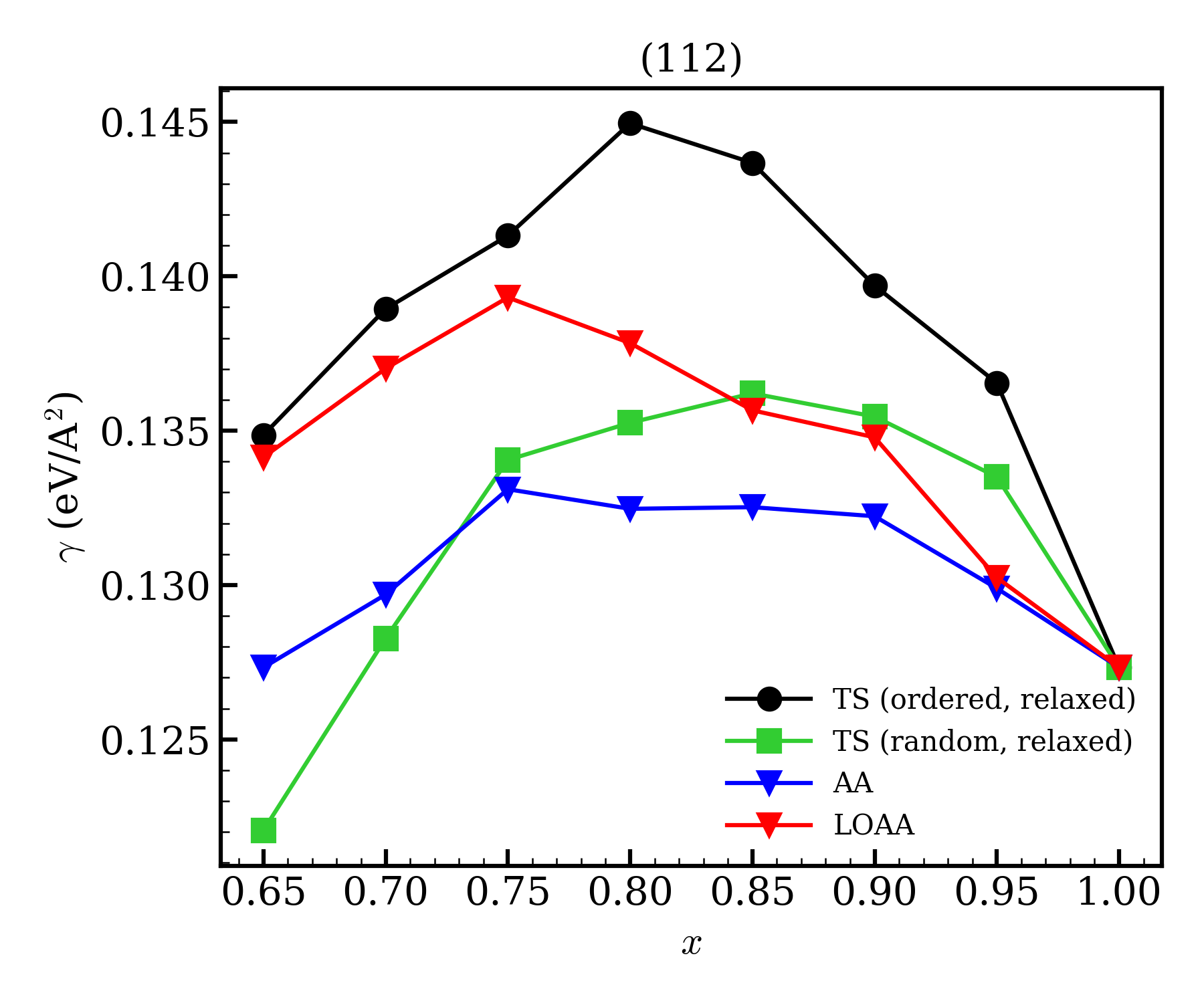}
\caption{}
\end{subfigure}
\caption{Comparison of TS, AA and LOAA surface energy calculations for 3D fcc EAM \nixal\ alloys. Surfaces (a) (100), (b) (110), (c) (111) and (d) (112). TS quantities evaluated for relaxed structures.
\newline
\textit{Alt text}: Line plots with four curves per plot depicting the relaxed TS ordered, relaxed TS random, AA, and LOAA surface energies for \nixal\ as $x$ increases from $0.65$ to $1.0$. The curves do not exhibit good agreement across the plotted range until $x=1.0$. }
\label{fig:NiAl-relax}
\end{figure}

For the 3D \nixal system, in the MS we examine the surface energies of the (100), (110), (111), and (112) surfaces for random and ordered TS samples compared with AA and LOAA results. These results are for unrelaxed systems where we do not allow atoms to move off of their perfect lattice positions, because we have seen that AA and LOAA cannot fully capture this relaxation effect. The surface energy plots for relaxed samples are included here in \fig{fig:NiAl-relax} to further demonstrate this point. Here, the bulk crystals as well as the surfaces were allowed to relax before surface energy was computed. Note that here, both random and ordered TS results are only averaged over five surfaces from a single sample. 

\bibliography{loaa}
\nocite{cowley:1965}
\nocite{norman:warren:1951}
\nocite{rao:curtin:2022}